\documentclass[iop]{emulateapj}

\usepackage[utf8]{inputenc}
\usepackage{graphicx}
\usepackage{amsmath}
\usepackage{mathrsfs}
\usepackage{natbib}
\usepackage{array}
\usepackage{hyperref}

\usepackage{xcolor}

\graphicspath{{figures/}}

\def\imagetop#1{\vtop{\null\hbox{#1}}}

\slugcomment{Submitted to ApJ}

\shorttitle{Concurrent effects of gas drag and planet migration on pebble accretion}
\shortauthors{C. Surville et al.}

\begin{document}

\title{Concurrent effects of gas drag and planet migration on pebble accretion}

\author{Cl\'{e}ment Surville, Lucio Mayer}
\affil{Institute for Computational Science, University of Zurich, Winterthurerstrasse 190, 8057 Zurich, Switzerland}
\email{clement.surville@physik.uzh.ch}
\and
\author{Yann Alibert}
\affil{Physikalisches Institut \& Center for Space and Habitability, Universität Bern, 3012 Bern, Switzerland}

\begin{abstract}
We study the effect of a migrating planet ($10<M_p<20$ Earth mass) on the dynamics of pebbles in a radiative disk using 2D two-fluid simulations carried out with the RoSSBi code. The combined action of the waves induced by the migrating planet and drag back-reaction on the gas produces a complex evolution of the flux of pebbles. The waves excite zonal flows which act as dust traps, and accumulate pebbles. Owing to drag back-reaction, a Kelvin-Helmholtz instability develops, generating turbulent dust rings containing several Earth masses of solids, where planetesimal formation is likely. The more massive the planet, the faster the dust rings develop. In competition with planet migration, this process triggers several dust rings, with a radial separation scaling as $1/M_p$. We discover a transition around $13$ Earth masses. Below that, pebbles are stopped at the inner side of the planet's orbit, but pebble accretion on the planet is sustained. Above that, drag back-reaction in the dust ring region flattens the pressure profile while the planet migrates, which limits further growth by pebble accretion. This new {\it pebble isolation mass} is about a factor of 2 lower than often reported in the literature. This reduces the overall formation timescale of Super-Earth planets, while favoring their survival after the disk's dissipation because further accretion is stifled. Finally, our results support an hybrid model for the formation of Jupiter via both pebbles and planetesimals accretion.
\end{abstract}


\section{ Introduction }
\label{Sect_Intro}

	For a long time, the classical and most accepted model to explain the formation of planets and giant planet embryos, the core accretion model, relied on the accretion of large bodies (hundred of meters to hundred of kilometer in size) called planetesimals. The timescale of growth given by this model is however longer than the constrains given by observations, and by the lifetime of disks (a few My). More recently, it has been proposed that accreted solids could be much smaller bodies, centimeter in size, called pebbles. Due to their size, the pebbles are strongly affected by the gas drag, and flow radially to the inner parts of the disk. If a planet core is growing in such a disk, the accretion of these flowing material could speed up the growth of the planet compared to what is typically observed in planetesimal-based models \citep{Lambrechts2012}.

	Despite it appears as a reliable answer to the problem of timescale, the pebble accretion model is too efficient and produces massive planets as a common outcome. The population of exoplanets as well as the Solar system itself do not have such a distribution of planet masses. Indeed, observational data show that a third of Sun-like stars hosts super-Earth planets \citep{Zhu2018}. Processes that reduce or stall the pebble accretion are necessary to explain or even mimic these observations. A promising way to avoid accretion onto the forming planets is the presence of dust traps, where the radial flow of the pebbles is reduced.

	When the planet is massive enough, the formation of a gap in the gas disk is proposed as a possible dust trapping region \citep{Lambrechts2012}. The planet mass at which the outer gap edge locks the solids is the so-called 'isolation mass' from pebble accretion. Simple models of isothermal disks suggest that this mass is about $20$ Earth masses for a MMSN disk \citep{Lambrechts2012, Lambrechts2014, Lambrechts2014a}, at a canonical distance of $5$ au from the star. While this scenario would produce more mini-Neptunes than without isolation process, it is still a too massive mass. In fact, a $20$ Earth mass embryo can still enter into a runaway accretion phase easily, before the disk dissipates, and thus produce a giant planet with a too high efficiency. More details about the physics of the isolation mass process are needed, and this is the topic of this paper.

	Another aspect related to the radial migration of pebbles and solids in the disk is the distribution of chondrites in the Solar system. There is evidence that the populations in the inner Solar system have a different chemical composition than the ones in the outer Solar system \citep{Leya2008, Trinquier2007}. We could explain this dichotomie by a separation due to a change in the flow of the solids, e.g. a trapping in the middle of the Solar nebula. The formation of Jupiter has been invoked as a source of such a separation. However, the dating of the chondrites suggests that such a separation should occur within less than a million year after the formation of the Sun, typically a couple of hundred thousand years. This is a strict constrain on the formation timescale of Jupiter. While an isolation from radial inflow of the solid can be achieved by a $20$ Earth masses (the classical isolation mass), the timescale needed to form such a planet can be too long. We seek for a process that can isolate the solids from migrating inward earlier in the life of the disk, for example if a less massive planet core can produce such an isolation.

	The paper is organized as follows. In Section \ref{Sect_Methods} we present the disk model and the physical processes taken into account. Section \ref{Sect_Results} gives an overview of the numerical results and highlights the main outcomes. The physical interpretations and implications on pebble accretion, isolation mass, and planet formation scenarios of these results are given Section \ref{Sect_Discussion}. Finally, Section \ref{Sect_Conclusions} assembles our findings and proposes a new consistant scenario which answers the above mentioned questions.

\section{ Disk properties and methods }
\label{Sect_Methods}

\subsection{ Gas and solid particles properties }

	We consider a 2D protoplanetary disk around a solar mass star (of mass $M_s$), which imposes an angular frequency at a distance $r$ from the center of mass $\Omega_k(r) \propto (r/r_0)^{-1.5}$. The gas surface density and temperature of the axisymmetric equilibrium state are $\sigma_0(r) \propto (r/r_0)^{\beta_{\sigma}}$ and $T_0(r) \propto (r/r_0)^{\beta_{T}}$, respectively. The gradients are chosen to fit better the observations than the classical MMSN model, by setting $\beta_{\sigma}=-1$, and $\beta_{T}=-0.5$. The isothermal disk scale height, $H_0=\sqrt{P_0/\sigma_0}/\Omega_k$, which normalizes the pressure, is $H_0(r_0)=0.05 \: r_0$. We finally define the reference radius $r_0=5 \: \text{au}$, and the gas surface density at this location ${\sigma_0(r_0)=152 \: \text{g.cm}^{-2}}$.
	
	The gas of the disk follows the ideal gas model, and its dynamical evolution respects the inviscid compressible Euler equations. The inviscid assumption is justified by the low level of turbulent angular momentum transport in that region of the disk. The thermodynamical evolution of the gas is given by the conservation of kinetic and internal energy. In order to explore deviation from the adiabatic assumption, we impose a source of dissipation of the internal energy in the form of thermal relaxation,
%
\begin{equation} 
	\partial_t P \propto \sigma \frac{\Omega_k(r)}{\tau_c} \left[T-T_0(r)\right],
\end{equation}
with $\tau_c$ a parameter without unity that defines how fast the temperature relaxes to the background equilibrium profile. An infinite value will converge to the adiabatic assumption, while a canceling value converges to locally isothermal assumption. In this study, we will consider long cooling timescales, with $\tau_c>1$. For convenience we can also define a cooling frequency, $\omega_c=\Omega_k(r)/\tau_c$. This thermal relaxation term is implicitly integrated in time in the code RoSSBi, so the numerical method is stable over the time step of the scheme even for large cooling frequencies.

	In addition to the gas, the protoplanetary disk contains solid material, in the form of spherical grains. We consider grain sizes smaller than the mean free path of the gas molecules, so that the Epstein drag law is valid. In this context, we can define a Stokes number for a group of grains as
%
\begin{equation} 
	St=\frac{\pi}{2} r_s \rho_s \sigma_g^{-1},
\end{equation}
with the grain radius $r_s$ in cm, $\rho_s=3 \: \text{g.cm}^{-3}$ being the internal density of the grain material, and $\sigma_g$ the local gas surface density. For the equilibrium profile of the gas, the choice of $r_s$ set the values of the Stokes number at $r_0$. The degeneracy of the Stokes number with the choice of $r_s$ and $ \rho_s$ allows to consider the results identical as long as $r_s \rho_s$ is the same. For example with $St=0.05$, the results of the simulations can be interpreted as with grains of $r_s=1.61 \: \text{cm}$ and $\rho_s=3 \: \text{g.cm}^{-3}$ or for a lighter material with $\rho_s=1 \: \text{g.cm}^{-3}$ and $r_s=4.83 \: \text{cm}$. In a nutshell, the choice of Stokes numbers used in this study corresponds to dust grains that fall into the category of the so-called pebbles. 

	The drag force, due to aerodynamical friction between gas and dust particles is expressed in the Epstein regime as 
%
\begin{equation} 
	\vec{f}_{aero}= - \sigma_p \Omega_k(r) St^{-1} (\vec{V}_p - \vec{V}_g),
\end{equation}
with $\sigma_p$ denoting the local dust surface density, $\vec{V}_p$ and $\vec{V}_g$ as the dust and gas velocity fields, respectively. With $St<1$ in the disk region of interest, the dynamics of the solid particles is well described by a pressure-less fluid, following the compressible Euler equations. Then gas and dust fluids are coupled together by the drag force, $\vec{f}_{aero}$ acting on the dust and $-\vec{f}_{aero}$ on the gas. The numerical scheme for the two-fluid interaction of the code RoSSBi is described in \cite{Surville2019}. Finally the initial distribution of pebbles follows the surface density profile
%
\begin{equation} 
	{\sigma_p(r)|}_{t=0}= 0.5 \% \: \sigma_0(r).
\end{equation}

\subsection{ Planet and star dynamics }
	The presence of a planet in the disk is modeled by a gravitational field giving at a position $\vec{r}$ a force in the form
%
\begin{equation}
	\vec{g}_{pla}(\vec{r}) = -\frac{G M_p}{{\left[|\vec{r} - \vec{r}_p|^2 + l^2(r_p) \right]}^{3/2}} \left(\vec{r} - \vec{r}_p \right).
\end{equation}
The planet position is $\vec{r}_p$ and its mass $M_p$. To correct this 2D gravitational field for the lack of the third dimension, a softening length is applied to the gravitational potential, given by
%
\begin{equation}
\label{Eq_Softening}
	l(r_p)= 0.6 \: H_0(r_p),
\end{equation}	
with $r_p$ the radial distance to the center of mass, or in other words the orbit of the planet. Both gas and the fluid of pebbles are affected by the gravitational field of the planet, while the planet is also affected by the gravitational field created by the disk. The distribution of mass of gas and solids creates a gravity on the planet given by

\begin{equation}
	\vec{g}_{disk}(\vec{r}_p) = -\iint \frac{G (\sigma_g + \sigma_p) dS }{{\left[|\vec{r}_p - \vec{r}|^2 + l^2(r) \right]}^{3/2}} \left(\vec{r}_p - \vec{r} \right).
\end{equation}

	The integral is running on the whole disk, and for consistency, the softening of the gravity is the same law as for the planet gravity (Eq. \ref{Eq_Softening}).

	Imposing a planet potential in the disk right from the beginning of the simulations can strongly perturb the disk, so it is common to smoothly load the mass of the planet during the first disk rotations. We define the mass loading function as $M_p(t)= M_p f_{load}(t)$, with $f_{load}(t)$ being
	
\begin{eqnarray}
	2 (t \omega_{load})^2, \: &\text{if}& \: t \omega_{load} < 1/2, \\
	1 - 2 (t \omega_{load}-1)^2, \: &\text{if}& \: 1/2 < t \omega_{load} < 1, \\
	1, \: &\text{if}& \: 1< t \omega_{load}.
\end{eqnarray}	

	This function is continuous and derivable, which are two necessary properties to obtain a stable numerical integration. The period of loading is of $10$ disk rotations, i.e. $\omega_{load}= \Omega_k(r_0)/10$.
	
	The evolution of the planet position, $\vec{r}_p$, is obtained by solving the equations of motion in a reference frame of origin the center of mass of the star/planet system. We neglect the effect of the disk mass on the position of the center of mass of the whole system. The position and velocity of the star, $\vec{r}_s$, and $\vec{V}_s$, are derived from the ones of the planet following
	
\begin{eqnarray}
	& M_s \vec{r}_s + M_p(t) \vec{r}_p = \vec{0}, \\
	& M_s \vec{V}_s + M_p(t) \vec{V}_p + \vec{r}_p \partial_t M_p(t) = \vec{0}, 
\end{eqnarray}
which are the conditions for the origin of the reference frame to be at rest. We finally solve the equations of motion of the planet using a second order Leapfrog method, over the time step of the hydrodynamical scheme. The Leapfrog scheme has the advantage to bound energy compared to Runge-Kutta methods, and to keep a good accuracy of the orbit of the planet. This is particularly true when using the time step of the hydrodynamical scheme, which is order of magnitude smaller than what could be used with the LeapfFrog scheme.
	Ultimately, the gravity of the star onto the disk is corrected from the change of the star position by an additional force included in the fluid equations
\begin{equation}
	\delta \vec{g}_{s}(\vec{r}) = -\frac{G M_s}{|\vec{r} - \vec{r}_s|^3} \left(\vec{r} - \vec{r}_s \right) + \frac{G M_s}{|\vec{r}|^3} \vec{r}.
\end{equation}
This correction, often called indirect terms of gravity, is significant for planet masses above $10$ Earth masses. We include it for self consistency of the gravitational interactions (only disk self gravity is neglected).

	We have implemented the dynamics of the planet and the star in a new version of the code RoSSBi. The details of the algorithm will be published in a future publication on the numerical method of the code RoSSBi.

\subsection{ Simulation runs }
	
	The simulations presented in this study were performed with the updated version of the code RoSSBi, and were run on the Swiss Piz Daint supercomputer. The disk domain covers the full azimuthal extent and ${[1/3, \: 2] \: r_0}$ in the radial direction. The radial distance is discretized in a logarithmic manner. Unless mentioned in the text, the numerical resolution is ${(N_r, \: N_{\theta})=(3072, \: 2048)}$.
	
	The main parameter we explore is the planet mass. We performed three main runs with a planet mass $M_p$ of $10$, $13$, and $20$ Earth masses. Our strategy was to find the sufficient minimum mass to change and stop the flux of pebbles beyond a certain radius in the disk. The pebble model is a sphere of radius $1.61 \: \text{cm}$ and of material density of $3 \: \text{g.cm}^{-3}$, giving at $r_0$ a Stokes number of $St=0.05$ for the initial condition.
	
	The cooling frequency of the thermal relaxation used in this study is ${\omega_c= 10^{-3} \Omega_k(r)}$ for the run with $10\:M_e$ and ${\omega_c= 10^{-2} \Omega_k(r)}$ for the runs with a bigger planet. Having a sufficient finite cooling is important because of the heating produced in the shocks of the spiral waves excited by the planet. In absence of cooling, the temperature in the disk would increase systematically up to several times the initial temperature profile. Because more massive planets excite stronger waves, we applied a faster cooling for the runs with $13$ and $20$ Earth mass planets. There is no physical argument behind this decision, we choose it to ensure stability and robustness of the numerical results. Finally, the effect of different cooling frequencies on the disk evolution is beyond the scope of this paper, and it will be tested more precisely in a future publication.

\section{ Results }
\label{Sect_Results}

	We present in this section the results of the simulations for the different cases. We follow the evolution of the disk during several thousands of disk rotations, thus we decided to show only two snapshots in order to describe the structures arising in the disk: when significant structures appeared, and at the end of the run. The physical quantities presented are: {\it{(i)}} the density of pebbles normalized to the initial disk profile, {\it{(ii)}} the Rossby number of the gas, defined by ${Ro = \vec{\nabla} \times [\vec{V}_g - \vec{V}_k(r)]/[2 \Omega_k(r)] \cdot \vec{e}_z}$, and {\it{(iii)}} the gas density scaled to the disk background, $\sigma_g/\sigma_0(r) -1$.
	
\subsection{ Results for a 10 Earth mass planet }

\begin{figure*}
	\begin{center}
	\begin{tabular}{ccp{15mm}}
	\scriptsize{$t=2000$ rot} & \scriptsize{$t=3785$ rot} & \\
	\imagetop{\includegraphics[height=7.1cm, trim=4mm 0mm 24mm 3mm, clip=true]{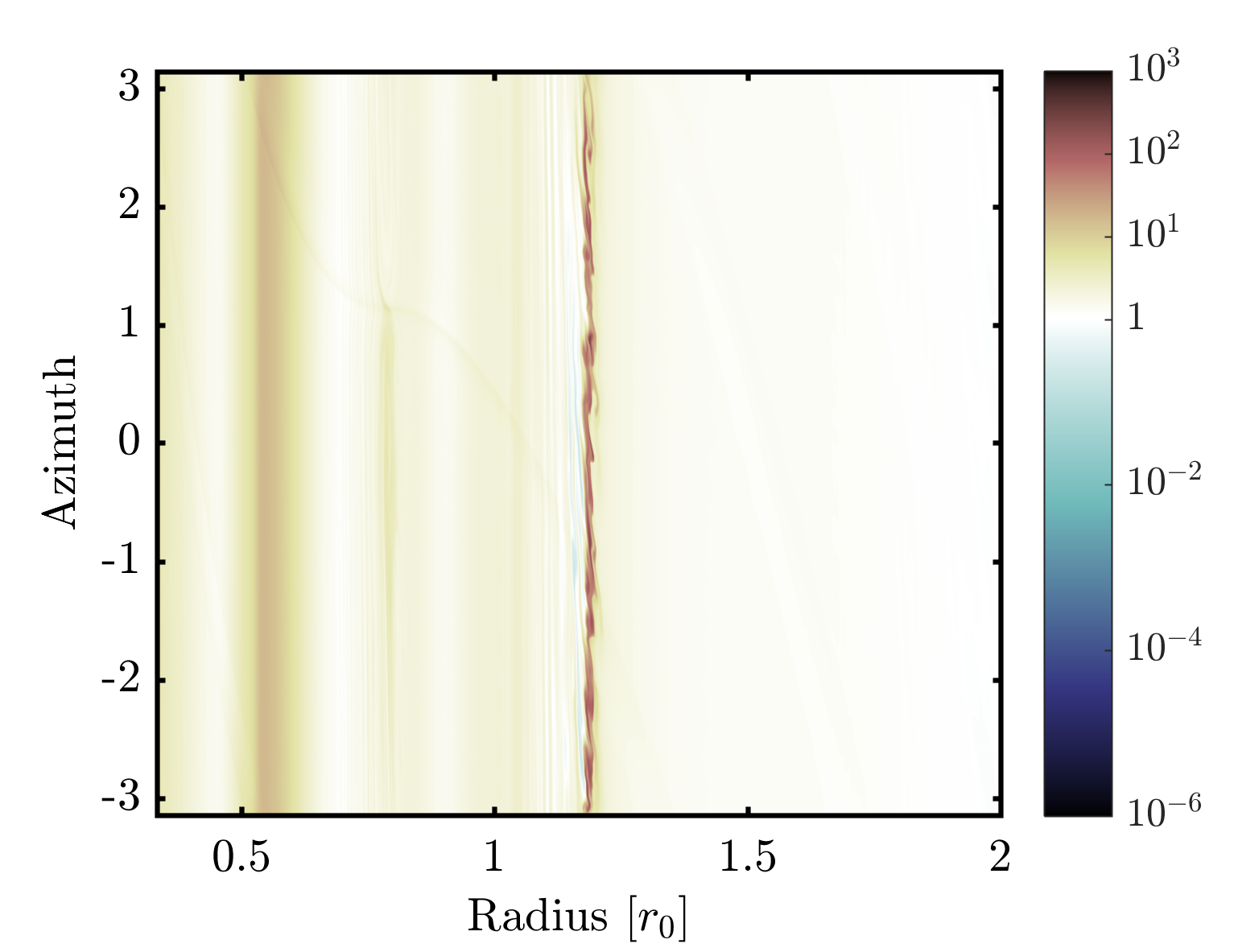}} &
	\imagetop{\includegraphics[height=7.1cm, trim=2mm 0mm 0mm 3mm, clip=true]{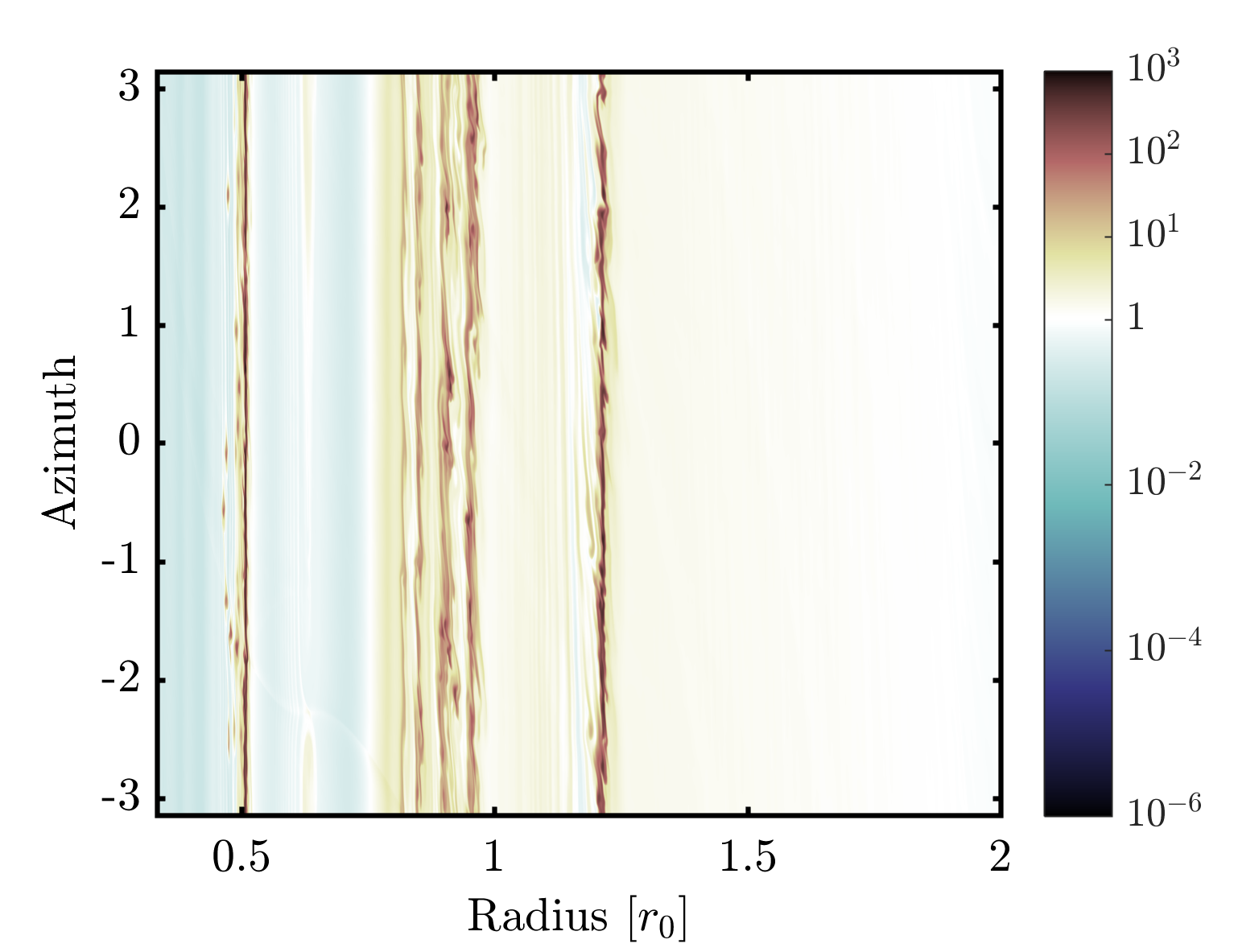}} &
	{\scriptsize{$\:$ \newline \newline (a)}} \\ 
	\imagetop{\includegraphics[height=7.1cm, trim=4mm 0mm 24mm 3mm, clip=true]{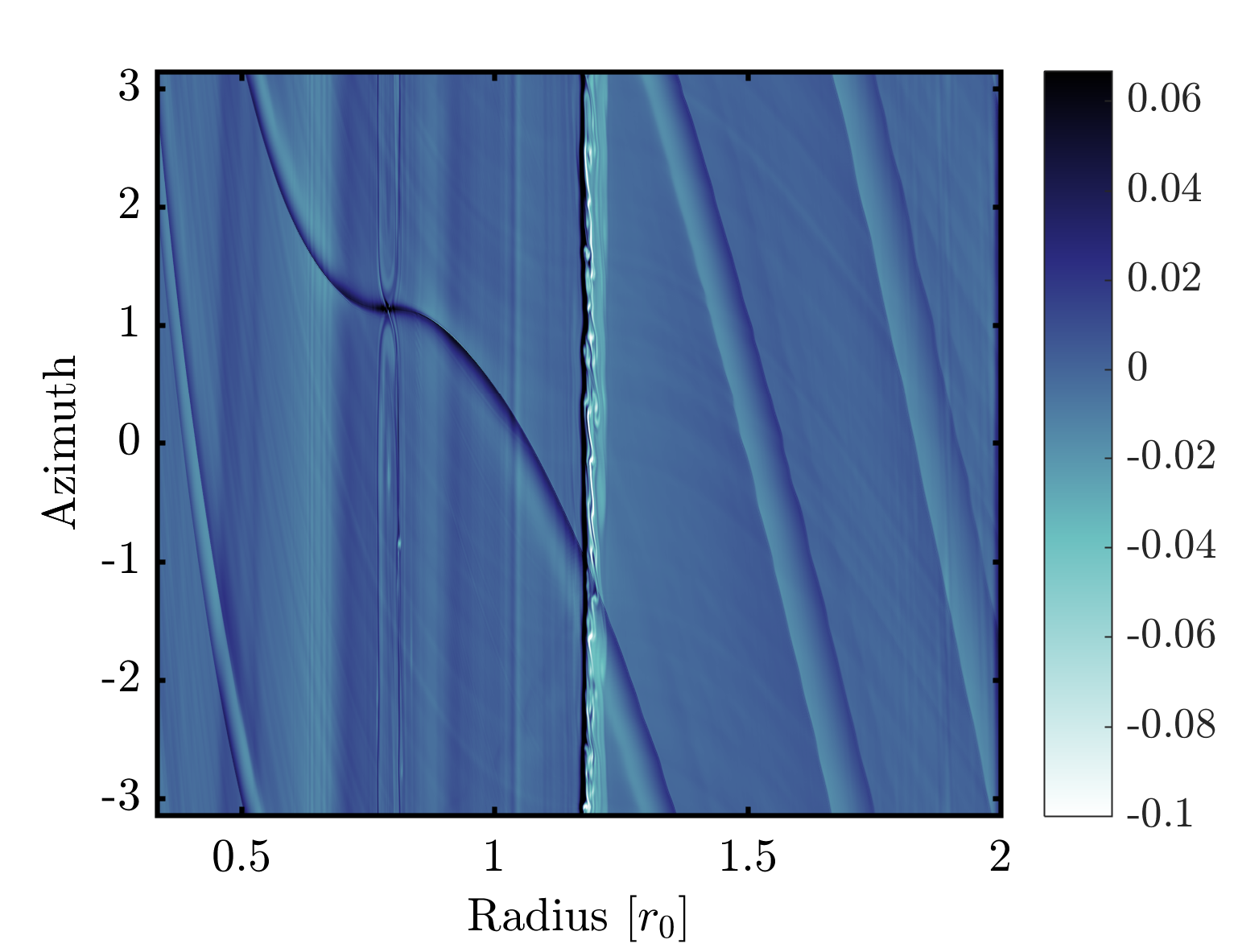}} &
	\imagetop{\includegraphics[height=7.1cm, trim=2mm 0mm 0mm 3mm, clip=true]{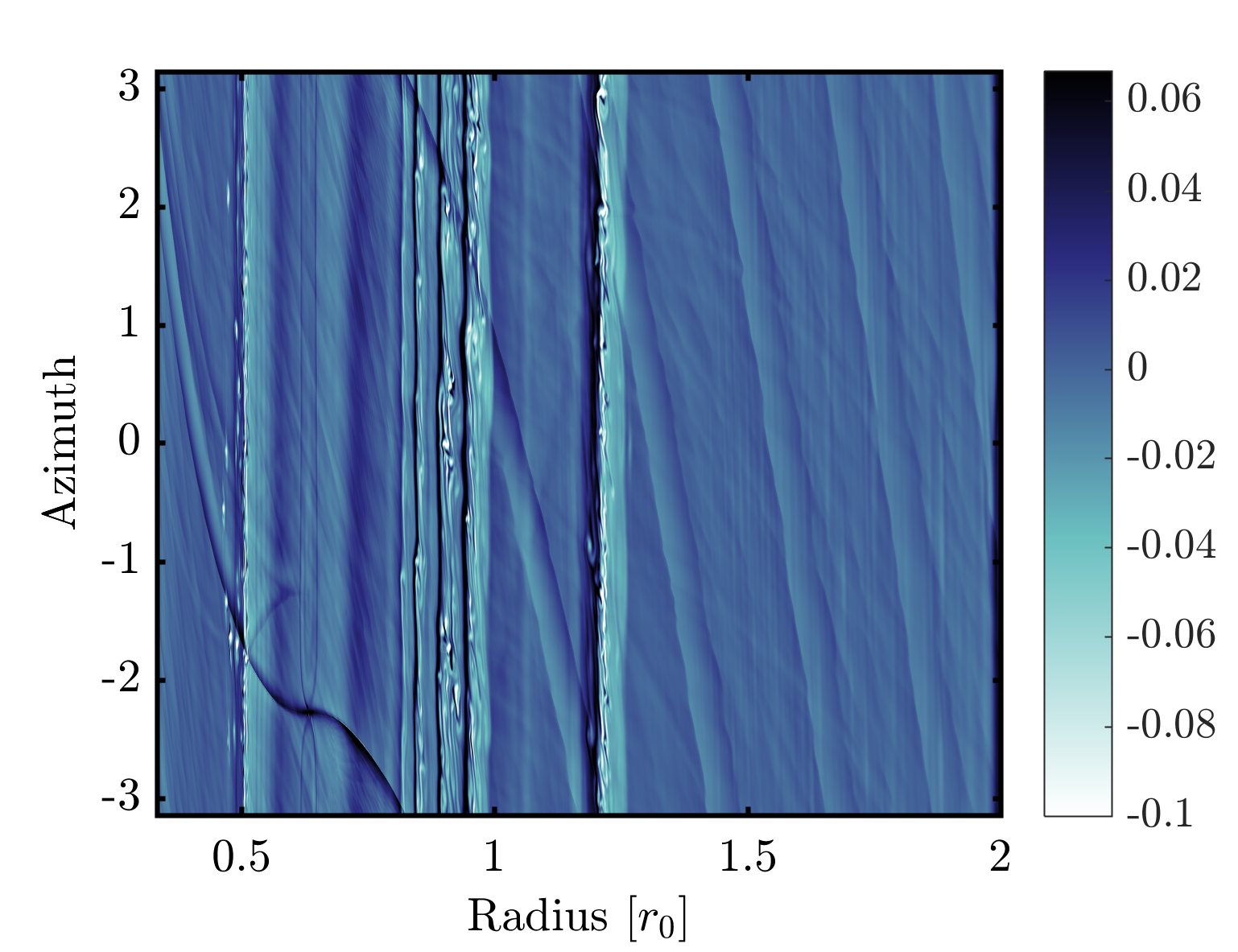}} &
	{\scriptsize{$\:$ \newline \newline (b)}} \\ 
	\imagetop{\includegraphics[height=7.1cm, trim=4mm 0mm 24mm 3mm, clip=true]{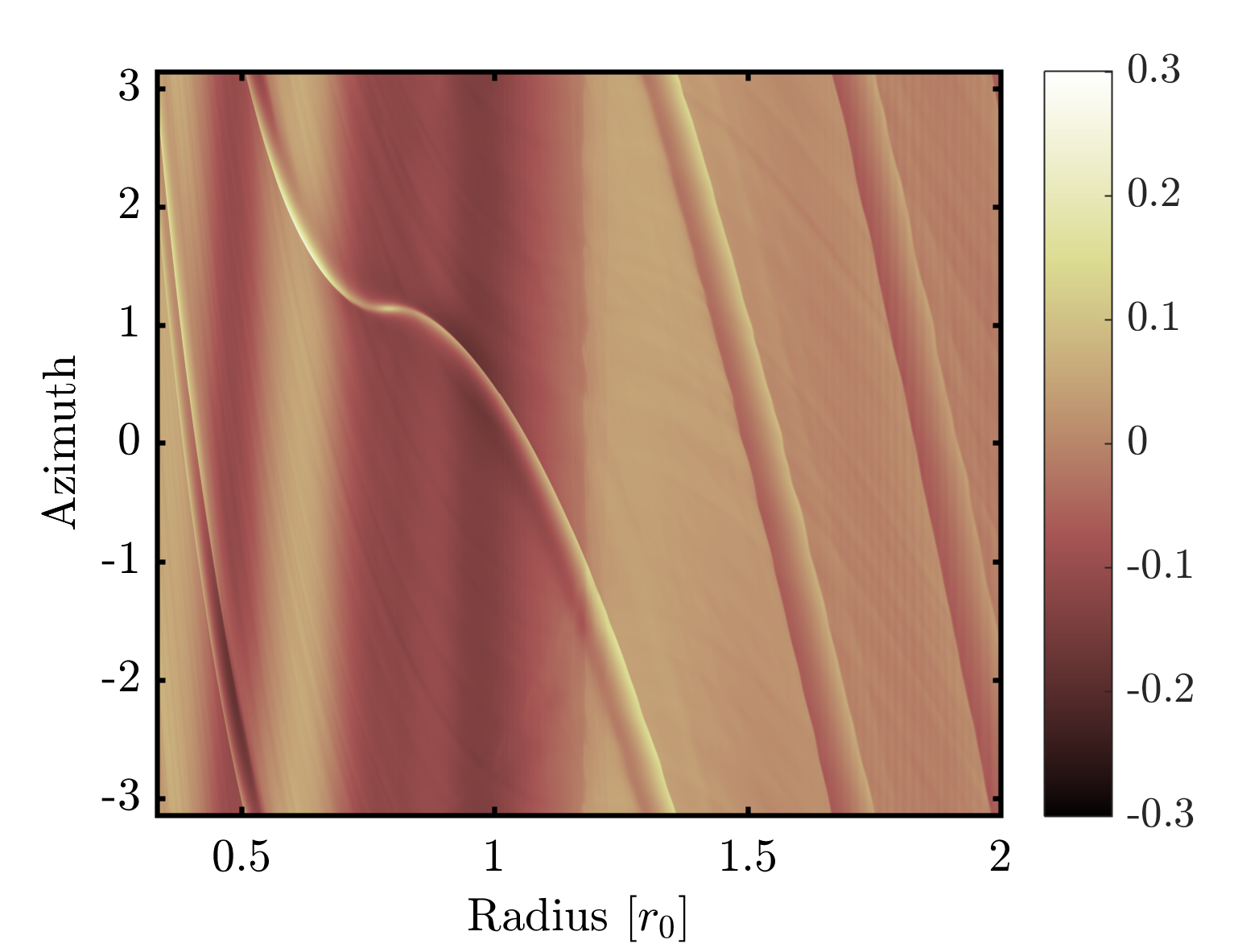}} &
	\imagetop{\includegraphics[height=7.1cm, trim=2mm 0mm 0mm 3mm, clip=true]{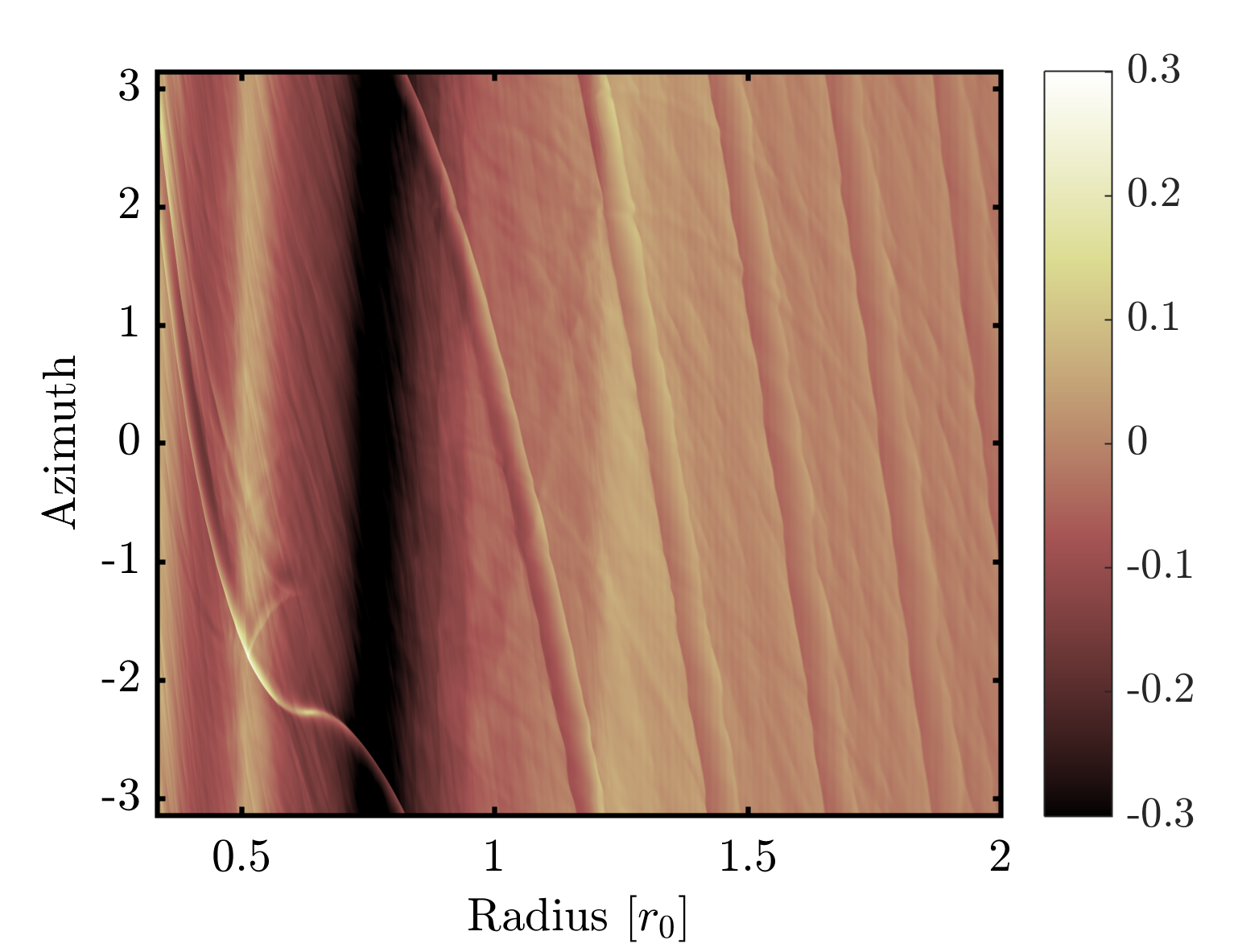}} &
	{\scriptsize{$\:$ \newline \newline (c)}} \\ 
	\end{tabular}
	
	\caption{\label{Fig_Mp_10_tau_p3_St_005} Disk structure for $M_p = 10\: M_e$, with $St=0.05$, and a cooling rate $\omega_c=10^{-3} \Omega_k$. From top to bottom: We show the dust density scaled to the initial condition (a), the gas Rossby number (b), and the gas density scaled to the disk background (c) respectively. The left column presents the state obtained after $2000$ disk rotations, and the right column the state after $3785$. }
  \end{center}
\end{figure*}

	During the first thousand disk periods of evolution, we observe a moderate change of the disk profile. The planet interacts with the disk by exciting sound waves in the form of spiral wakes, which exchange energy and momentum with the disk. As a consequence, the density of gas is reduced in the co-rotation region by a few percent. This reduction increases in time on a long timescale. Simultaneously, the motion of the gas is affected around this gap, and a region of negative vorticity is growing at the outer border of the gap. In this region around ${r=1.2\: r_0}$, the dust accumulates and the development of a turbulent dust ring starts at $t=1500$ disk rotations.
	
	The state of the disk is shown Figure \ref{Fig_Mp_10_tau_p3_St_005}, on the left, just after the formation of this structure, at ${t=2000}$ rotations. As seen at the top, the density of pebbles is larger than the background by a factor of hundreds in this ring, and the local dust-to-gas ratio is larger than $0.5$. The gas vorticity, shown as the Rossby number on the middle panel, traduces the turbulence sustained in the flow, in a form of patches of strong vorticity ($R_o<-0.1$). The dusty eddies are not clearly separated, but rather organized in a narrow band. It is striking to observe a step in the gas density, bottom panel, at the location of this ring. The gas density contrast between the two sides of the dust ring is around $10 \%$. Despite its relatively low mass, the planet carved a gap in the disk region with ${0.7 <r<1.2 \: r_0}$, where the gas density is reduced by $\sim 10 \%$ from the disk background. At that time, it is not obvious that the density of pebbles is modified in the gap region. 
	
	Interestingly, we observe the presence of an additional gap in the gas around $r=0.5 \: r_0$. This gap is excited by the inner arm of the wake of the planet. The existence of this structure was mentioned in \cite{Bae2017} but in the regime of massive planets ($\sim 100 \: M_e$), and under the isothermal assumption. Moreover, these calculations were done in a disk containing only gas. However, they investigate the same low-viscosity regime as here. In this paper, we show for the first time the existence of such a gap under the presence of pebbles and a light planetary embryo. Between the planet and this gap, the density of pebbles is more than 10 times larger than the disk background, showing that solids also accumulate at the border of this secondary gap.

	Later on, the system is affected by planet migration. The migration regime is linear for such a mass of the embryo, and is sufficiently fast to be visible over the duration of the simulation. For example, at ${t=2000}$ rotations, the orbit of the planet is $r_p=0.8 \: r_0$. This migration is responsible for the repetition of accumulation of pebbles at different radii, because the location of the gap edge is moving with the planet. As a consequence, several dust rings have formed at the end of the run.
	
	On the right of Figure \ref{Fig_Mp_10_tau_p3_St_005}, we show the picture of the disk at ${t=3785}$ disk rotations. Four additional dust rings are present within $0.8<r<1 \: r_0$, where the density of pebbles is enhanced almost by a factor of $10^3$ above the disk background. The one at ${r=1.2\: r_0}$, formed at $t=1500$ disk rotations, is still alive, and did not loose material over more than $2000$ rotations. The four additional generations have formed successively within that time period, from outside to inside, along with the inward migration of the planet. Like for the first dust ring, the flow reveals some turbulence generated in the gas in the form of vorticity (middle panel). Interestingly, there is a quiet region, $1<r<1.2 \: r_0$, where no dust ring has formed. This effect is due to the competition between the planet migration rate, and the timescale needed to trigger the dust ring. We will discuss this aspect in the Section \ref{Sect_Discussion}.
	
	The secondary gap, formed at $r=0.5 \: r_0$, is also a favorable region for triggering a dust ring containing several dusty eddies. The dust-to-gas ratio in this structure is above unity, despite its location in a region depleted from pebbles. In fact, in the domain $r<0.8 \: r_0$, the density of pebbles is reduced hundred times below the initial profile. The existence of such a dust trapping region is shown for the first time, and may have implications that we discuss Section \ref{Sect_Discussion}.
	
	Finally, we observe that the shape of the gap in the gas density is different between the two snapshots (bottom line). While a reduction of only $10 \%$ over the gap was created at ${t=2000}$ rotations (left panel), a much deeper gap is visible at ${t=3785}$ rotations (right panel). Where $0.7<r<0.8 \: r_0$, a depth below $30 \%$ the disk background was created, while the reduction is in average at $\sim 10 \%$ in the rest of the gap. There is a clear asymmetry between the profiles before and after the orbit of the planet. 
	It is striking that the step at $r=1.2 \: r_0$ is still in place, and is associated with the dust ring. Similarly, the bump in gas density at $r=0.5 \: r_0$, where the pebbles can accumulate and form a dust ring, is significantly narrower at the end of the run (right panel) than at ${t=2000}$ rotations (left panel). Unfortunately, we could not continue the run much longer due to computing time limitations. 

\subsection{ Results for a 13 Earth mass planet }

\begin{figure*}
	\begin{center}
	\begin{tabular}{ccp{15mm}}
	\scriptsize{$t=900$ rot} & \scriptsize{$t=2400$ rot} & \\
	\imagetop{\includegraphics[height=7.1cm, trim=4mm 0mm 24mm 3mm, clip=true]{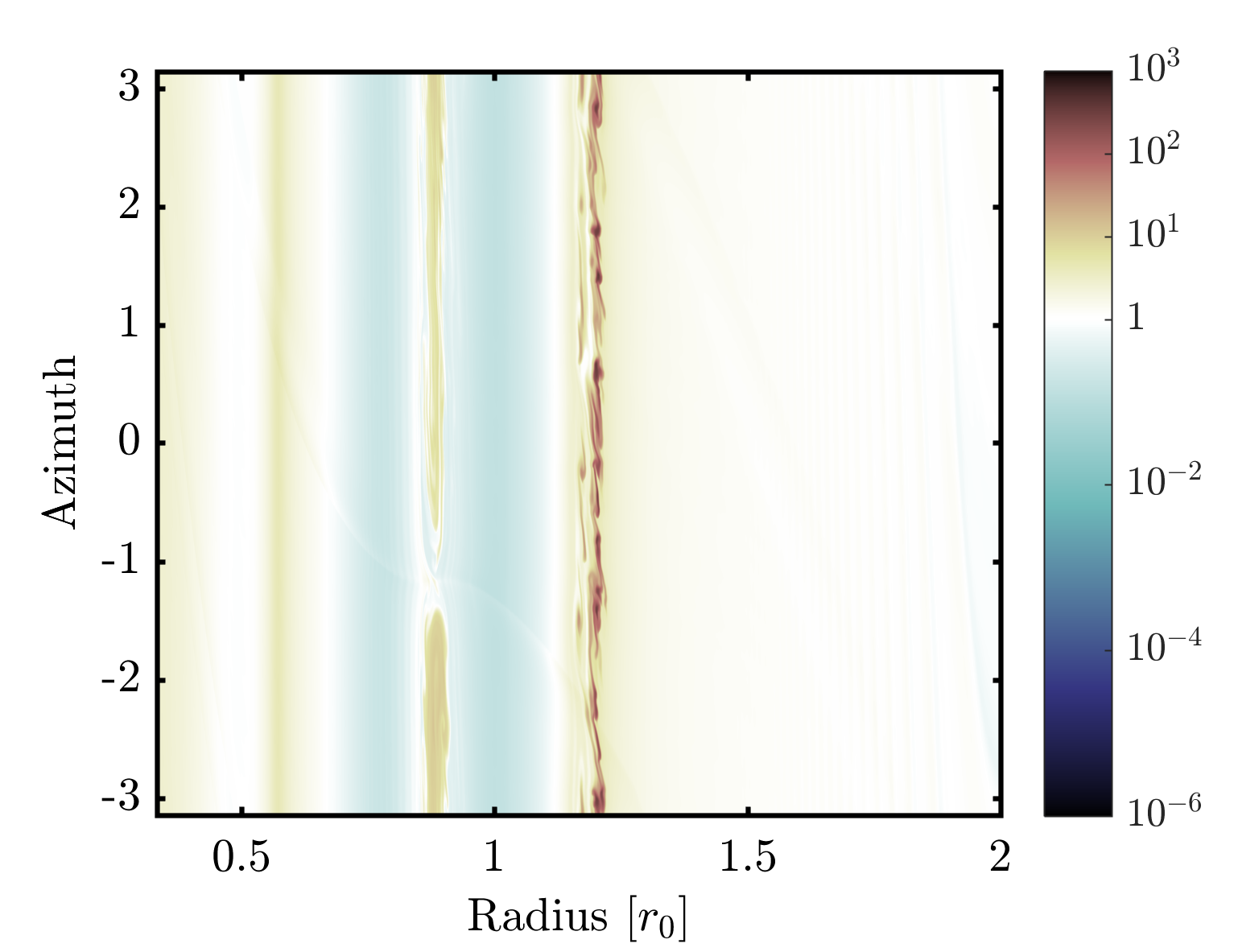}} &
	\imagetop{\includegraphics[height=7.1cm, trim=2mm 0mm 0mm 3mm, clip=true]{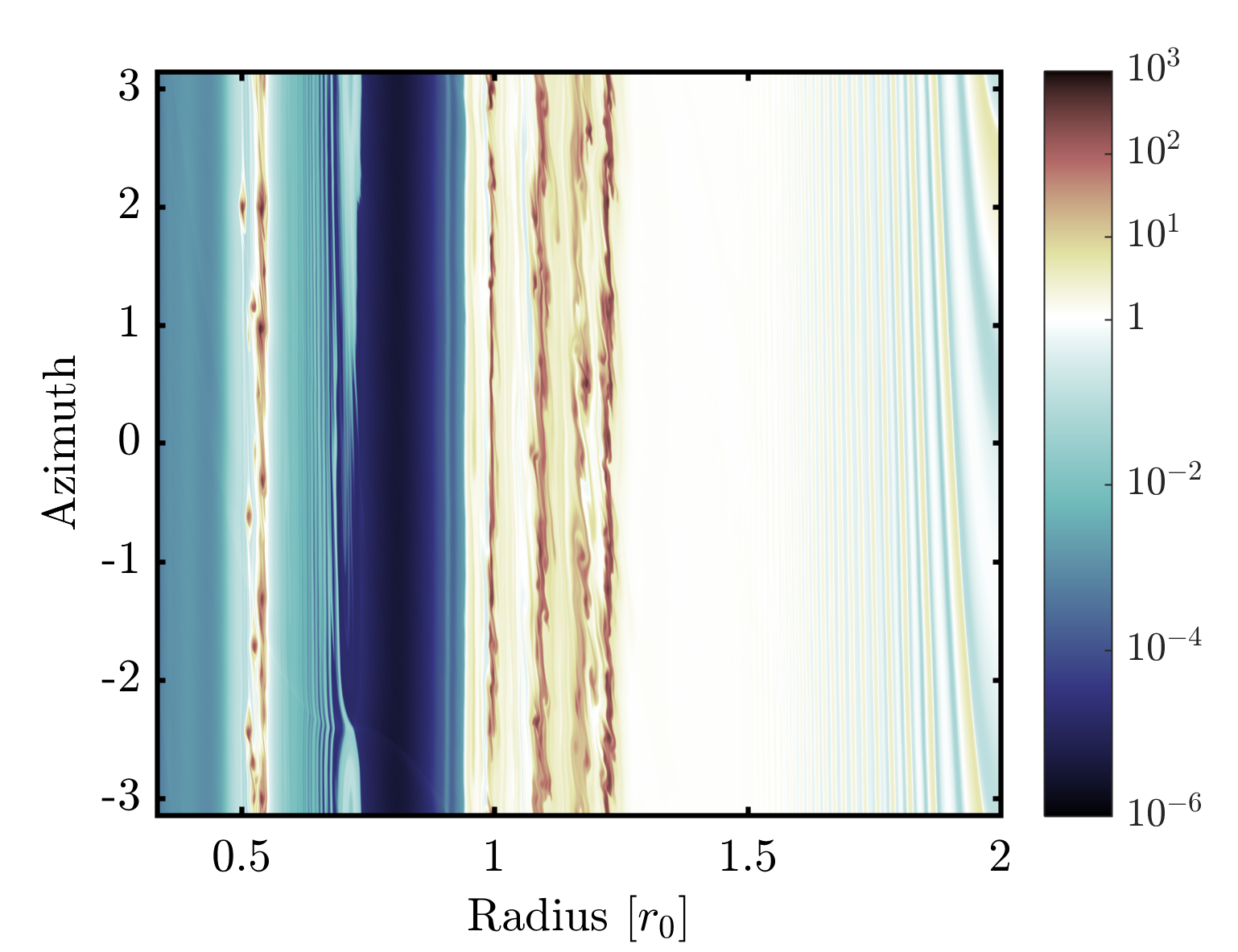}} &
	{\scriptsize{$\:$ \newline \newline (a)}} \\ 
	\imagetop{\includegraphics[height=7.1cm, trim=4mm 0mm 24mm 3mm, clip=true]{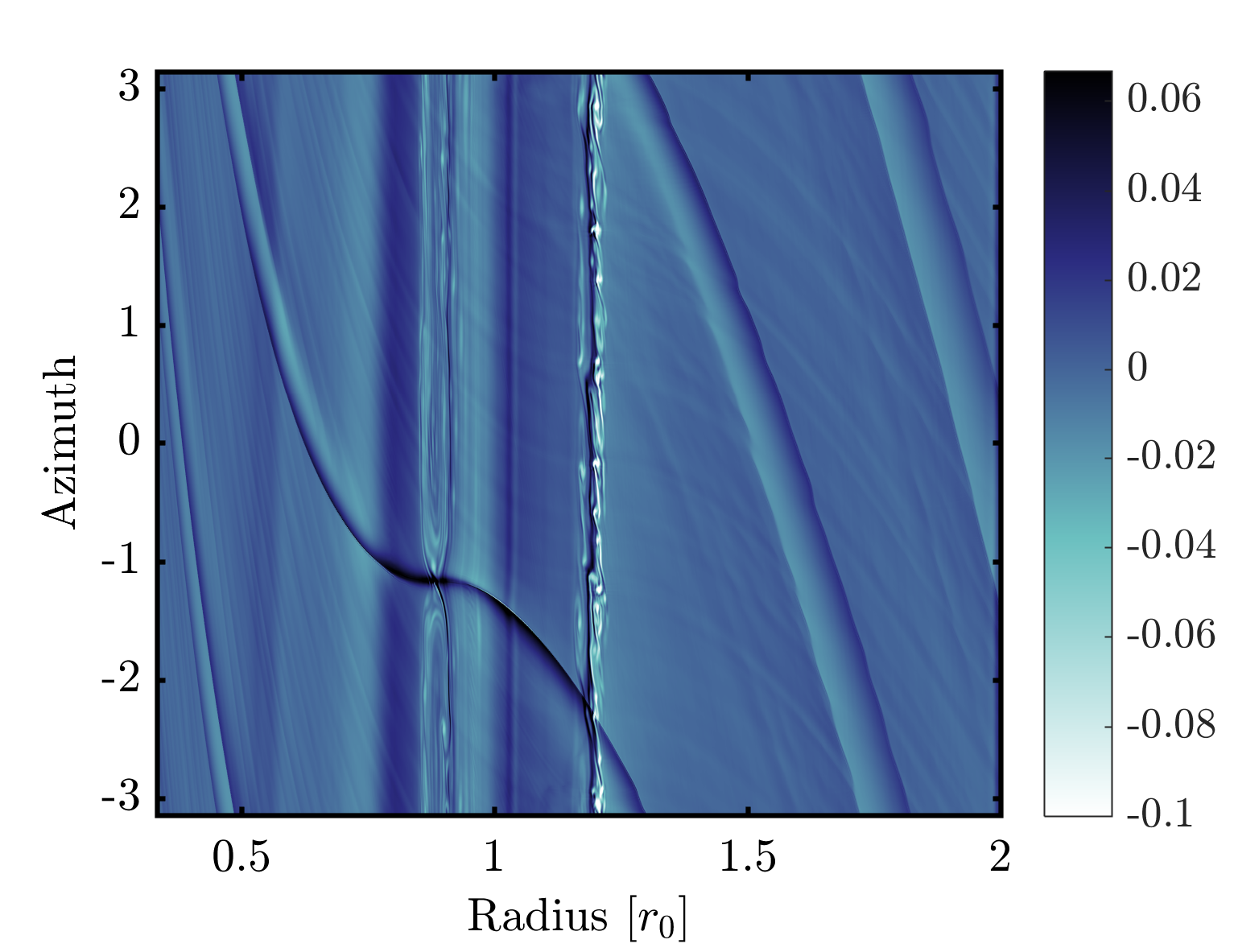}} &
	\imagetop{\includegraphics[height=7.1cm, trim=2mm 0mm 0mm 3mm, clip=true]{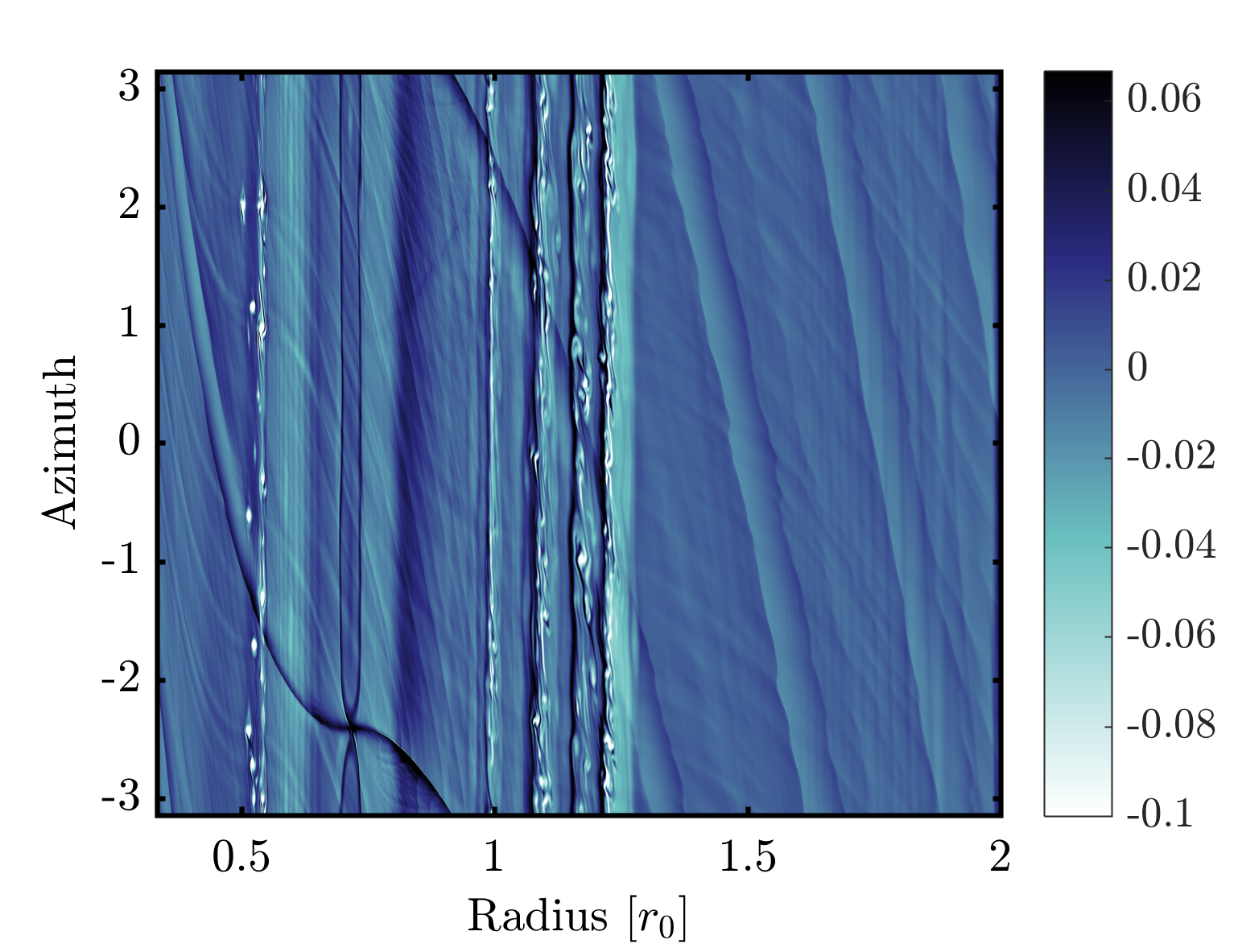}} &
	{\scriptsize{$\:$ \newline \newline (b)}} \\ 
	\imagetop{\includegraphics[height=7.1cm, trim=4mm 0mm 24mm 3mm, clip=true]{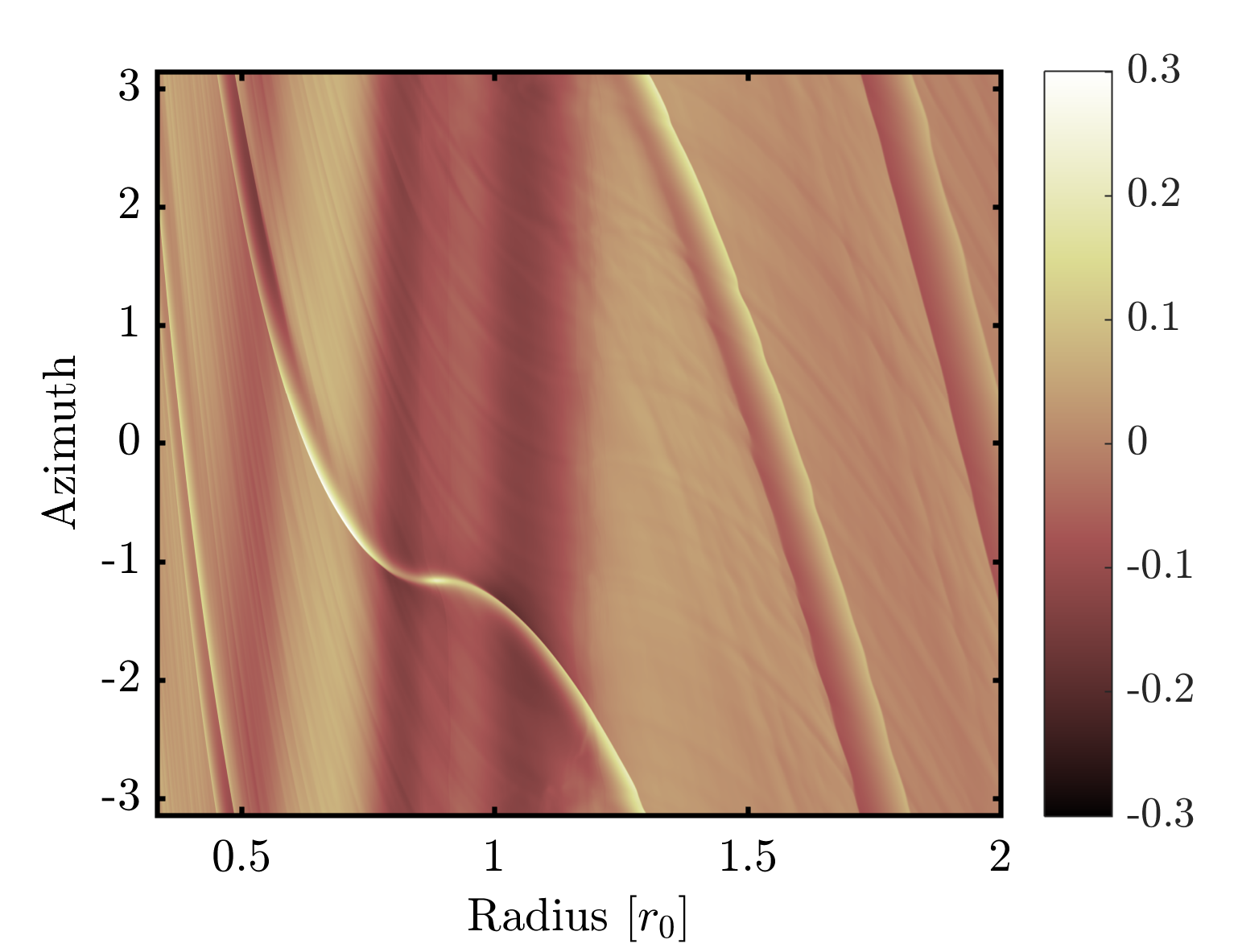}} &
	\imagetop{\includegraphics[height=7.1cm, trim=2mm 0mm 0mm 3mm, clip=true]{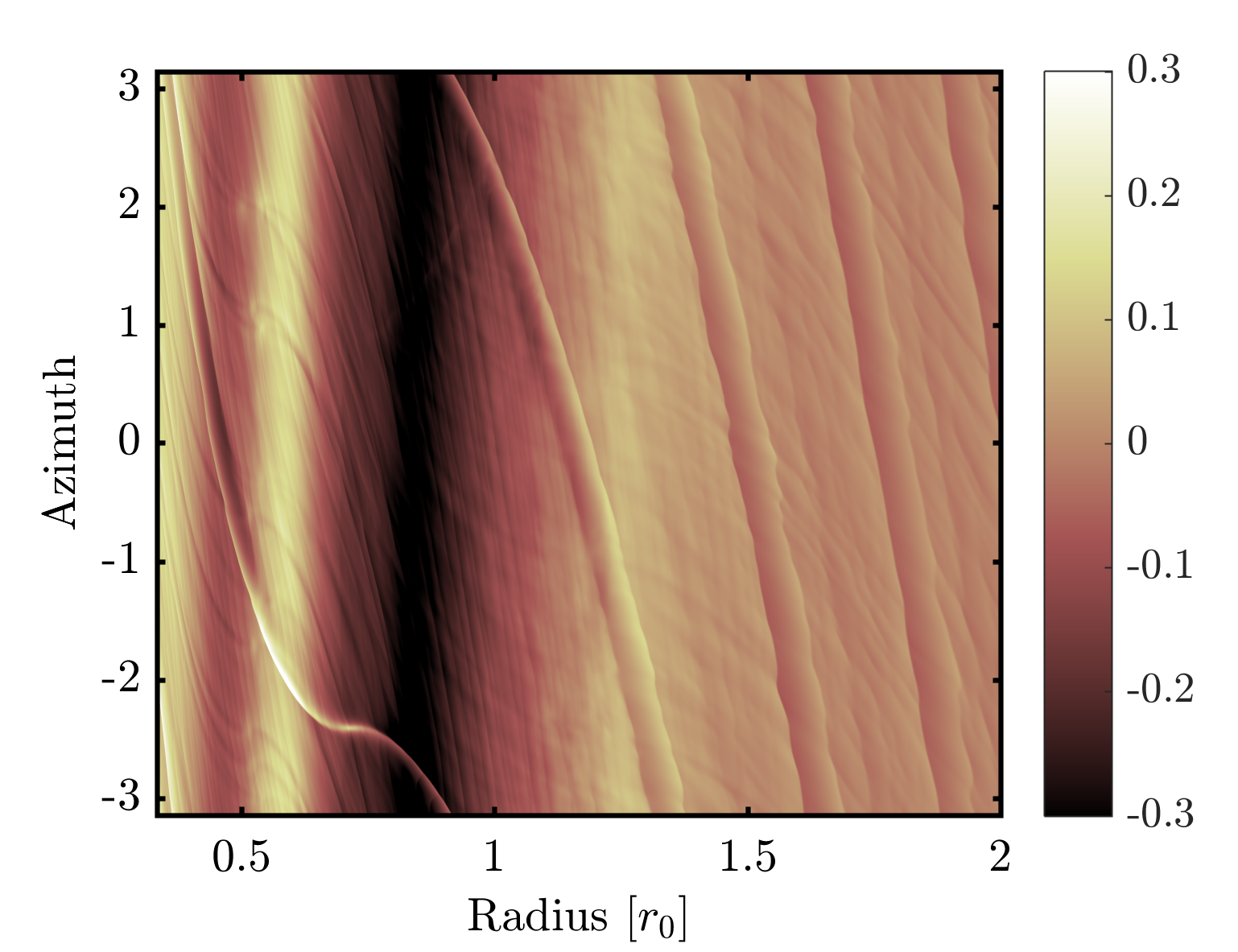}} &
	{\scriptsize{$\:$ \newline \newline (c)}} \\ 
	\end{tabular}
	
	\caption{\label{Fig_Mp_13_tau_p2_St_005} Disk structure for $M_p = 13\: M_e$, with $St=0.05$, and a cooling rate $\omega_c=10^{-2} \Omega_k$. From top to bottom: We show the dust density scaled to the initial condition (a), the gas Rossby number (b), and the gas density scaled to the disk background (c). The left column presents the state obtained after $900$ disk rotations, and the right column the state after $2400$. }
  \end{center}
\end{figure*}

	The results obtained with a 13 Earth mass planet are shown Figure \ref{Fig_Mp_13_tau_p2_St_005}. The evolution of the disk during the first hundreds of orbits is similar to the results obtained with $M_p= 10 \: M_e$, but structures appear faster. In particular, it takes only $800$ disk rotations to trigger a dust ring on the outside of the planet orbit, at $r=1.2 \: r_0$. On the left panels, we show the snapshot of the disk after the formation of the dust ring at $t=900$ disk rotations. Comparing to the case with $10$ Earth masses, it is striking that an increase of only $30\%$ of the planet mass changes the distribution of pebbles so clearly. In fact, a depletion of more than $10$ times the initial values is visible in the gap carved by the planet, whereas no significant change of the dust density was visible after the first dust ring formation in the run with $M_p = 10\: M_e$. Concerning the gap in the gas density, its depth is also about $10-15\%$ under the disk background. The secondary gap at $r=0.5\: r_0$ is formed, but less solids are accumulated at that time at the border of this gap.
	The distribution of pebbles in the dust ring at $r=1.2\:r_0$ is slightly different than in the previous case. About ten individual dense eddies are visible, where density is several hundred times larger than the initial value. The gas vorticity (middle panel) shows the anticyclonic property of these eddies, as white spots on the map. In the run with $M_p= 10 \: M_e$, the density in the dust ring was smoother and more even in space. 
	
	On the right of Figure \ref{Fig_Mp_13_tau_p2_St_005}, we show the picture of the disk at ${t=2400}$ disk rotations, at the end of the run. If we focus on regions of pebble accumulation, the evolution was similar to the previous case. New generations of dust rings have been formed along with the planet inward migration. Three new rings are distributed in the region with $1<r<1.2 \: r_0$ with a surprising regularity. There is no empty space like in the run with $M_p = 10\: M_e$, which shows that these rings could be easily triggered despite the change of the disk's local conditions due to planet migration. This is counter-intuitive as one could expect a wider separation between the rings formed with a more massive planet, because it migrates faster. However, this effect is balanced by a more effective accumulation of the solids, and thus a quicker triggering of the dust ring formation.
	
	The region at the border of the secondary gap, near $r=0.5 \: r_0$, is also a site of accumulation of pebbles. Indeed the highest densities of solids are found in the eddies generated in this dust ring. These eddies are more separated from each other than the ones in the outer region. They form rather a chain of eddies than really a dust ring. They even have individual motion, as some of them detach from the site of formation with an inward migration. However, around this region, and in particular where $r<0.5\:r_0$, the density of pebbles is reduced by at least two orders of magnitude. The reduction of the amount of solids in the inner disk part is more efficient than in the case with $10\: M_e$ planet. 
	
	Concerning the gap open by the planet, we obtain a similar outcome as in the case with $M_p = 10\: M_e$. While the gas density (bottom panel) is reduced by $30\%$ at the maximum in the gap, on the outer side of the planet orbit, the reduction of the density of pebbles is of several orders of magnitude (top panel). From the orbit of the planet, $r=0.7\: r_0$, up to $r=0.95\: r_0$, there is $10^5$ times less solids than at the beginning of the run. Several factors contribute to this reduction. The solids are ejected from the vicinity of the planet by torque and dragged with the gas during the formation of the gap. However, the gap can be refilled by the solids which flow inward from the outer parts of the disk. If this flow reduces or even cancels, then the depletion in the gap is accelerated, and the density of solids drops down. This important issue is one of the focuses of this paper and will be addressed in Section \ref{Sect_Discussion}.

\subsection{ Results for a 20 Earth mass planet }

\begin{figure*}
	\begin{center}
	\begin{tabular}{ccp{15mm}}
	\scriptsize{$t=500$ rot} & \scriptsize{$t=3000$ rot} & \\
	\imagetop{\includegraphics[height=7.1cm, trim=4mm 0mm 24mm 3mm, clip=true]{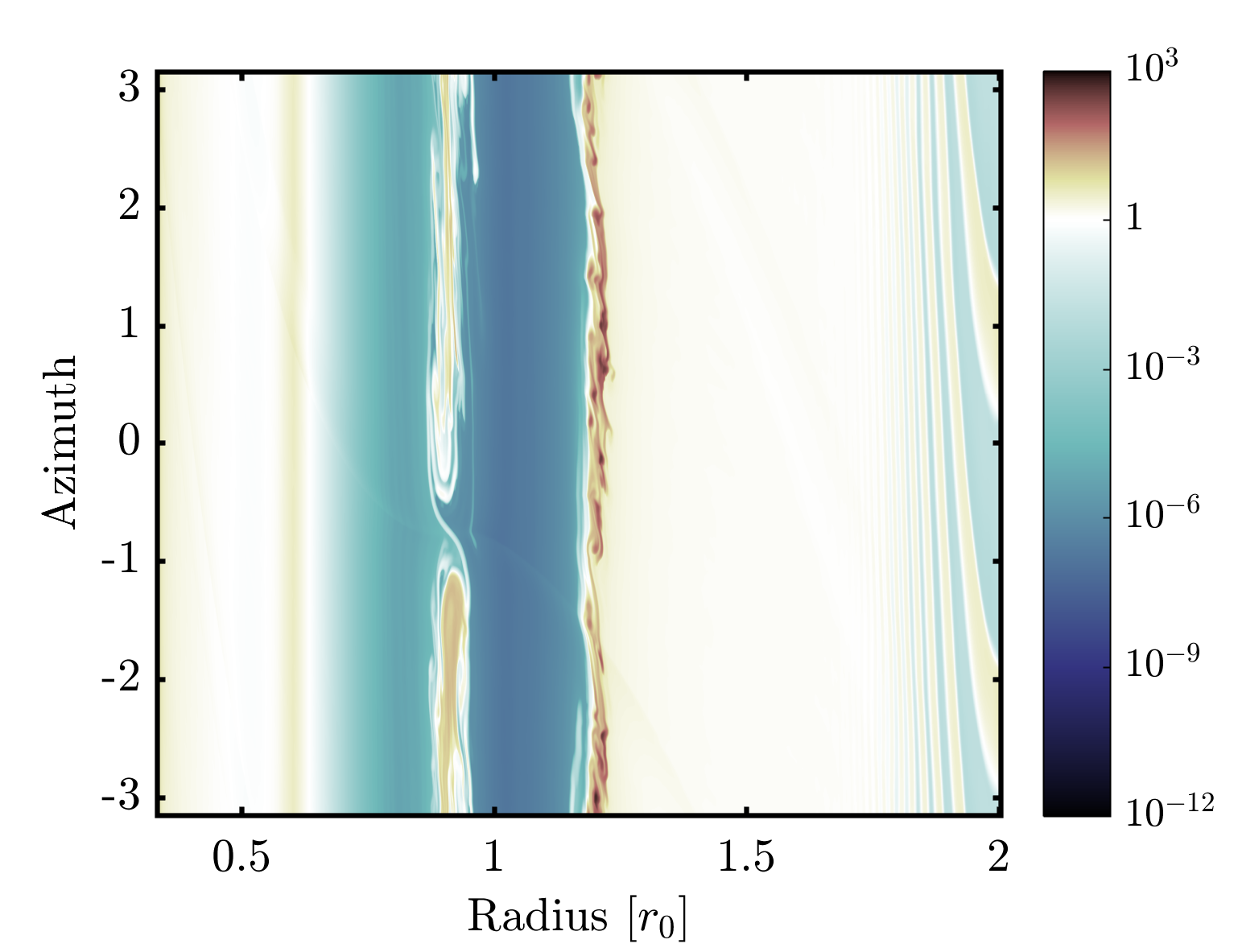}} &
	\imagetop{\includegraphics[height=7.1cm, trim=2mm 0mm 0mm 3mm, clip=true]{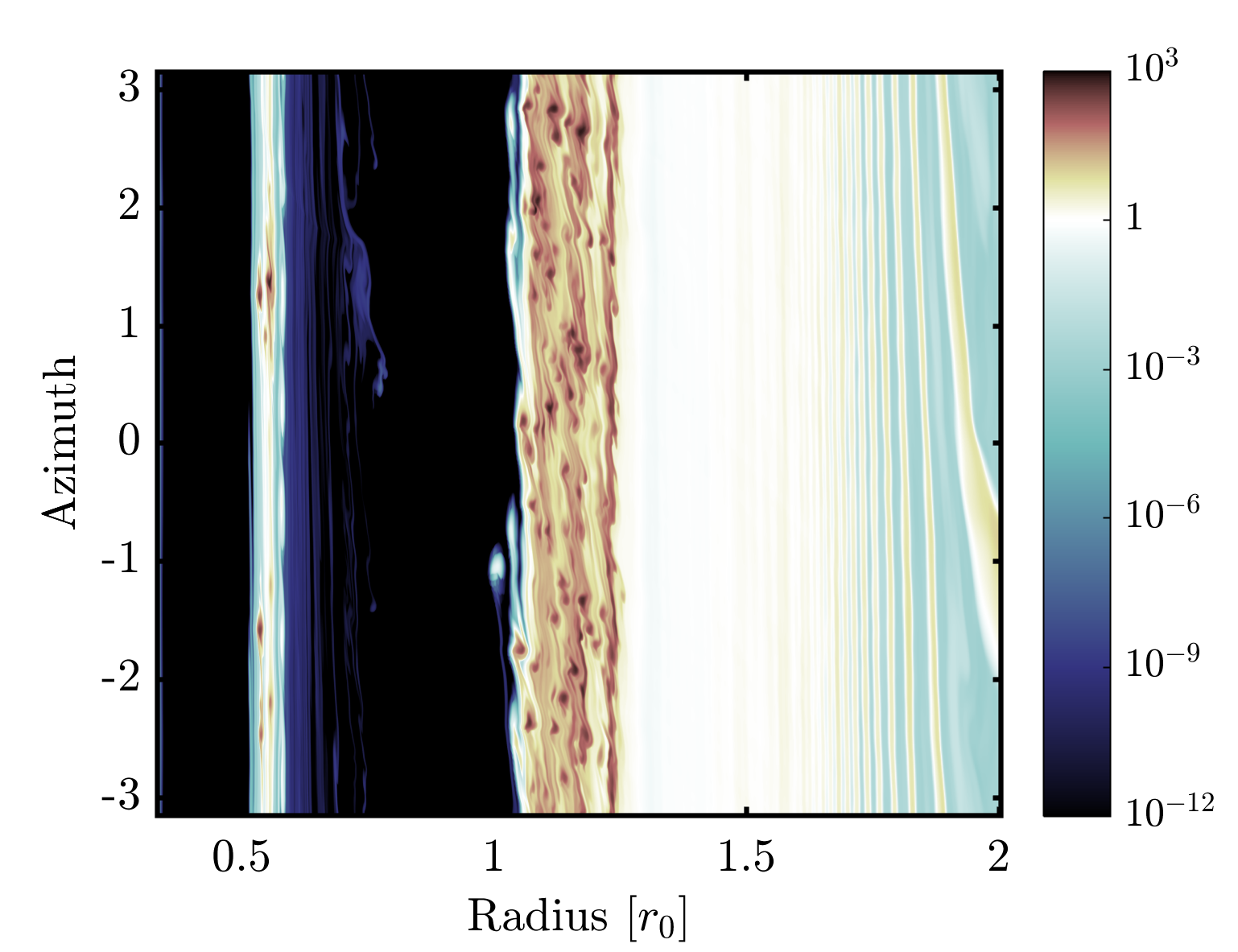}} &
	{\scriptsize{$\:$ \newline \newline (a)}} \\ 
	\imagetop{\includegraphics[height=7.1cm, trim=4mm 0mm 24mm 3mm, clip=true]{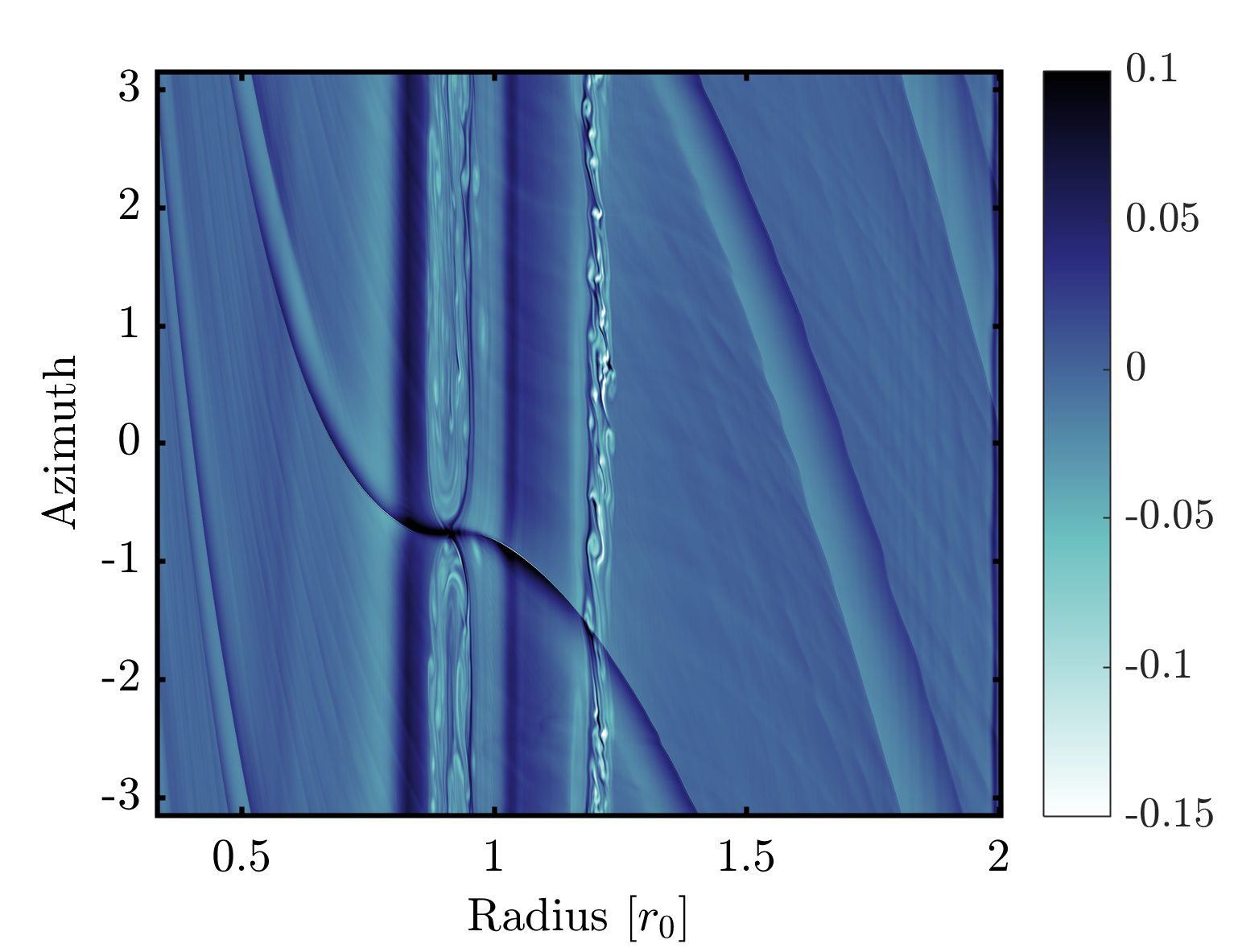}} &
	\imagetop{\includegraphics[height=7.1cm, trim=2mm 0mm 0mm 3mm, clip=true]{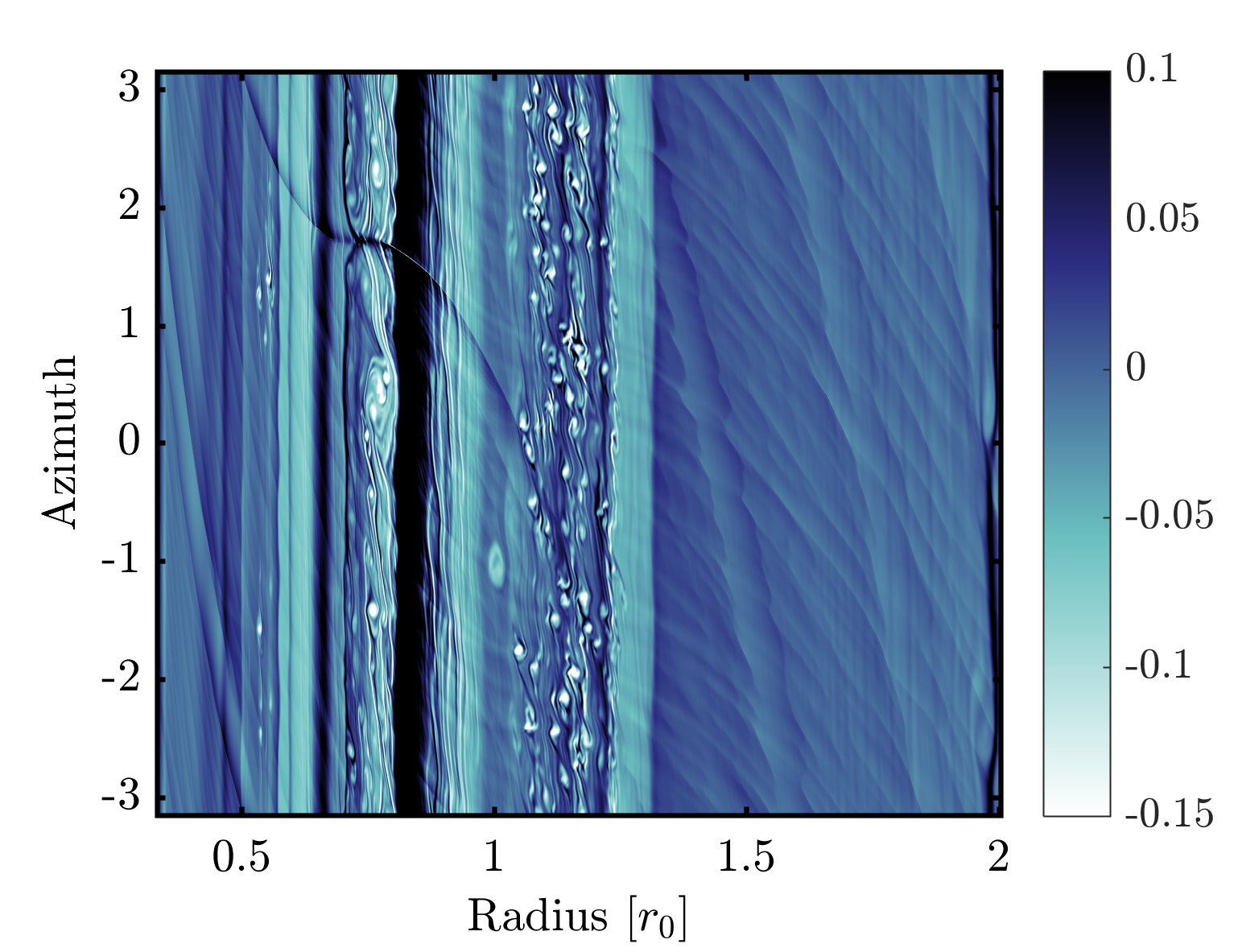}} &
	{\scriptsize{$\:$ \newline \newline (b)}} \\ 
	\imagetop{\includegraphics[height=7.1cm, trim=4mm 0mm 24mm 3mm, clip=true]{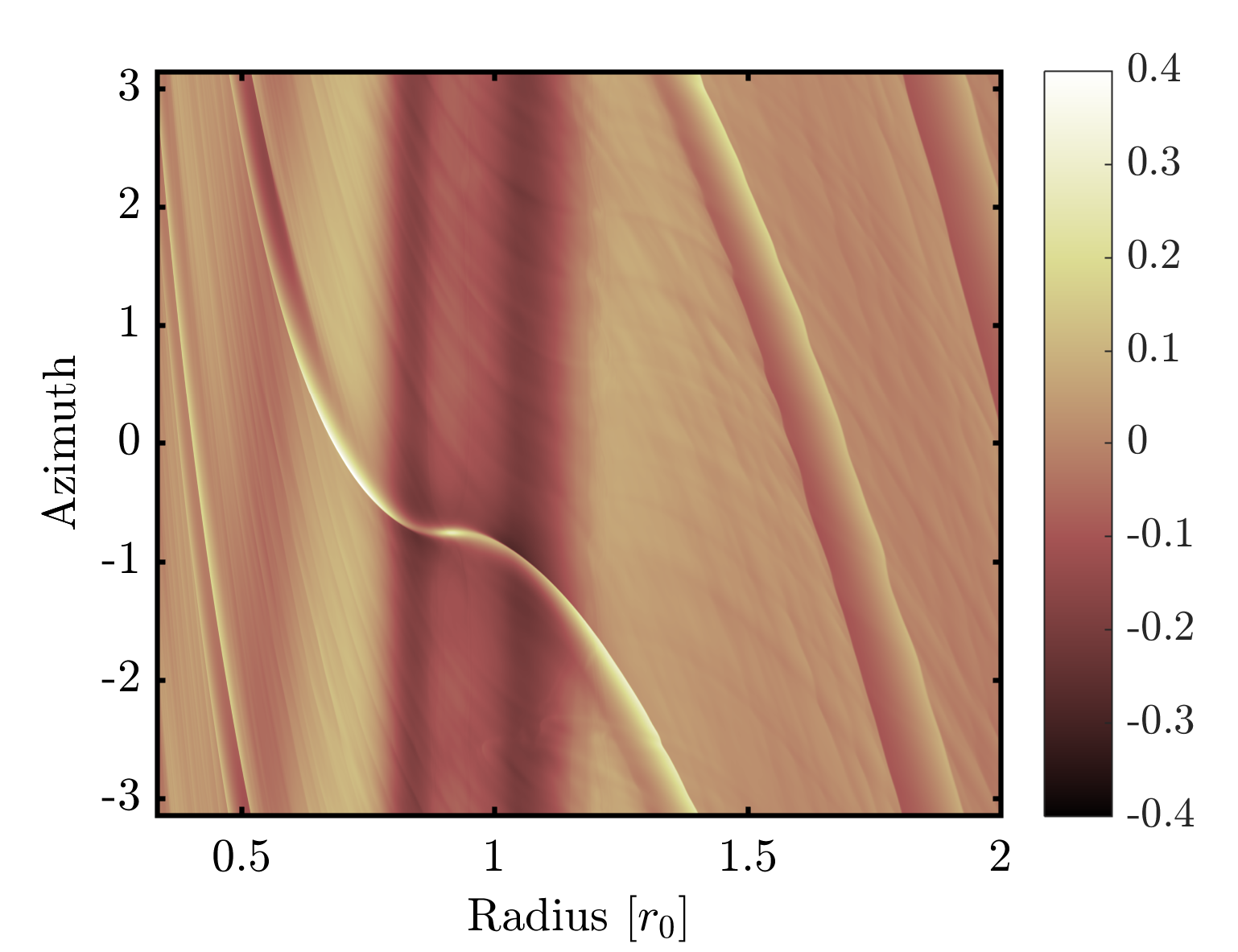}} &
	\imagetop{\includegraphics[height=7.1cm, trim=2mm 0mm 0mm 3mm, clip=true]{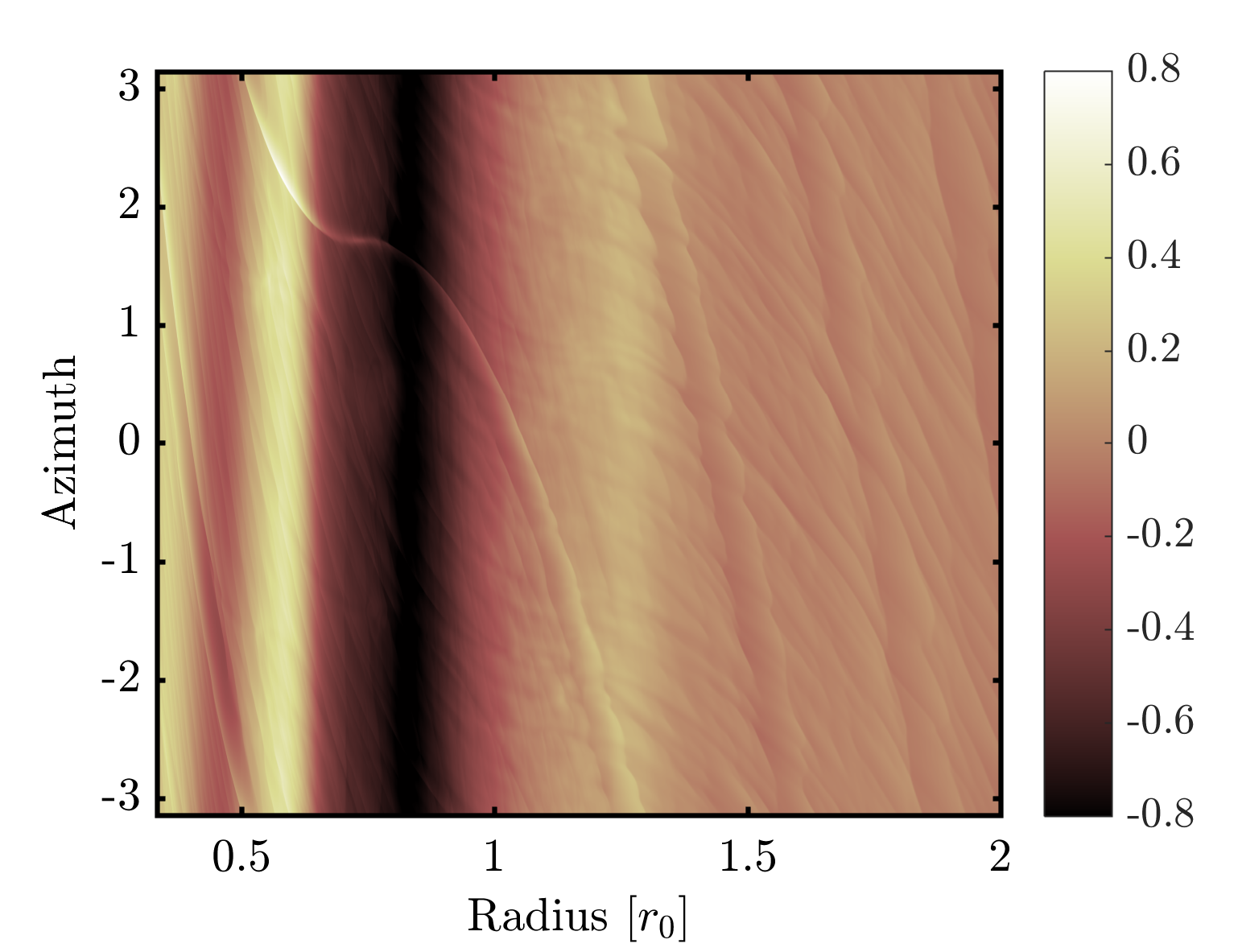}} &
	{\scriptsize{$\:$ \newline \newline (c)}} \\ 
	\end{tabular}
	
	\caption{\label{Fig_Mp_20_tau_p2_St_005} Disk structure for $M_p = 20\: M_e$, with $St=0.05$, and a cooling rate $\omega_c=10^{-2} \Omega_k$. From top to bottom: We show the dust density scaled to the initial condition (a), the gas Rossby number (b), and the gas density scaled to the disk background (c). The left column presents the state obtained after $500$ disk rotations, and the right column the state after $3000$. The colormap of the gas density at $t=500$ rotations (bottom left) is twice smaller than at $t=3000$ rotations (bottom right) for convenience. }
  \end{center}
\end{figure*}

	In the last run we use a planetary embryo of $20$ Earth mass, which is the typical value assumed to stop the pebbles at the gap edge \citep{Lambrechts2014, Bitsch2018}. We show the results of the full two-phase evolution Figure \ref{Fig_Mp_20_tau_p2_St_005}. During the first hundreds of disk rotations, the planet carves quickly the disk, and creates a gap in the gas. The pebbles are also affected by this effect, through the action of the drag force. After about $400$ rotations, the density of pebbles is reduced by a factor of $100$. The accumulation of pebbles at the outer edge of the gap triggers the formation of a dust ring, like in the other cases.
	
	We show the disk profile after the formation of the ring on the left of Figure \ref{Fig_Mp_20_tau_p2_St_005}, at $t=500$ disk rotations. In the active ring formed at $r=1.2\:r_0$, the density of pebbles is $100 - 1000$ times larger than the initial value. This ring is slightly wider than in the previous cases with smaller planets, and there are more individual eddies as we can see on the map of vorticity, middle panel. At the inner part of the disk, around $r=0.6\:r_0$, the pebbles also accumulate and the density is $10$ times larger than the initial condition.
	
	Between these two structures, the pebbles are rarefied. The surface density is $10^{-3}-10^{-4}$ times the one at the beginning of the run. Comparing with the gas density, bottom panel, this reduction is not exactly superimposed with the gap. This gaseous gap extends between $0.8$ and $1.15 \:r_0$, and the edges correspond to the highest reduction of gas density. This reduction is about $10\%$ in the middle of the gap, but close to $20\%$ near the edges. We warn that for practical reasons, the colormap of the bottom left panel is twice smaller than the one shown bottom right, so the colors cover the range $[-0.4 , 0.4]$.
	
	During the evolution of the disk, the planet migrates inward and the gap follows this migration. As a consequence, we observe the formation of several dust rings as in the other cases. In the gap region, the density of pebbles continues to reduce for two reasons. First the planet depletes gas from the gap and the solids are dragged along with the gas. Secondly, the flux of pebbles from the disk's outer parts is very limited and new material cannot refill the gap. We stopped the simulation after $3000$ disk rotations.
	
	We show on the right of Figure \ref{Fig_Mp_20_tau_p2_St_005} the disk at the end of the run. On top, the pebble surface density reveals the evolution we described. Compared to the results with a smaller mass of the embryo, the region beyond the outer edge of the planetary gap contains several very dense dusty eddies, rather than narrow rings. It is confirmed by the map of the Rossby number, middle panel, where we see individual anticyclonic vortices. In the previous runs, azimuthal stripes of vorticity were associated with the dust rings. Here, in this very active flow, the density of pebbles is larger than $1000$ times the initial profile. The dust-to-gas ratio is locally larger than $10$ in many of the eddies. The gas density is not particularly enhanced in this region (bottom panel). However, sound waves are excited by the eddies and perturb at a low level the gas density up to the outer disk boundary.
	
	At $r=0.5\:r_0$, the accumulation of the pebbles also triggered an active dust ring, which transformed into individual eddies. At the end of the run, five of them survive. Because pebbles are quickly depleted from the gap carved by the planet, a smaller amount of pebbles could be confined in this secondary dust trap than in the other runs. However, the density of pebbles is still in the range $10^{-1}-10^3$ times the initial profile in the region $0.5<r<0.6\:r_0$. In comparison, the density in the rest of the disk region with $r <1\:r_0$ is smaller than $10^{-10}$ times the background.
	
	The gap carved by the planet in the gas over the $3000$ disk rotations of evolution is very deep (bottom panel). Outside the planet orbit, up to $r=0.9\: r_0$, the gas density is reduced by $80\%$. Because of the migration, the planet moves inward inside the gap. During the course of time, the planet/disk interaction pushes gas outside the gap, amplifying the asymmetry between the two sides of the gap. This effect also is visible with smaller planet embryos. At the inner part of the disk, gas accumulates and creates a bump in the surface density of amplitude $50\%$ over the background at $r=0.6\:r_0$.
	
\bigskip

	In the light of these results, we point out some general features of the dynamics of a giant planet embryo in the disk:
\begin{itemize}
\item pebbles accumulate at $r=1.2\:r_0$, which triggers the formation of a dust ring where dust-to-gas ratio is larger than unity,
\item planet migration displaces this favorable region inward, thus several active rings can form in a wide portion of the disk,
\item after a certain duration - shorter for more massive embryos - the planetary gap and the inner disk are depleted from solids,
\item a secondary region of dust trapping and active ring formation exists at $r\sim 0.5\:r_0$.
\end{itemize}
	The explanation and implications of these observations will be given in the Section \ref{Sect_Discussion}.

\subsection{ Planet migration comparison }
\label{Sect_Result_Migration}

\begin{figure}[t]
	\begin{center}
	\includegraphics[height=6.5cm, trim=0mm 0mm 0mm 0mm, clip=true]{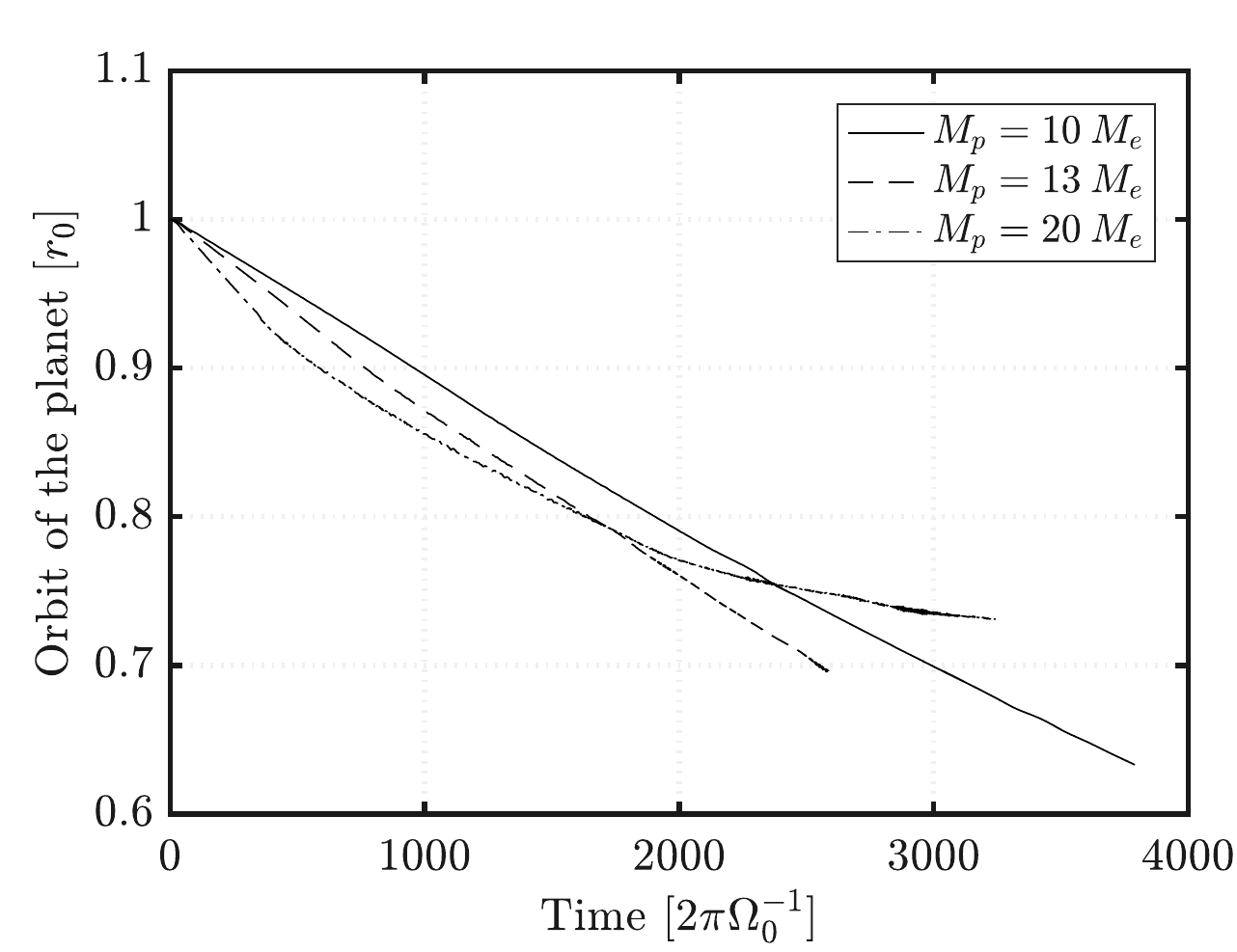}
	\caption{\label{Fig_Orbit_planet_compar} Evolution of the orbit of the planet, for the runs with $M_p = 10$, $13$, and $20\: M_e$ (solid, dashed, and dashed-dot lines respectively). The constant type-I migration rate is a good model for the two less massive planets, while two significant inflections of the migration are visible for $M_p = 20\: M_e$ at $t=400$ and $t=2000$ disk rotations. }
  \end{center}
\end{figure}

	We observe significant planet migration in the different runs. Indeed, the displacement of the planet has a strong influence on the disk evolution and on the distribution of the pebbles. Therefore, we must be confident in how accurately we evolve the planet orbit.
	
	The numerical scheme is a second order LeapFrog integrator, which is proven to keep the Keplerian orbit of the planet very accurately. The total energy error is bounded to a very low level. We have tested our implementation of the LeapFrog method and obtained the expected accuracy. The only dissipation of energy and momentum of the planet is due to the gravity exerted by the disk on the planet. This effect is the driver of the planet migration. It is a well understood problem and years of research have been devoted to analyze this process, in particular the Type-I migration regime which is what corresponds to our study \citep{Lin1986, Rafikov2002, Paardekooper2010a, Ayliffe2011}.
	
	We plot the orbital distance from the planet to the center of mass of the system as function of time Figure \ref{Fig_Orbit_planet_compar}. Because we vary the mass of the embryo in the different runs, the corresponding masses are identified by the type of lines. In all cases, the migration is directed inward and the orbit of the planet varies smoothly. It is particularly true for the $M_p=10 \: M_e$ run, where the migration rate of the planet is nearly constant.
	
	 For the $M_p=13 \: M_e$ run, migration is faster in proportion to the larger mass of the embryo. We measure a small inflection of the curve at $t=900$ disk rotations when the migration velocity is reduced by a few percent. This effect is due to the formation of the dust ring triggered at that time. In fact we include the gravitational interaction with the dust fluid, thus the mass of pebbles contained in the active ring produces a small amount of torque directed outward on the planet and reduces the inward migration. Until the end of the run, the migration conserves the same rate.
	 
	 In the last run, with $M_p=20 \: M_e$, we observe the same effect at $t\sim 450$ disk rotations when the ring pebbles at $r=1.2\:r_0$ is triggered. The curve of the radial orbit of the planet presents an inflection at this time, with a reduction of the migration rate of $\sim 50\%$. This rate is then conserved until $t=2000$ disk rotations when another reduction of the migration occurs. This second inflection can be explained by the complete depletion of pebbles from the co-orbital region and from the gap. The details of these changes of migration rate require further analysis which is beyond the scope of this study, and will be covered in a future publication. The main result is that, over long periods of time, the inward planet migration is regular (Type-I regime) and significant in the three runs.
	 
	 To finish, we compare the migration rate measured during the first hundreds of orbits, i.e. before any inflection of the orbital position, with the analytical estimate of the Type-I migration regime. In \cite{Tanaka2002}, a detailed linear analysis of the migration is given in the isothermal limit, but could be used under finite cooling rate models. Indeed the location of the shocks is too much away from the planet to affect the migration (we recall that Type-I migration is due to the gravity of the wake due to sound waves excited in the vicinity of the planet). They obtain the following migration rate:
	 
 \begin{equation}
 \frac{d r}{dt} = -(2.7-1.1\:\beta_{\sigma})^{-1} \frac{M_p}{M_s} r \Omega_k \frac{\sigma r^2}{M_s} \left( \frac{c_s}{r\Omega_k}\right)^{-2}.
 \end{equation}
	 
	 In this equation, $r$ is the orbit of the planet, $M_s$ the mass of the central star, $\Omega_k$ the Keplerian orbital frequency, $\sigma$ the gas surface density, and $c_s$ the sound speed of the gas. We recognize in $c_s/(r\Omega_k)$ the ratio of the disk scale height to the radial location in the disk, $H/r$. The thermodynamical model of the gas can affect the estimate of the sound speed by a factor under unity, .e.g. $\sqrt{\gamma}$ in an adiabatic disk.
	 
	 Applying this analytical estimate to our setups, we obtain $dr/dt= {-1.95, \: -2.54, \: -3.89 \times 10^{-5}}$ in units of $r_0 \Omega_k(r_0)$ at the reference radius, for planet masses $M_p=10, \: 13, \: 20 \: M_e$ respectively. The values of the migration rate measured in the simulations are $dr/dt= {-1.65, \: -2.1, \: -3.27 \times 10^{-5}}$ in the same units. As a result, the inward migration of the planets in our runs is within $16\%\pm1\%$ in agreement with the Type-I model. This small discrepancy is probably due to the presence of a thermal relaxation in our method. As a conclusion, the orbital evolution of the planet is accurately followed, and we can be confident that the impact of this process on the disk is well resolved.

\section{ Discussion }
\label{Sect_Discussion}

	The results of the simulations show a complex series of events leading to the formation of multiple dust rings and the possible stopping of the flux of pebbles. We discuss here the details of the main processes and their role in the disk evolution and planet formation scenarios.

\subsection{ Planet-disk interaction: gas structures }
\label{Sect_Gas_structures}

	As mentioned previously, the dynamics of the dust and the possible formation of high density regions, is dependent on the gas flow. Solids smaller than centimeter size have a Stokes number well bellow unity, typically $0.01-0.1$, at less than $10$ au from the star. As a results, they have a certain inertia to the gas dynamics, and the effect of drag forces takes a certain delay, typically in $1/St$ local rotations. As a result, the gas dynamics affected by the planet/disk interaction is the dominant process that will generate structures in the solid phase.

	The presence of a planet in the disk triggers perturbations of mass, pressure, and momentum of the gas through different channels. First, the gravity force that the planet exerts on the disk changes the momentum of a gas parcel directly; it is the principle of motion. This effect weakens as the distance between this parcel and the planet increases, and becomes weaker than the pressure forces for example. The direct effect of planet gravity has a sweet spot at a few disk scale height away from the planet. This acceleration is partly responsible for the gap formation \citep{Takeuchi1996, Crida2006, Kanagawa2015, Malik2015}. Material before the planet orbit is slightly decelerated compared to the unperturbed motion, while the material after the planet orbit is accelerated. Thus a positive shear (i.e. positive vorticity) is added to the steady state of the disk, resulting in a lower pressure region because of geostrophic balance.

	We note in passing that a little confusion may arise concerning the formation of a gap. The conventional simplified picture is that massive planets ($M_p>50 \: M_e$) open gaps and migrate slowly and even not at all, while light planets do not open gaps and migrate efficiently (in type-I regime). In fact, as described in the aforementioned references, the carving of the gap is progressive, as function of the planet mass. So the 'gap opening' regime depends on the definition of the shape of a gap, e.g. its depth in pressure (or density). Indeed we notice a weak gap in gas density and pressure, even for planet masses below $20 \: M_e$, while Type-I migration still occurs, precisely because the depth of the gap is much smaller than that produced by for example a $50 \: M_e$ planet (see Figure \ref{Fig_P_Rho_ave}). As a last detail, we recall that isothermal disks would produce a stronger gap in gas density than weakly radiative disks, such as in models with a finite cooling rate like here.

	 Concerning the solids, the gravity of the planet is weaker than the drag force, when $St<1$, and almost negligible. This has been tested in a simulation where the planet gravity was switched off in the dust phase. The results where almost identical to the normal case. Hence, pebble dynamics is mainly dominated by the flow conditions of the gas. However, in the gap, these conditions favor the depletion of solid grains from the planet vicinity. 

	The second major effect the planet produces in the disk is the emission of sound waves. As a permanent disruptor of the flow, the planet excites inertial waves that form the spiral wakes we see in the models. This process has a linear regime when planet mass is below $10-20$ Earth masses typically \citep{Ogilvie2002, Li2009, Dong2011a, Kley2012, Paardekooper2014}. In this case, the amplitude of the waves is directly proportional to the mass of the planet, and related to the disk scale height. This dependence on planet mass is still valid for bigger planets, while non-linear relation with the scale height appear in the massive objects. In turn, the gravity the waves exert on the planet is the engine of planet migration. In the linear regime of wave emission, planets are in type I migration. As mentioned previously, the models in this paper are in this linear regime. Bigger planets enter more complex migration laws, mainly because a gap is carved.

	The last important effect is vorticity generation by wake/disk interaction. Close to the planet, i.e. within a scale height, the waves are well described by a linear model, and the torque they create on the planet is described by the type-I migration regime. However, the propagation of these waves in the disk is a source of deviation from this description. At further distance from the planet orbit, i.e. several disk scale heights, the local properties of the disk and of the wake itself are different and the conditions for the generation of a shock wave can be achieved. However, this non-linear effect has a weak influence on the gravitational torque exerted on the planet, as the dominant component comes from the neighborhood of the planet, where the waves are linear. Where the wake becomes a shock wave, it changes the entropy of the surrounding gas. But even before that, when the wake velocity reaches a significant fraction of the local sound speed, the energy exchange between the wake and the background gas deposits heat and momentum into the flow. This additional momentum is a source of negative vorticity of the local flow. This is a sustained process, as the wave generation by the planet is constant. The rate of vorticity amplification depends on the amplitude of the waves, the entropy gradient of the disk (i.e. the disk profile, see \cite{Rafikov2002, Richert2015, Lyra2016}), and the dissipative properties of the flow. In particular, the heat dissipation and gas thermodynamics have an important impact on the vorticity perturbation. It is difficult to quantify the influence of each of these processes, and we cannot provide an analytical equation of the vorticity amplification. However, different trends can be drawn:
%
\begin{itemize}
\item vorticity increases with planet mass. This amplification depends on the wave's amplitude, which is in the linear regime $\propto M_p$, and additionally on the energy exchange with the shock, which is nonlinear;
\item the region of highest vorticity amplification is located at $r\sim 0.5$ and $\sim 1.2$ the planet orbit, because shocks are forming there; 
\item vorticity increases in hotter regions of the disk, due to the baroclinic amplification of vorticity \citep{Lyra2011a, Klahr2014, Gomes2015};
\item planet migration attenuates the amplification of vorticity, because there is less pile-up of energy at the same location in the disk (see Section \ref{Sect_Multiple_rings}).
\end{itemize}

	These effects, in particular the excitation of gaps and rings in the gas flow at different locations in the disk have been recently studied in \cite{Dong2017}, and \cite{Dong2018}. While their models are isothermal, they show the formation of a bump in density at the region equivalent to $r=0.5 / r_0$ in our models. The confirmation of our findings by a different numerical method lends support to the robustness of these effects.

	Finally, the perturbation of the flow by the different processes listed above affects the dynamics of solid grains. In particular, it explains the depletion that occurs in the co-rotation region (i.e. the gap forming region), and the accumulation of dust in negative vorticity regions. The rate at which the dust density increases is proportional to the Stokes number of the grains and to the local Rossby number of the gas, in agreement with the capture of dust in vortices \citep{Surville2016}. Therefore, we reach more quickly higher dust densities in the runs with massive planets, and the formation of dust rings occurs earlier.

\subsection{ Formation and evolution of the dust rings }
\label{Sect_Ring_formation}

\begin{figure}[t]
	\begin{center}
	\begin{tabular}{c}
	\includegraphics[height=6.5cm, trim=0mm 0mm 0mm 0mm, clip=true]{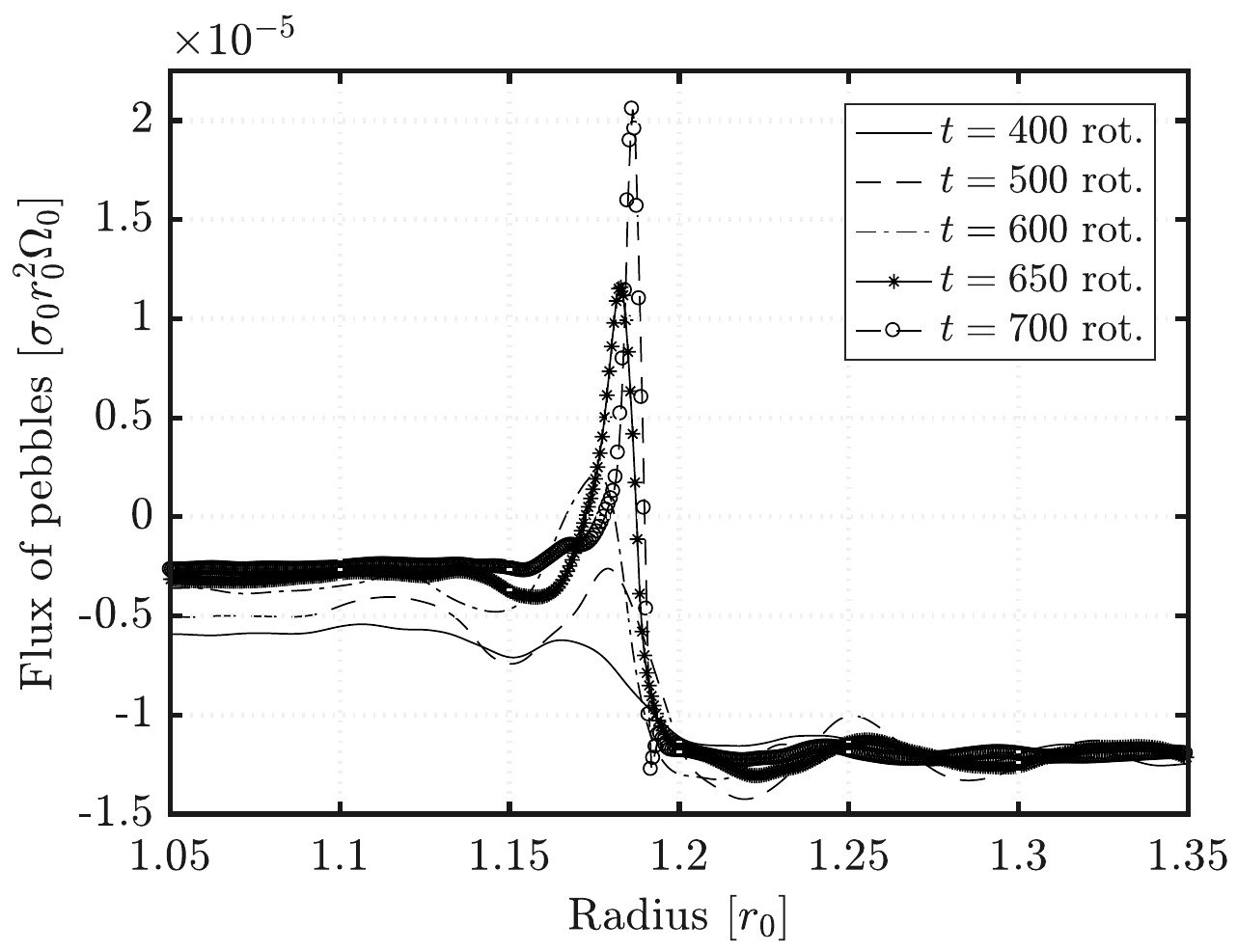} \\
	\includegraphics[height=6.5cm, trim=0mm 0mm 0mm 0mm, clip=true]{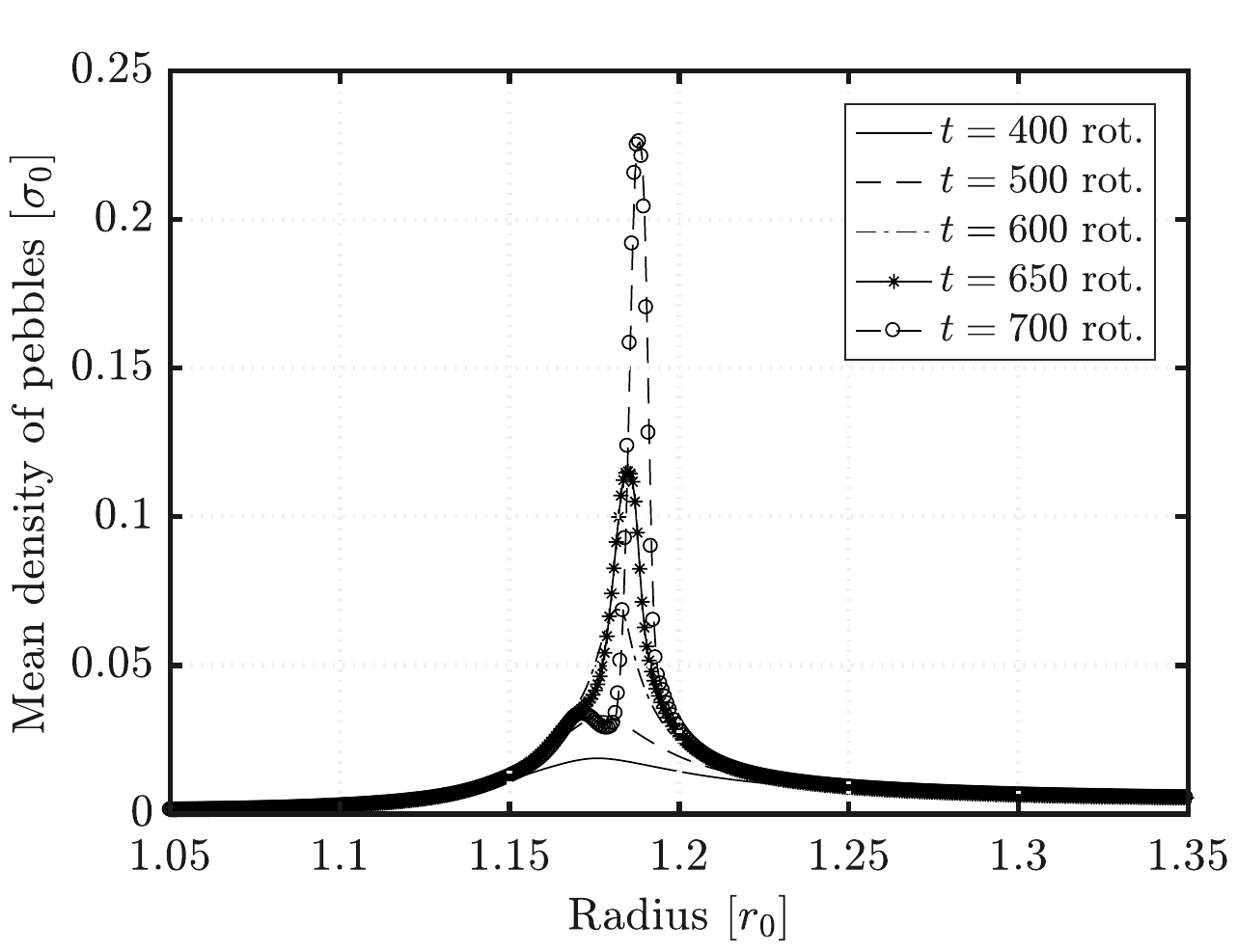}
	\end{tabular}
	\caption{\label{Fig_Flux_in_dust_ring} Formation of the dust ring at $r=1.2\:r_0$ for $M_p=13\: M_e$. The azimuthal average of the radial flux of pebbles (top) and of the density (bottom) are shown at different times (lines). }
  \end{center}
\end{figure}

\begin{figure*}
	\begin{center}
	\begin{tabular}{ccp{15mm}}
	\imagetop{\includegraphics[height=5.7cm, trim=0mm 12mm 0mm 3mm, clip=true]{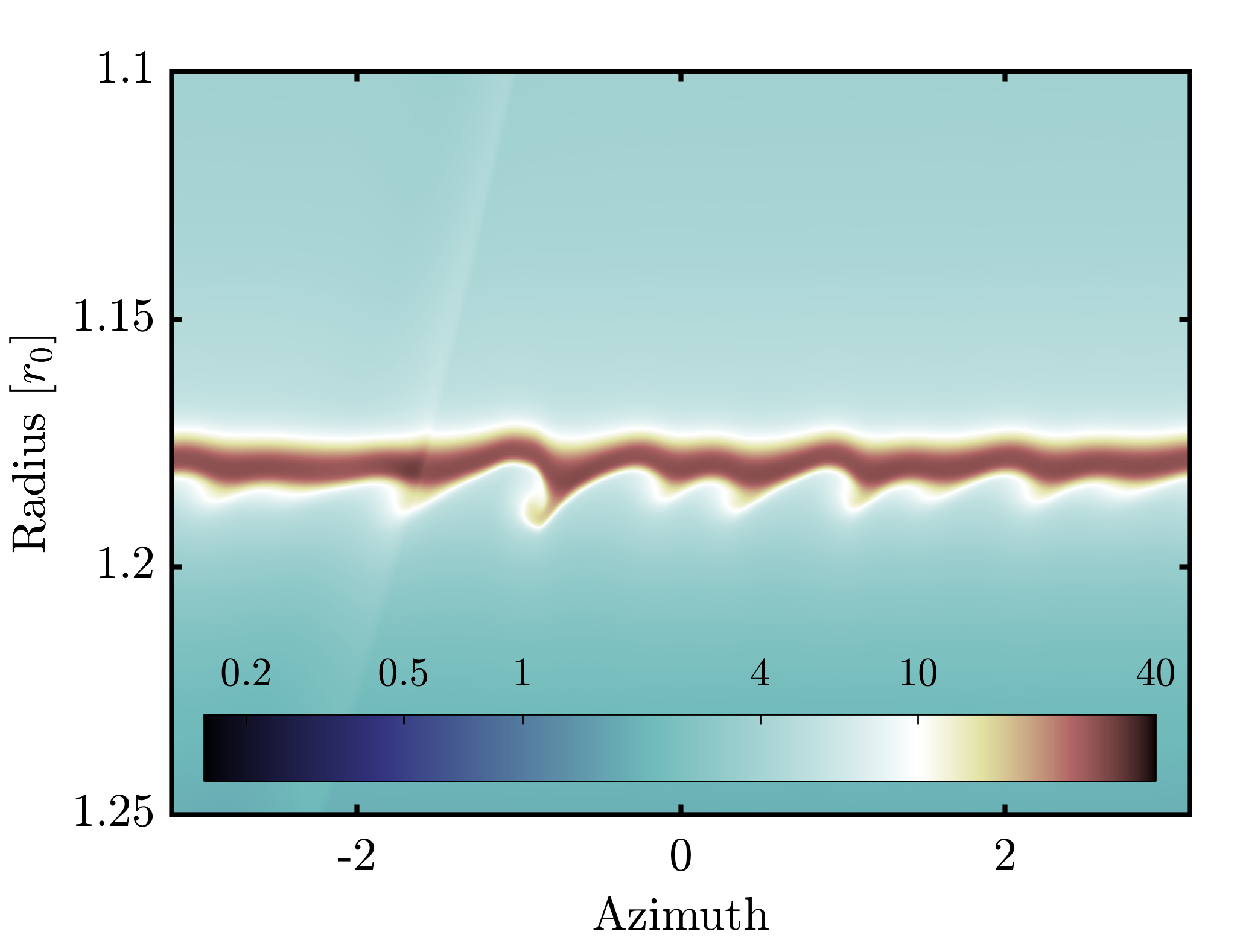}} &
	\imagetop{\includegraphics[height=5.7cm, trim=17mm 12mm 2mm 3mm, clip=true]{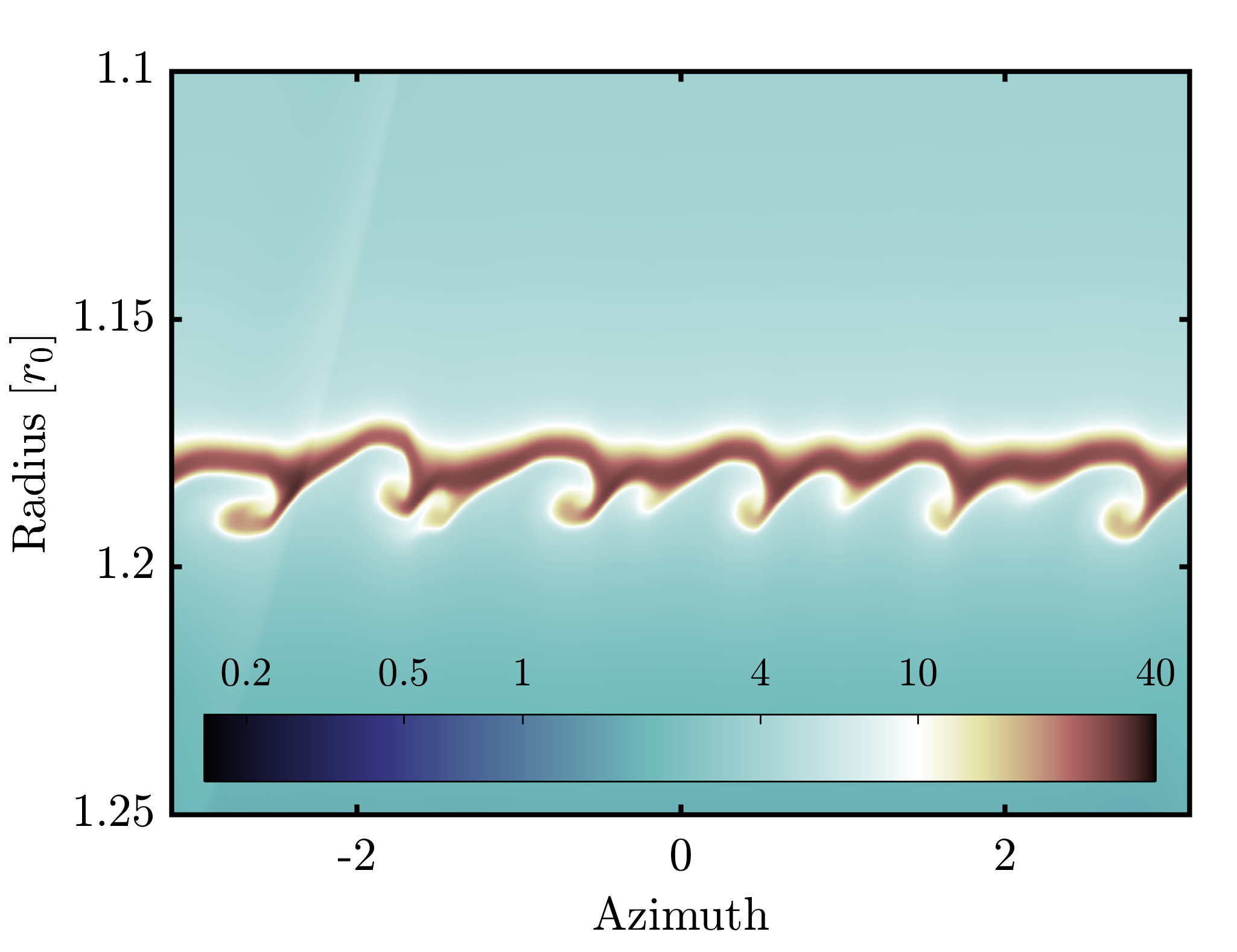}} &
	{\scriptsize{$\:$ \newline \newline (a), (b)}} \\
	\imagetop{\includegraphics[height=6.5cm, trim=0mm 0mm 0mm 3mm, clip=true]{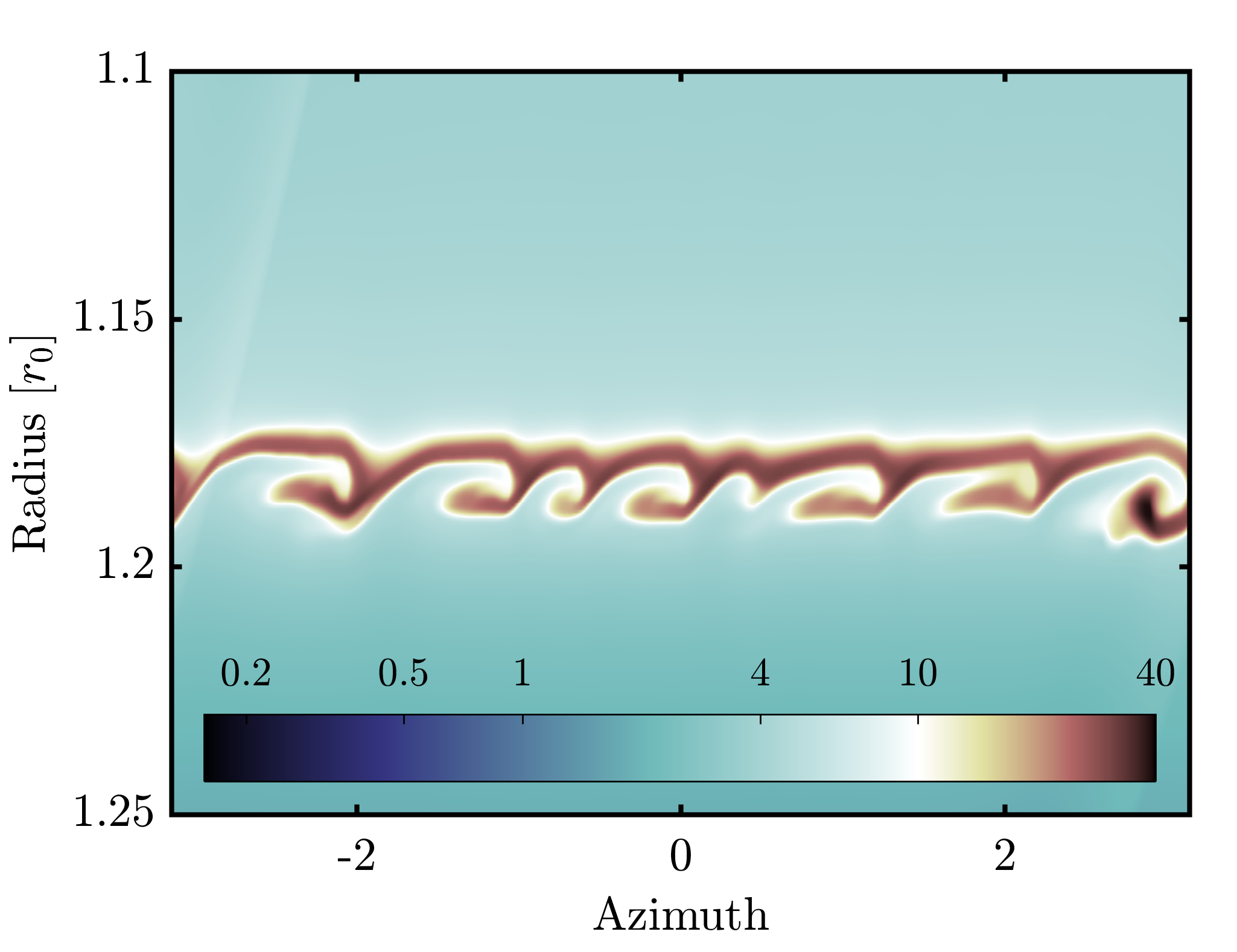}} &
	\imagetop{\includegraphics[height=6.5cm, trim=17mm 0mm 2mm 3mm, clip=true]{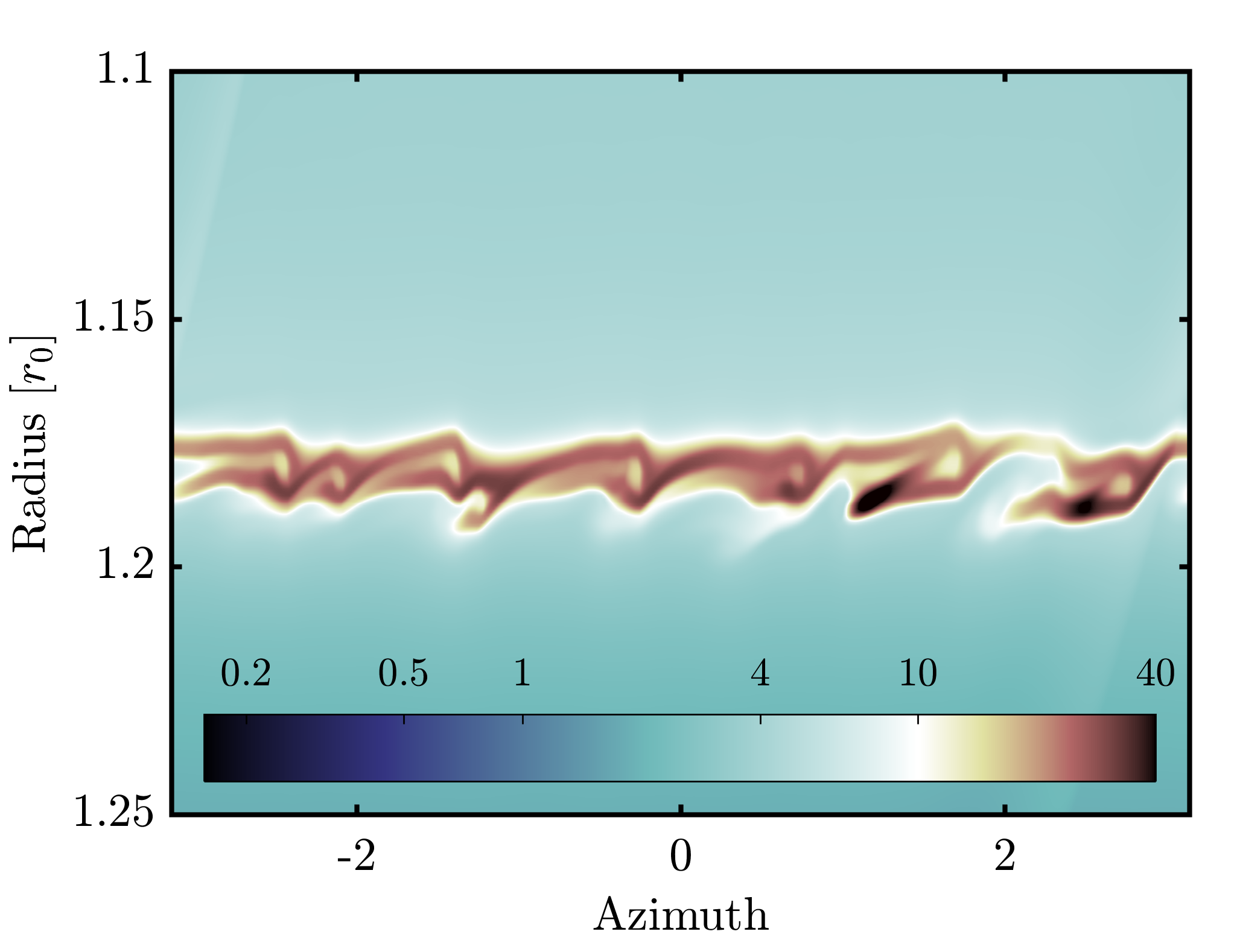}} &
	{\scriptsize{$\:$ \newline \newline (c), (d)}}
	\end{tabular}
	
	\caption{\label{Fig_Ring_Insta_Mp_10_tau_p3_St_005} Development of the instability of the dust ring at $r=1.2\: r_0$ for the run with $M_p=10\:M_e$. The density of pebbles is shown at $t=1465$, $1475$, $1485$, and $1495$ disk rotations, from (a) to (d) respectively.}
  \end{center}
\end{figure*}

	The formation of turbulent dust rings is observed in all the simulations, even when the planet mass is as low as $10\: M_e$. These rings result from two successive processes: {\it{(i)}} the accumulation of solids in a certain disk region, where the drag force favors an increase of the dust fluid density, {\it{(ii)}} the development and non linear saturation of a shear instability of the gas, when the dust/gas mixture is solid rich, i.e. when $\sigma_p/\sigma_g>\sim 0.5$. It is important to underline that the second process, which also generates turbulence in the gas, is only possible when the drag back reaction onto the gas is computed. It is the result of a momentum exchange between the two phases due to the drag force. 
	
	To understand the accumulation of pebbles in a ring structure, we may first assume an axisymmetric disk structure. If the dust fluid density and velocity field are continuous, one obtains from the conservation of mass:
\begin{equation}
	\partial_t \sigma_p = - \frac{1}{r} \partial_r (r \sigma_p V_r),
\end{equation}
	with $V_r$ the radial velocity of the pebbles. As a result, the density of solids is increasing where the flux of pebbles $r \sigma_p V_r$ is locally decreasing. The azimuthal integral of this flux is thus a crucial function to understand the radial dynamics of the pebbles in the disk.

	Figure \ref{Fig_Flux_in_dust_ring} shows the profiles of the radial flux of pebbles (top) and of the surface density (bottom), averaged over the azimuthal dimension. We focus on the region close to $r=1.2\: r_0$ for the run with $M_p=13 \: M_e$. At $t=400$ rotations (solid line), we observe a negative slope of the radial flux of pebbles in the region $1.17<r/r_0<1.2$, where the solids accumulate. A significantly higher density of pebbles is visible over the same region (bottom panel). At $t=500$ rotations (dashed line), the slope of the radial flux is twice steeper than at $t=400$, and a peak in the profile of the flux has formed at $r=1. 18 \: r_0$. As expected, the surface density has increased and there is a bump in the profile of the density of pebbles reaching four times the initial density. Later on, the correlation between density and the radial flux amplifies the steepness of the slope of this flux at ${r=1. 18 \: r_0}$. Over a narrow ring around this location, the flux even become positive. Quickly, the surface density increases exponentially and reaches $\sim 20$ times the initial density at $t=700$ disk rotations.

	The non linear dependence of the surface density evolution with the radial flux of solids is thus responsible for the formation of dense and narrow ring of pebbles in the disk, with a quasi axisymmetric profile. The region at $r=1.2 \: r_0$ is favorable for this formation as well as $r=0.5\: r_0$. The further evolution of this ring is shown Figure \ref{Fig_Ring_Insta_Mp_10_tau_p3_St_005}. These panels show four snapshots of the run with $M_p=10\:M_e$, and focus on the surface density of pebbles around $r=1.2 \: r_0$ (notice the different axis compared to previous figures). These snapshots cover a relatively short period of time, from $1465$ to $1495$ disk rotations. On the first two maps, top row, we observe the grows and saturation of azimuthal modes. The dominant mode is $m=10$, and grows at the outer edge of the ring (bottom part of the map). At this location, the contrast of surface density of pebbles was $40/4=10$ between the ring and the disk. Between $t=1485$ and $t=1495$ rotations, the waves break to a chain of eddies, although in this case a remnant of the initial ring is still present.

	The evolution of the ring instability is visually similar to the behavior of the Kelvin-Helmholtz Instability (KHI), with the growth of waves and breaking into vortices. Indeed, the high density contrast at the outer edge of the dust ring produces a large contrast in the gas azimuthal velocity, because of the drag back reaction. Therefor the gas is susceptible to this instability, and solids are dragged along with the gas and follow the patterns of the KHI. The details of this instability of the ring deserves a deeper study, which is beyond the scope of this paper. Recently, \cite{Huang2020} showed that such an instability could occur at the boundary between accreting and dead zones, where dust accumulates. They also see the formation of vorticity rolls when drag back-reaction is strong (see Figure 8 of this reference). Their interpretation of the underlaying process is similar to the one proposed here, despite the different trigger of the instability, ie. accretion vs. planet-disk interaction.

	The excitation of these turbulent dust rings was reported in global 2D models by \cite{Yang2020}, in isothermal disks, with a very different numerical method (the PENCIL code, a finite difference scheme, and Lagrangian dust particles). However they investigate more massive planets (equivalent to $50 \: M_e$ in our model), and do not resolve the instability for $St<0.1$, due to the diffusion of their scheme. We thus push the boundary of the parameter space of the planet mass and grain size much lower values, which indicates that dust ring formation is a common and general process in disks. Finally, \cite{Carrera2020} show that 3D pressure bumps are also favorable for dust ring formation, and eventual planetesimal formation.

	As a summary, this ring instability can be triggered easily, several times during the disk evolution, and at different places. It is an efficient process to increase the density of pebbles by several orders of magnitude, and to form dusty eddies. However, because this process results from the KHI that grows in the disk, these dust rings can form only in models that resolve correctly the drag interaction between gas and dust, and in particular the back reaction onto the gas.


\subsection{ Planet migration and multiple dust ring formation }
\label{Sect_Multiple_rings}

\begin{figure}[t]
	\begin{center}
	\begin{tabular}{c}
	\includegraphics[height=6.5cm, trim=0mm 0mm 0mm 0mm, clip=true]{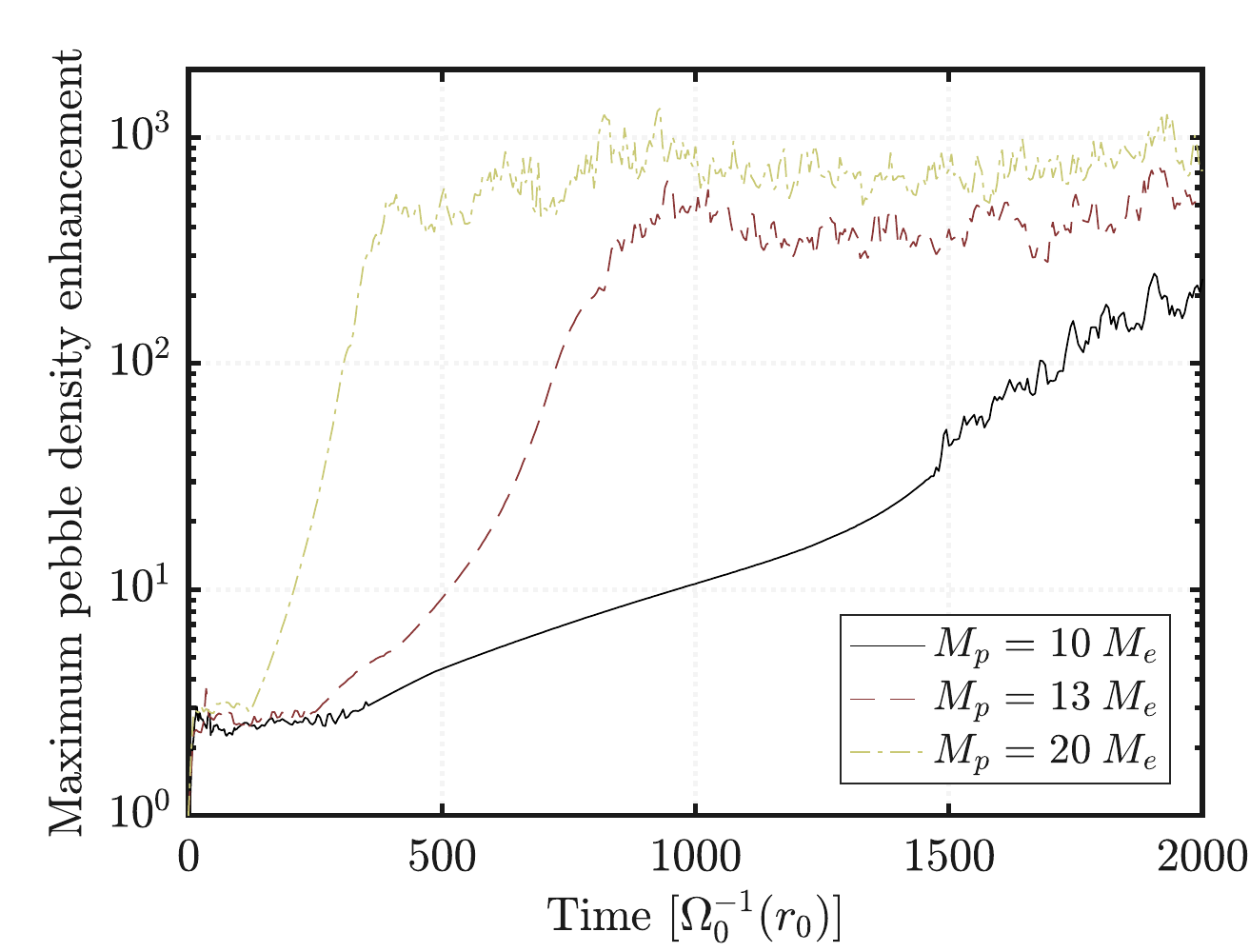} \\
	\includegraphics[height=6.5cm, trim=0mm 0mm 0mm 0mm, clip=true]{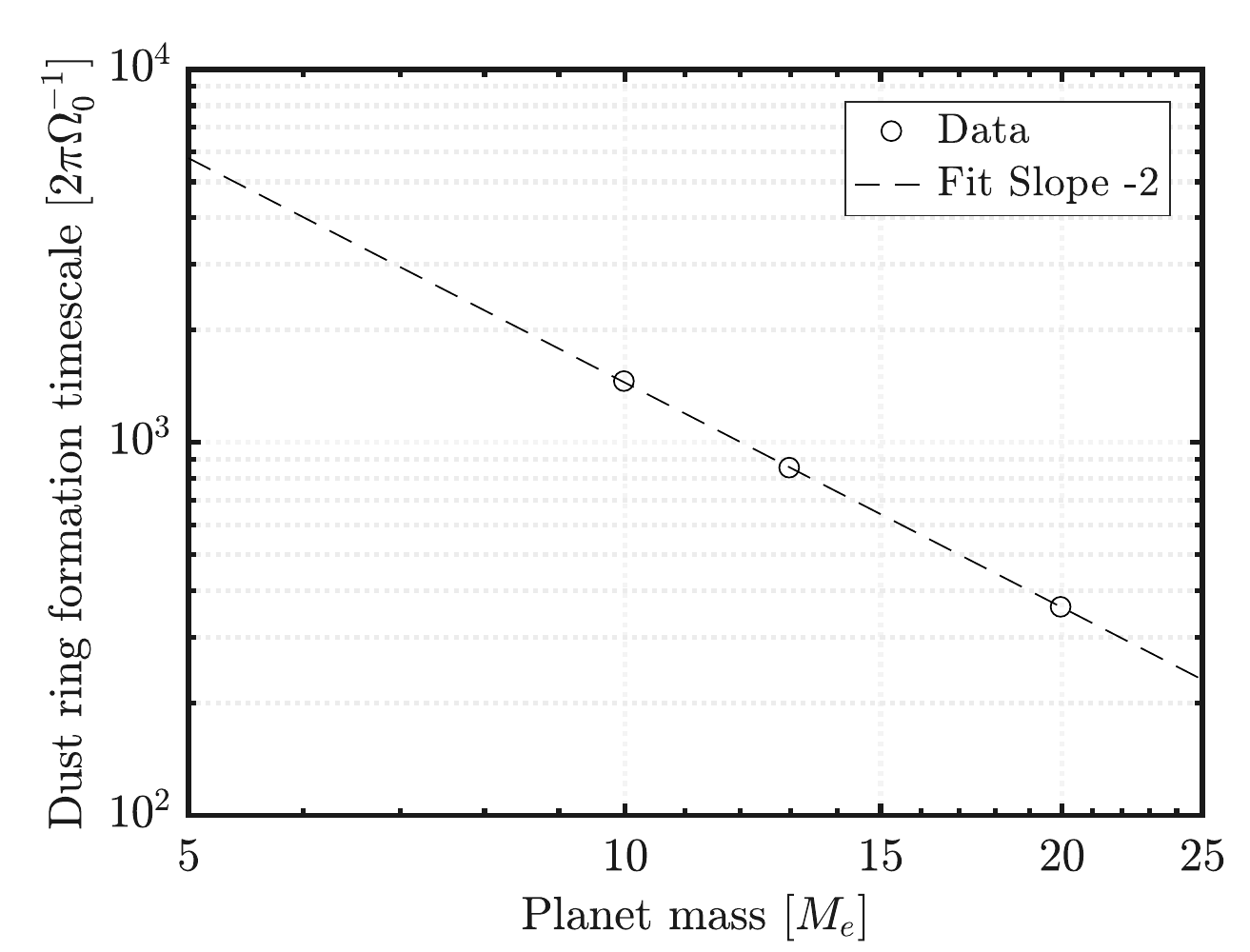}
	\end{tabular}
	\caption{\label{Fig_Dust_ring_formation_timescale} Top: Dust accumulation where the first dust ring forms, for the different planet masses. The exponential increase of the density of pebbles during the beginning of the evolution is followed by a saturation of the density enhancement. The transition indicates the typical timescale of the dust ring formation. Bottom: Dependence of the dust ring formation timescale on the mass of the planet, in log-log scale. The dots correspond to the result given by the simulations. The black line is the best fit to the data, $f(M_p) = 145 \times 10^3 M_p^{-2}$, in unit of Earth mass and disk rotations. }
  \end{center}
\end{figure}

\begin{figure*}
	\begin{center}
	\begin{tabular}{ccp{15mm}}
	\scriptsize{Dust density} & \scriptsize{Rossby number} & \\
	\imagetop{\includegraphics[height=6.8cm, trim=4mm 0mm 2mm 3mm, clip=true]{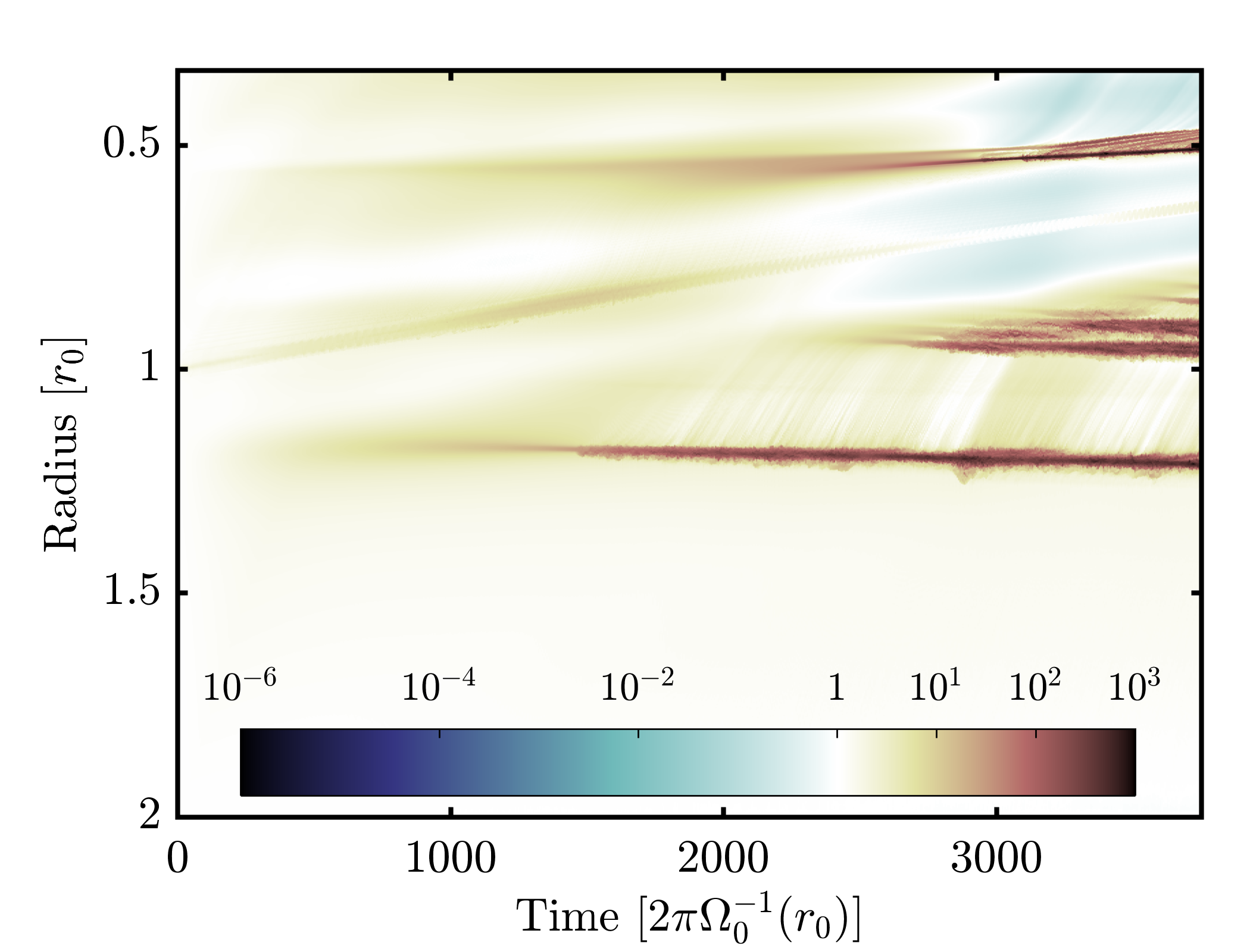}} &
	\imagetop{\includegraphics[height=6.8cm, trim=10mm 0mm 2mm 3mm, clip=true]{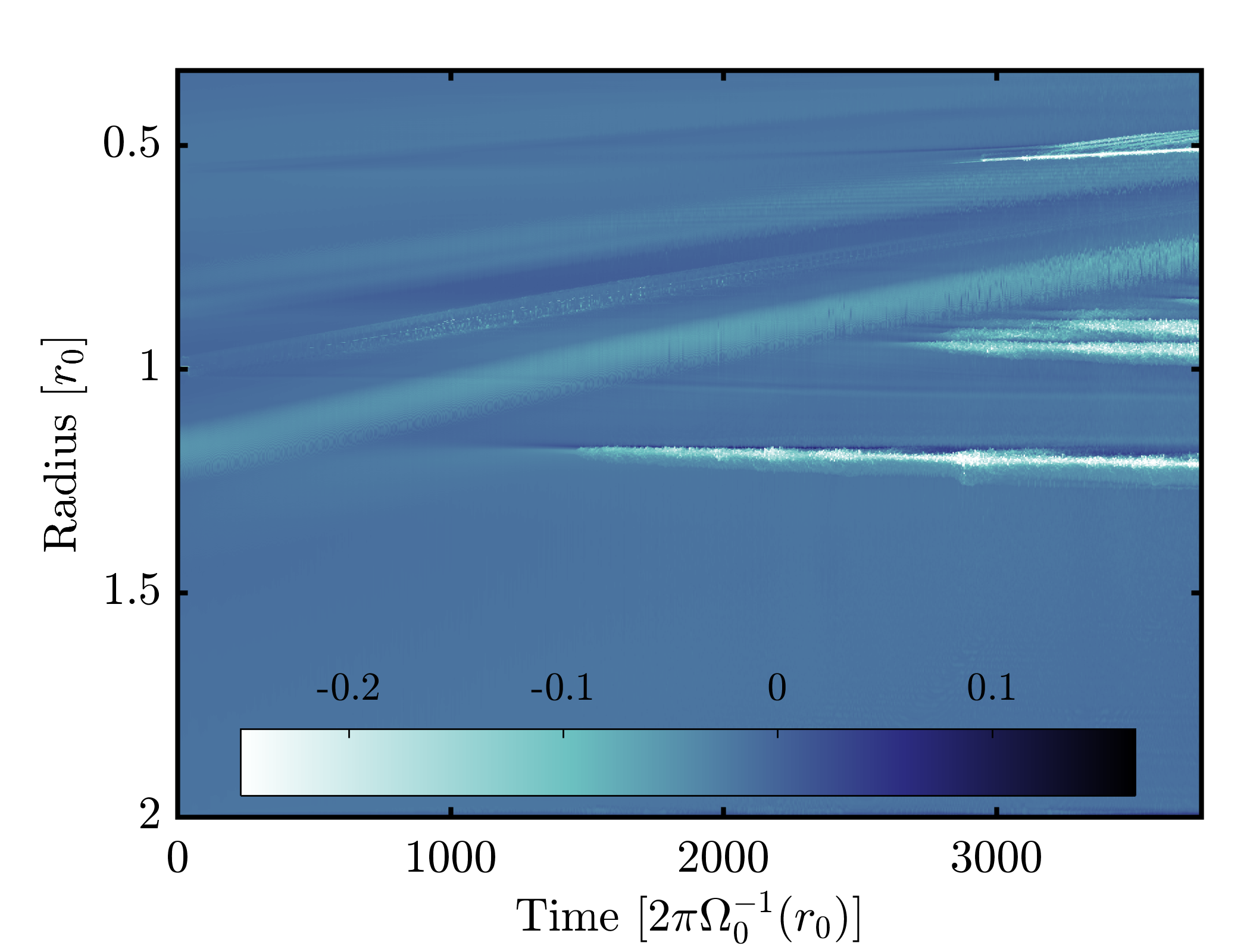}} &
	{\scriptsize{$\:$ \newline \newline (a)}} \\ 
	\imagetop{\includegraphics[height=6.8cm, trim=3mm 0mm 3mm 3mm, clip=true]{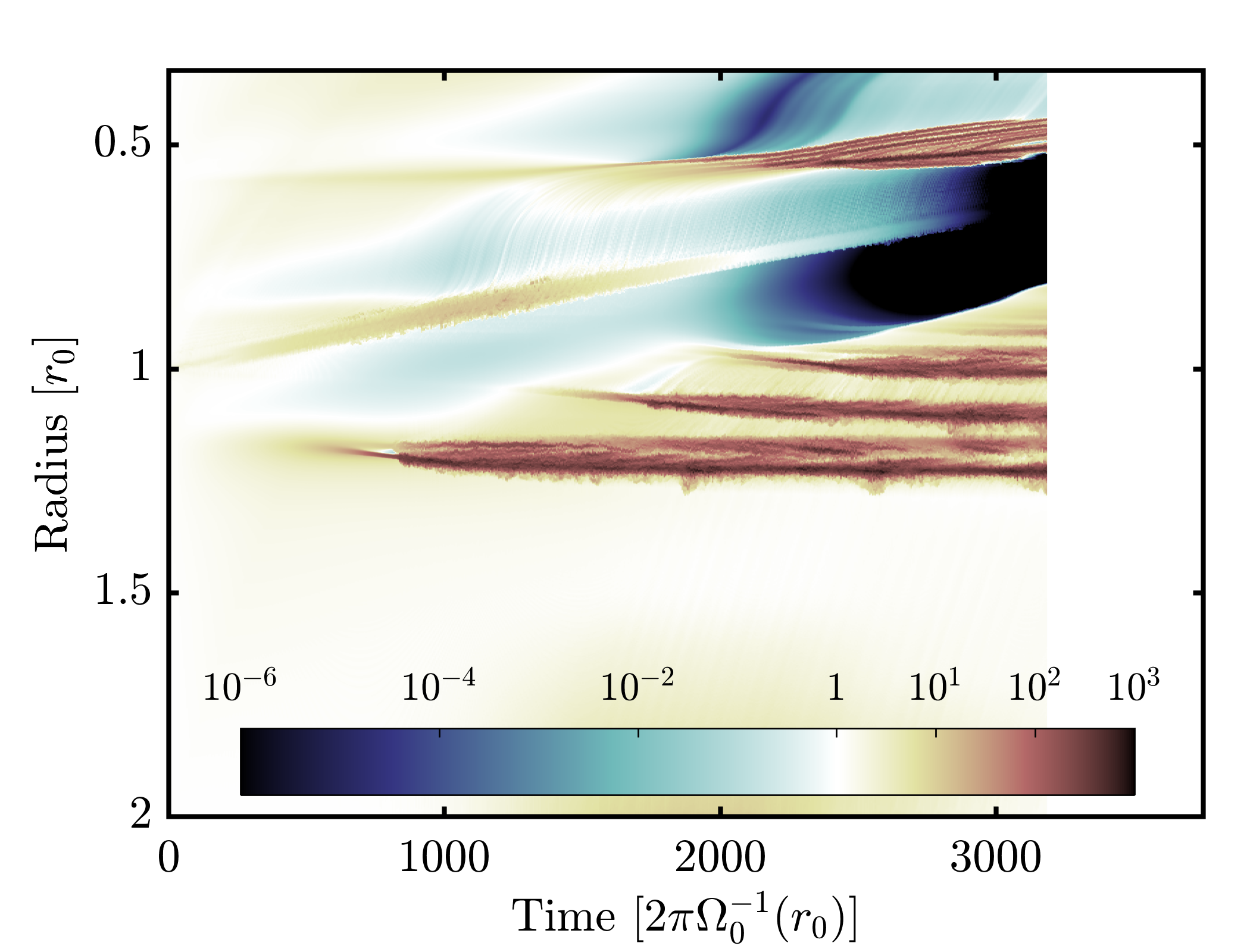}} &
	\imagetop{\includegraphics[height=6.8cm, trim=10mm 0mm 2mm 3mm, clip=true]{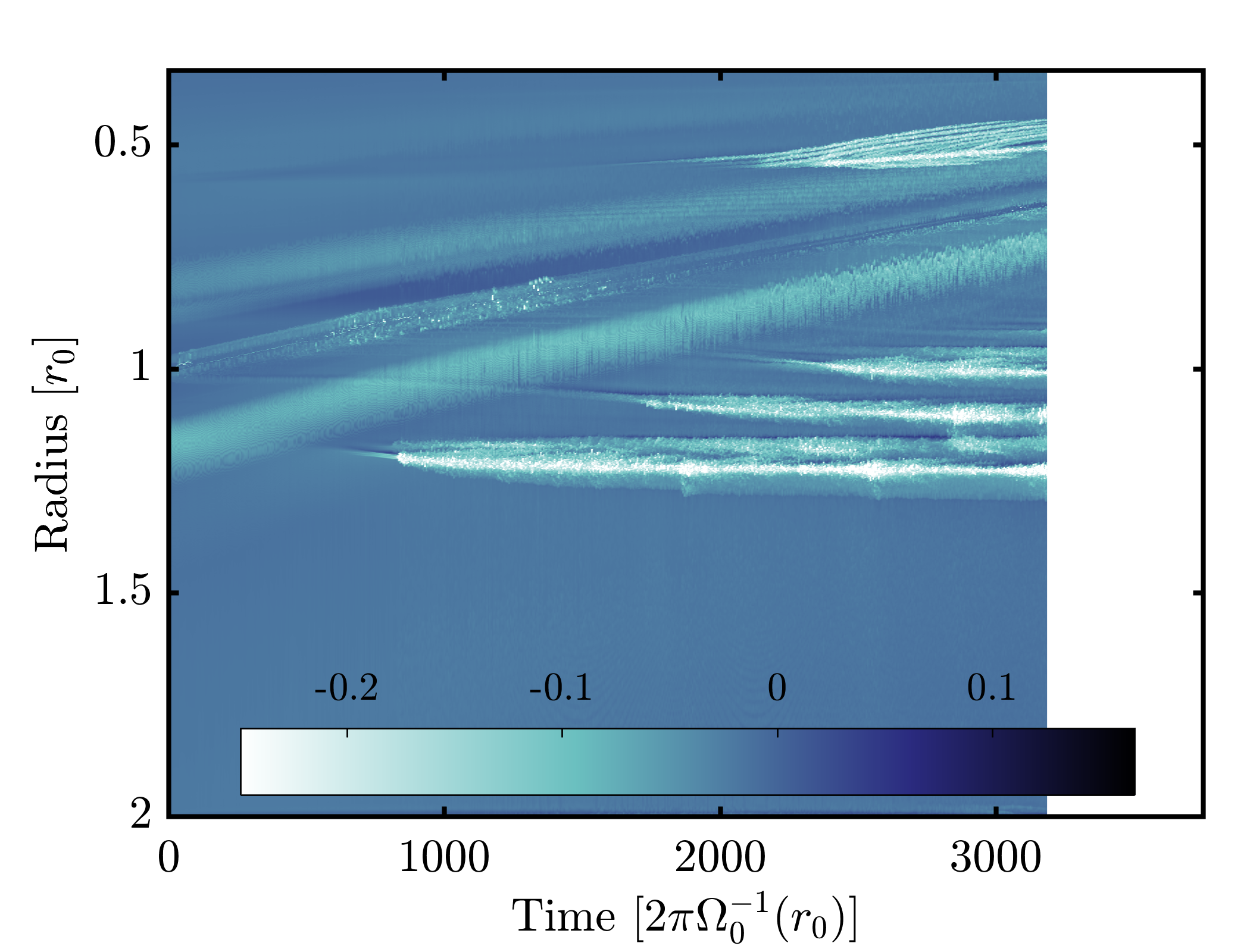}} &
	{\scriptsize{$\:$ \newline \newline (b)}} \\ 
	\imagetop{\includegraphics[height=6.8cm, trim=4mm 0mm 2mm 3mm, clip=true]{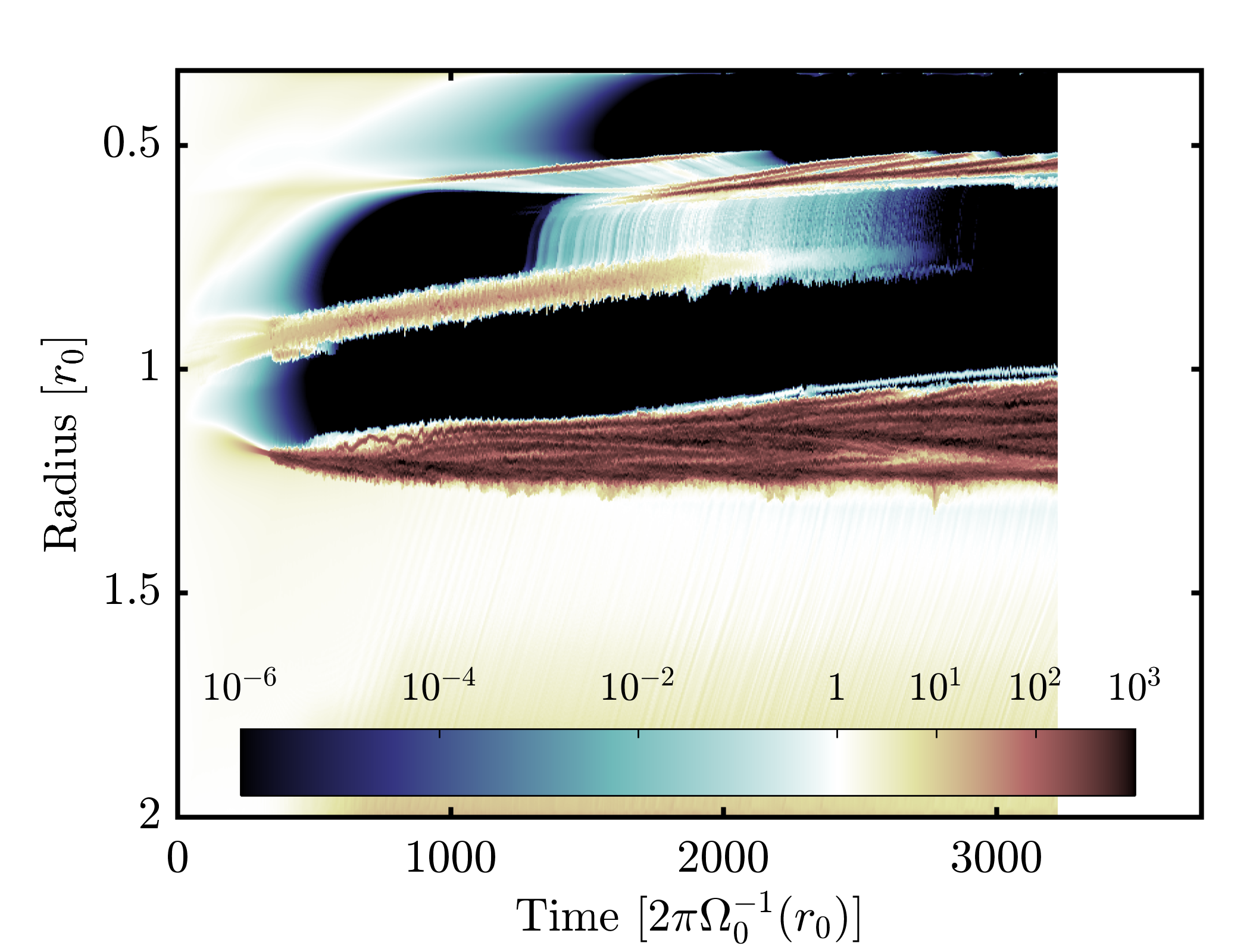}} &
	\imagetop{\includegraphics[height=6.8cm, trim=10mm 0mm 2mm 3mm, clip=true]{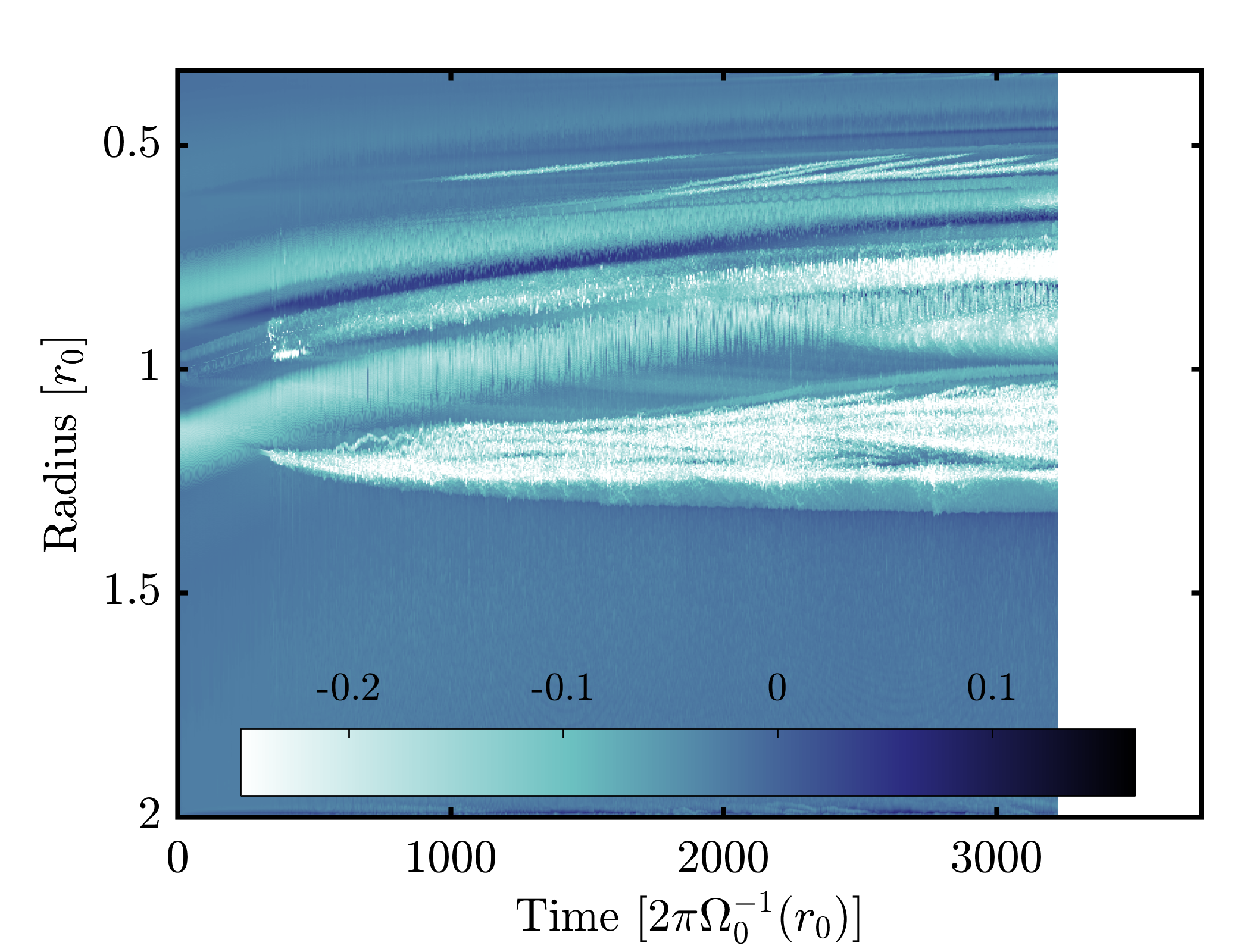}} &
	{\scriptsize{$\:$ \newline \newline (c)}} \\ 
	\end{tabular}
	
	\caption{\label{Fig_Time_evo_rho_rossby} Time evolution of the dust density and of the gas Rossby number, left and right columns respectively. Dust density is defined as the maximum value of the enhancement ($\sigma_p/\sigma_p(t=0)$) over the azimuthal direction. The Rossby number plotted is the minimal value over the azimuthal direction. Results of the different masses of the embryo $M_p=10, \: 13$ and $20\: M_e$ are shown top (a), middle (b), and bottom (c) rows, respectively. }
  \end{center}
\end{figure*}

	As presented in Section \ref{Sect_Results}, we observe the persistence of several dust rings at the end of the simulations, for the different planet masses. But we observe several forms of dissimilarity between the three runs: the number of dust rings that are excited after a given time of evolution, the separation between the different rings, and the degree of turbulence inside the rings. The understanding of these aspects relies on the evolution of the disk.

	The first structure that arises in the disk is the dust ring that forms at $r=1.2 \: r_0$. As described in the previous section, it is the result of a KHI triggered in the gas phase. The establishment of the condition for the instability is however dependent on the planet mass, as concerns the timescale. We show Figure \ref{Fig_Dust_ring_formation_timescale} the comparison of this timescale as function of the planet mass. Top panel records the evolution of the maximum of the surface density enhancement in the pebble phase during the first $2000$ disk rotations. It is clear that there is an exponential growth of the density in the three cases, with a growth rate faster when the planet mass is larger. When the instability of the ring is triggered, we observe fast oscillations of the density, due to the formation of small scale eddies. We can thus measure the time taken to trigger the instability. We obtain $1450$, $850$, and $360$ disk rotations for planet masses $10$, $13$, and $20 \: M_e$ respectively. 

	When we report these timescales as function of the mass of the embryo, bottom panel, we observe a simple correlation:
\begin{equation}
\label{Eq_Ring_timescale}
	\tau_{ring}(M_p) \sim 145 \times 10^3 \left(\frac{M_p}{M_e}\right)^{-2},
\end{equation}
	in unit of disk rotations $2\pi \Omega_0^{-1}$.

	The accuracy of the fit is within a few percent. It is striking that such a simple relation could be obtained from high resolution non linear simulations, thus supporting the view that the accumulation of solids is driven by the details of planet-disk interaction. All else being equal, it would take longer than a million year to form a dust ring with an Earth mass planet according to this model. We conclude that this process is relevant only for Super Earths and more massive planets, namely planets with masses larger than $4-5 \: M_e$.

	The coefficient in front of the power law depends mainly on the Stokes number of the dust grains. In this study, the Stokes number is initially set to $0.05$ at $r_0$, and we do not show results with other values. However, the results we obtained in \cite{Surville2016}and \cite{Surville2019} show that the timescale of capture of dust grains inside a vortex -- i.e. a negative vorticity region like the 'excitation band' we observe here -- is proportional to the $2 St / (1 + St^2)$. A corollary is that the vorticity excitation by the planet disk interaction is proportional to $M_p^2$, if we follow the model of dust capture in vortices. This should be confirmed by a deeper study.

	The other dependence of the coefficient is on disk properties, in particular the disk scale height, and the thermodynamics. However, this aspect is directly linked to the vorticity excitation in the disk by the planet wake. This aspect is beyond the scope of this paper, and will be covered in a future publication.

	After this first aspect of the evolution of the disk, we focus on the formation of the additional rings. We present Figure \ref{Fig_Time_evo_rho_rossby} the evolution as function of time of the density of pebbles in the disk, and of the vorticity of the gas (expressed as a Rossby number), which quantifies some degree of turbulence in the flow. In the case with a $10$ Earth mass embryo, top row (a), we see the two regions of accumulation of dust at $r=1.2$ and $r=0.5 \: r_0$. At $r=1.2 \: r_0$, a first dust ring forms, following the process described above, at $t= 1450$ disk rotations. From the vorticity map, we see that suddenly the Rossby number jumps below $-0.2$, while it was around $-0.1$ before the instability. This degree of turbulence is sustained until the end of the run. As time goes, we identify the planet migration, which results in the radial displacement of the region where the Rossby number is the most negative, which we can call the 'excitation band'. It corresponds to the outer edge of the 'gap' excited by the planet. During the time necessary to accumulate enough dust to trigger the KHI of a dust ring, i.e. $\sim 1500$ disk rotations for this planet mass and this Stokes number, the planet has migrated inward, and the 'excitation band' has moved to $r \sim 1 \: r_0$. Then, about $1500$ orbits later, before $t=3000$ disk rotations, a second dust ring forms at this radial distance.

	This sequence of dust ring formation is also visible in the run with $M_p=13 \: M_e$, middle row (b). As the planet mass is larger, the perturbation of vorticity in the 'excitation band' is also larger, which induce a shorter time to trigger the ring instability, i.e. $\sim 850$ disk rotations. After the formation of the first dust ring, at $r=1.2\: r_0$, the planet has migrated inward displacing the 'excitation band' to $r=1.05 \: r_0$. This is why, $\sim 850$ disk rotations later, a second dust ring is formed. The process continues a third time, with the formation of the third dust ring at $t\sim 2400$ disk rotations, which started to be excited when the 'excitation band' was at $r \sim 1$, i.e. at $t=1700$ disk rotations.
	In the last run with $M_p=20 \: M_e$, bottom row (c), the first dust ring formed within $360$ disk rotations. Because this formation time is much shorter than the other case, the displacement of the 'excitation band' is reduced, despite a faster migration speed. The migration speed is proportional to $M_p$, while the dust ring formation timescale is proportional to $M_p^2$.

	Now, if we consider that during this timescale, the planet is migrating inward with a rate close to the prediction of Type I migration, as show Section \ref{Sect_Result_Migration}, we can estimate the orbital displacement of the planet at the end of the dust ring formation. Using the empirical value of Equation \ref{Eq_Ring_timescale} and the values of the migration rate above mentioned, we obtain $\Delta r = 0.15$, $0.11$, and $0.07 \: r_0$ for $M_p = 10$, $13$, and $20 \: M_e$ respectively. This result compares well with the orbital displacement of the planet during the formation of the dust ring. It gives also the right trend concerning the ring separation we observe, i.e. as $1/M_p$, and the values compare roughly well to the ring separations measured in the different runs. However, we do not observe any separation in the run with $M_p = 20 \: M_e$. We will explain why later.

\begin{figure}[t]
	\begin{center}
	\includegraphics[height=6.5cm, trim=0mm 0mm 0mm 0mm, clip=true]{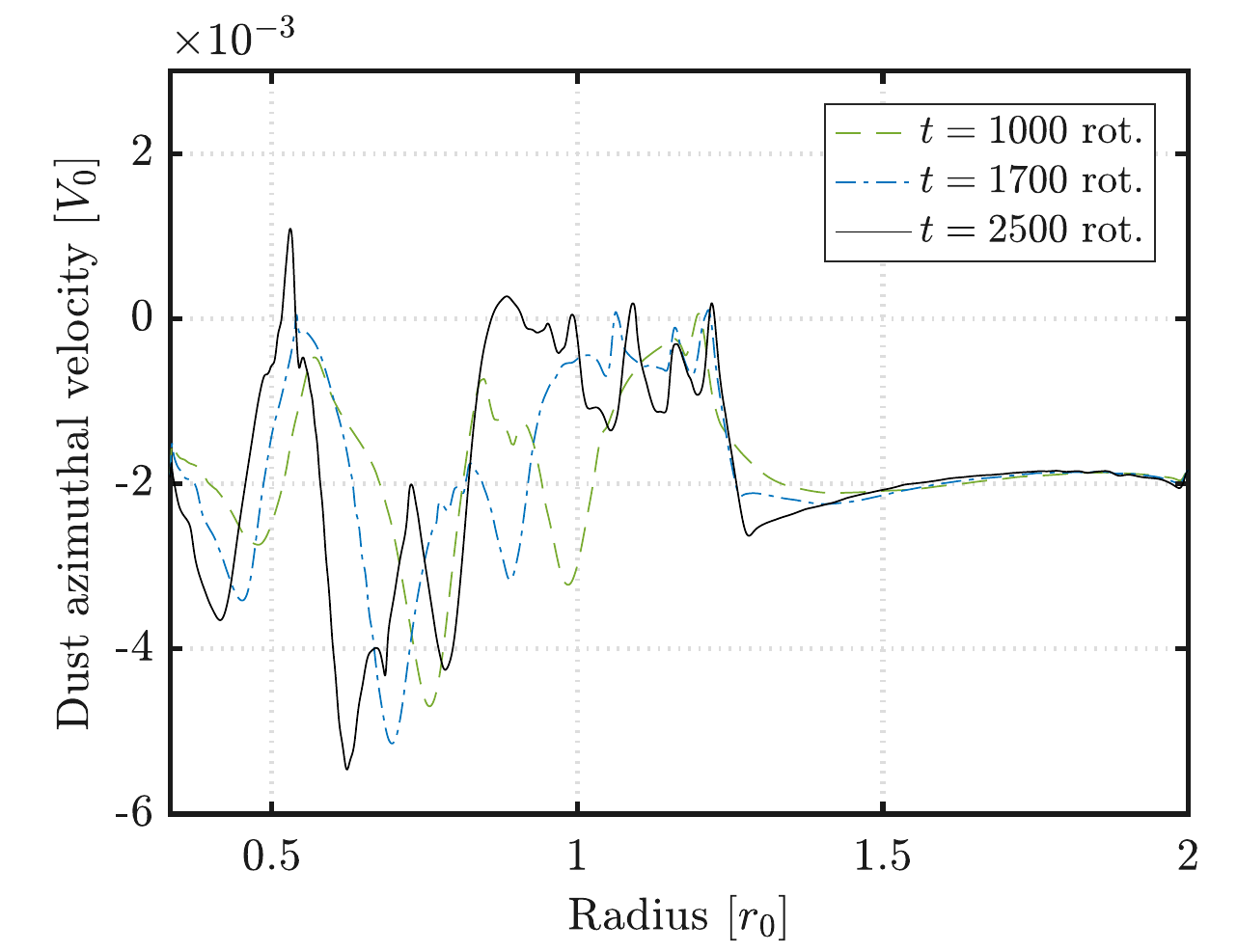}
	\caption{\label{Fig_Dust_Vy_ave} Azimuthal average of the dust azimuthal velocity, $V_{\theta}-r \Omega_k(r)$, in the run with $M_p=13 \: M_e$. Values at $t=1000$, $1700$, and $2500$ disk rotations are shown with different lines. }
  \end{center}
\end{figure}

	The formation of multiple dust rings due to the radial migration of the planet is one important new result of this study. The second one is the persistence of these rings during the disk evolution. In particular, the persistence of the degree of turbulence and of the orbital location of the rings, as it can be seen Figure \ref{Fig_Time_evo_rho_rossby}. The last aspect is due to the fact that there is a turbulent equilibrium between gas and dust velocities around the Keplerian motion. As a result, there is not mean force displacing the ring radially. However, the flow is not axisymmetric and has a high degree of turbulence. This can be seen Figure \ref{Fig_Dust_Vy_ave} in the case $M_p=13 \: M_e$. At the location of the dust rings, we see that the azimuthal average of the azimuthal velocity is close to Keplerian, i.e. close to zero in the plot. As time goes on, we can see that pics at zero appear, in particular in the region ${1<r<1.3 \: r_0}$. At $t=2500$ disk rotations, we clearly see that the additional pics still exist and the outermost are almost at the same location as they were formed.

	In a nutshell, we discovered a competing process where the formation of dust rings due to the KHI triggered by the dust drag happens in a timescale ${\propto 1/M_p^2}$, while the migration of the planet acts as displacing the rings by a separation ${\Delta r \propto 1/M_p}$. These results suggest that knowing the Stokes number of the dust grains and some local disk conditions (like gas density) allows to infer, in principle, the mass of the planet from the separation of the dust rings. Several observations of dust rings in disks are obtained nowadays in mm wavelengths \citep{ALMAPartnership2015, Andrews2018} as well as in the visible part of the spectrum (through scattered light) \citep{Garufi2014, Muro-Arena2020, Engler2020}. The spatial resolution currently available is still too coarse to resolve the separation seen in our models ($\Delta r=0.5$ au typically), but this might be possible in the future.

	While the rings are sustained at the same location for very long time, the flux of solids in the disk is not necessarily stopped. This is the topic of the next section.

\subsection{ Reduction of the flux of pebbles }
\label{Sect_Pebble_flux}

\begin{figure*}
	\begin{center}
	\begin{tabular}{ccp{15mm}}
	\scriptsize{Gas surface density} & \scriptsize{Gas pressure} & \\
	\imagetop{\includegraphics[height=6.5cm, trim=0mm 0mm 2mm 3mm, clip=true]{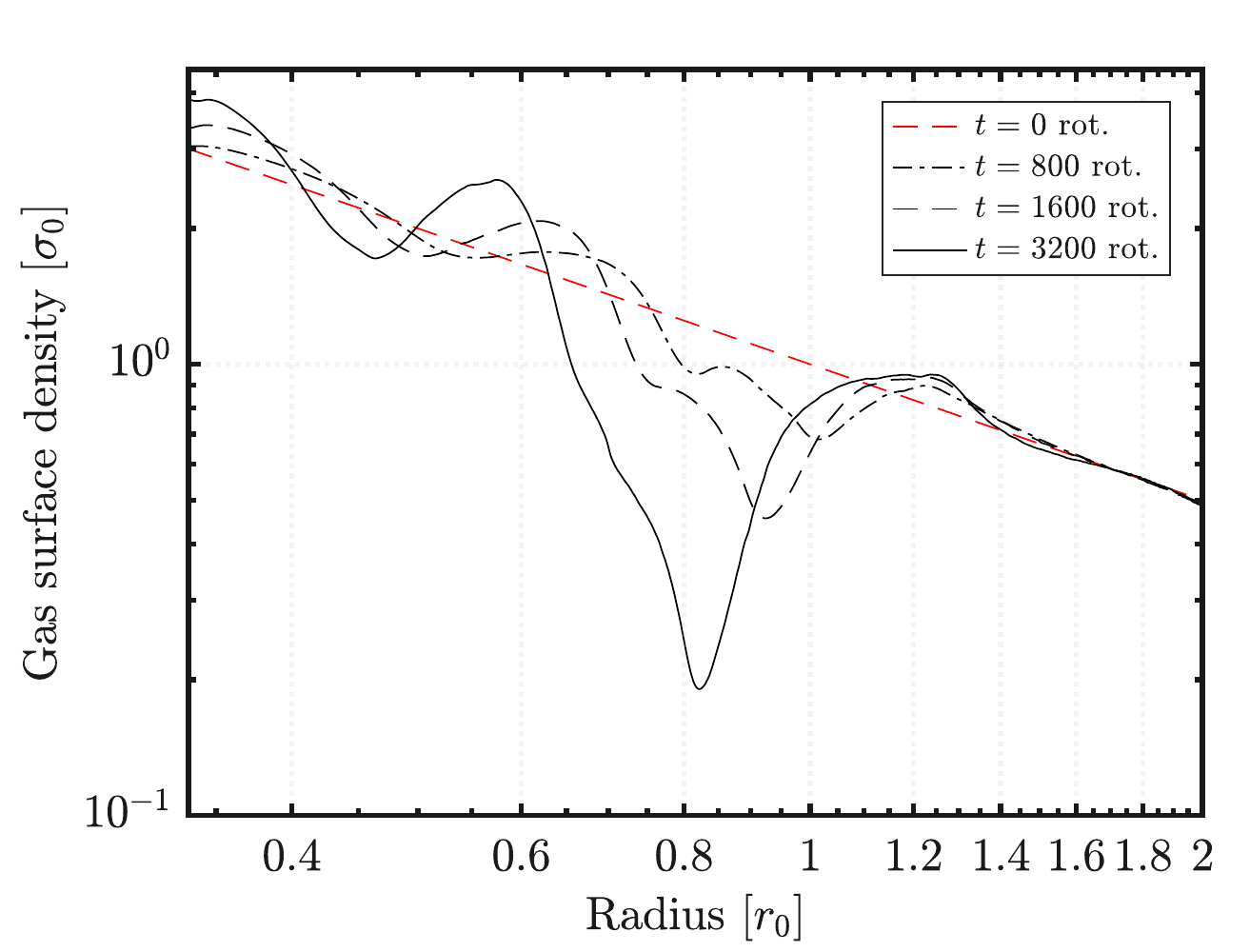}} &
	\imagetop{\includegraphics[height=6.5cm, trim=0mm 0mm 2mm 3mm, clip=true]{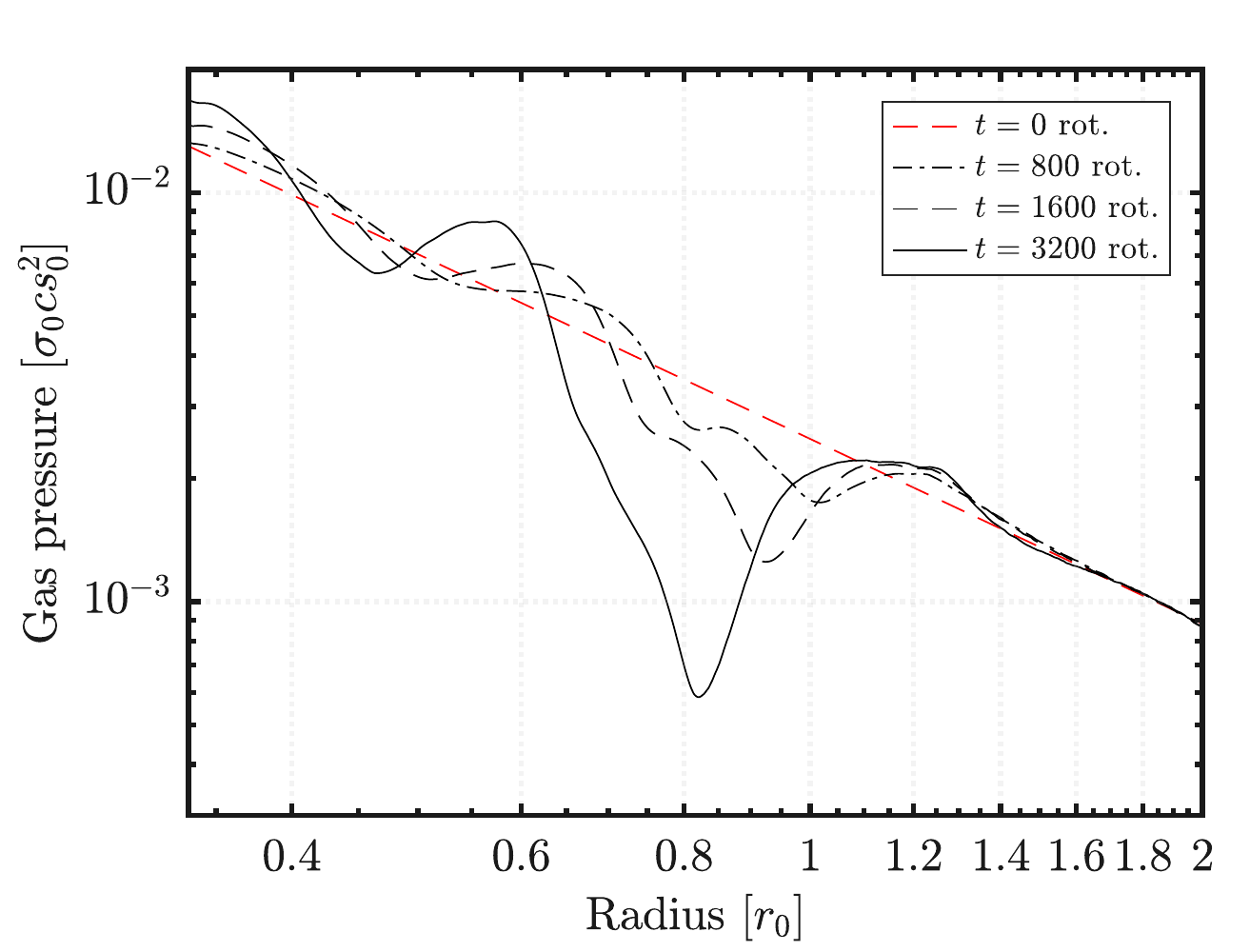}} &
	{\scriptsize{$\:$ \newline \newline (a)}} \\
	\imagetop{\includegraphics[height=6.5cm, trim=0mm 0mm 2mm 3mm, clip=true]{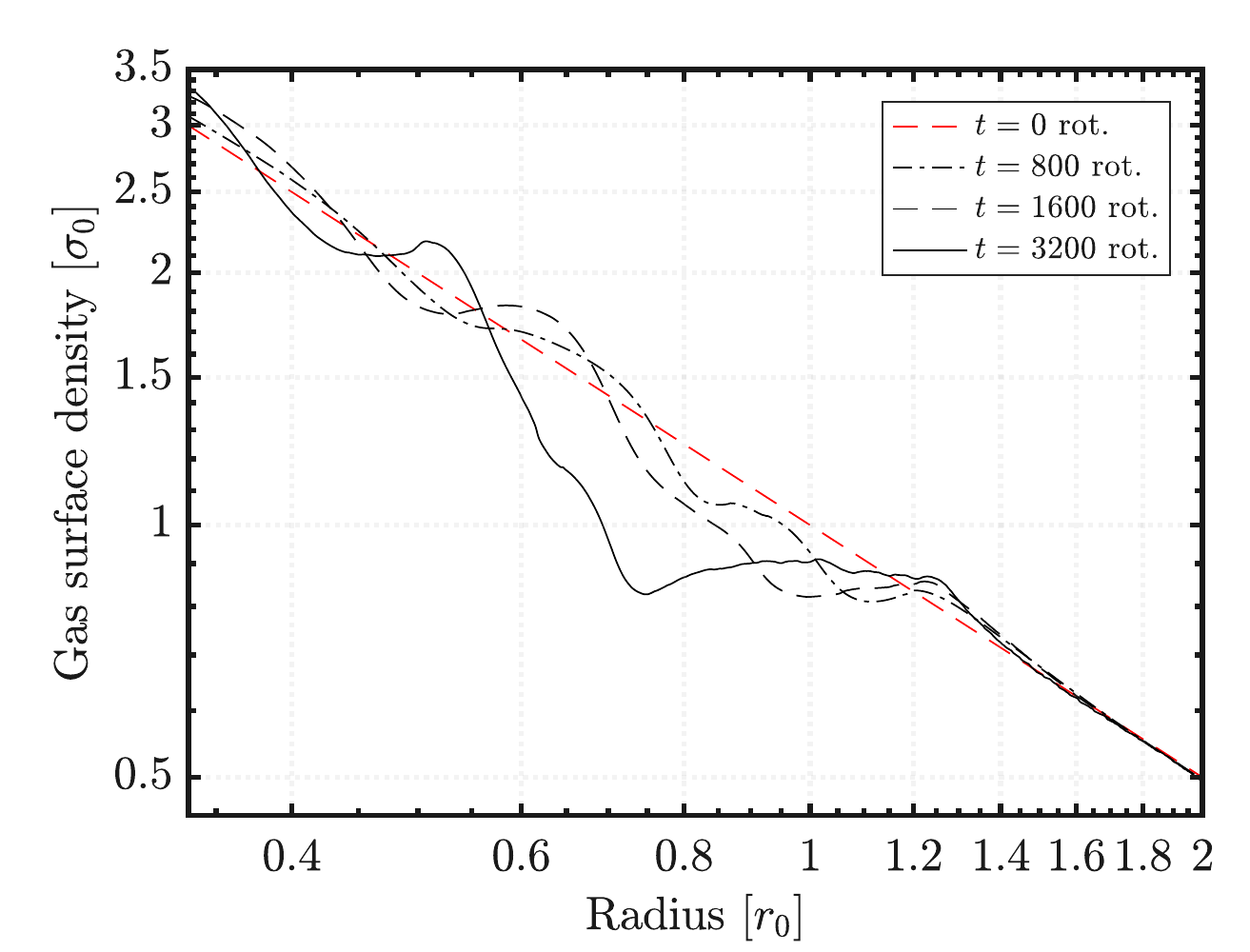}} &
	\imagetop{\includegraphics[height=6.5cm, trim=0mm 0mm 2mm 3mm, clip=true]{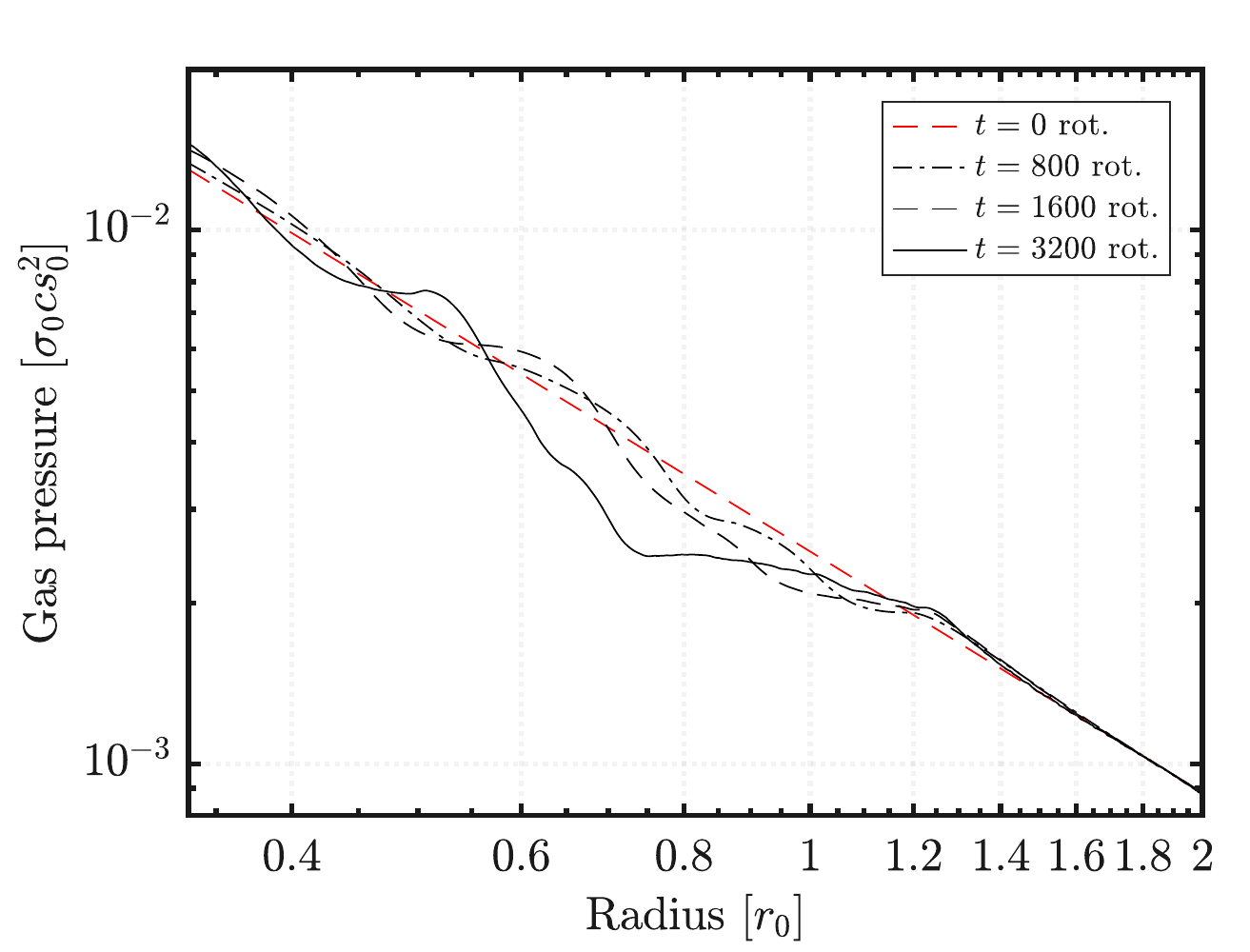}} &
	{\scriptsize{$\:$ \newline \newline (b)}} \\
	\imagetop{\includegraphics[height=6.5cm, trim=0mm 0mm 2mm 3mm, clip=true]{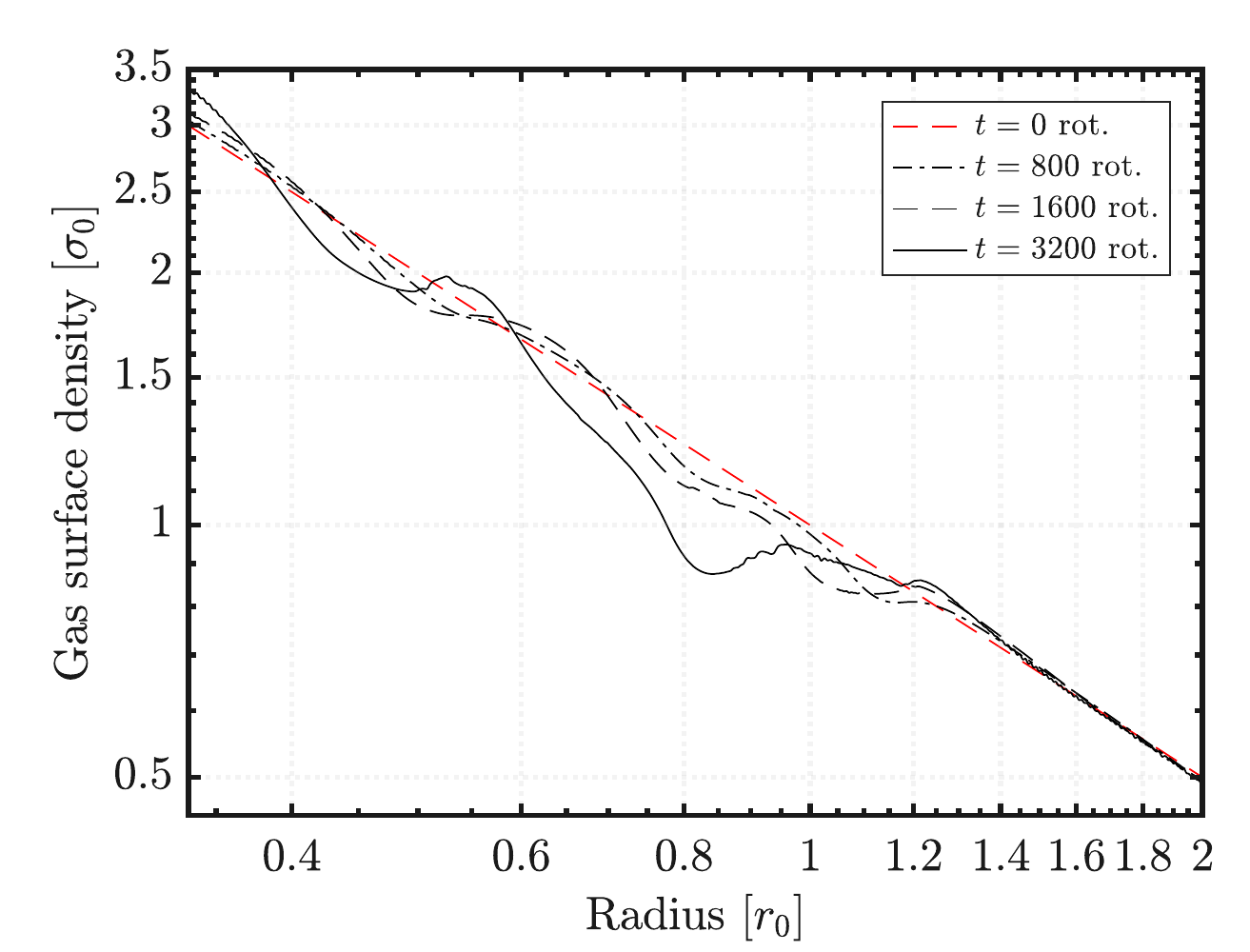}} &
	\imagetop{\includegraphics[height=6.5cm, trim=0mm 0mm 2mm 3mm, clip=true]{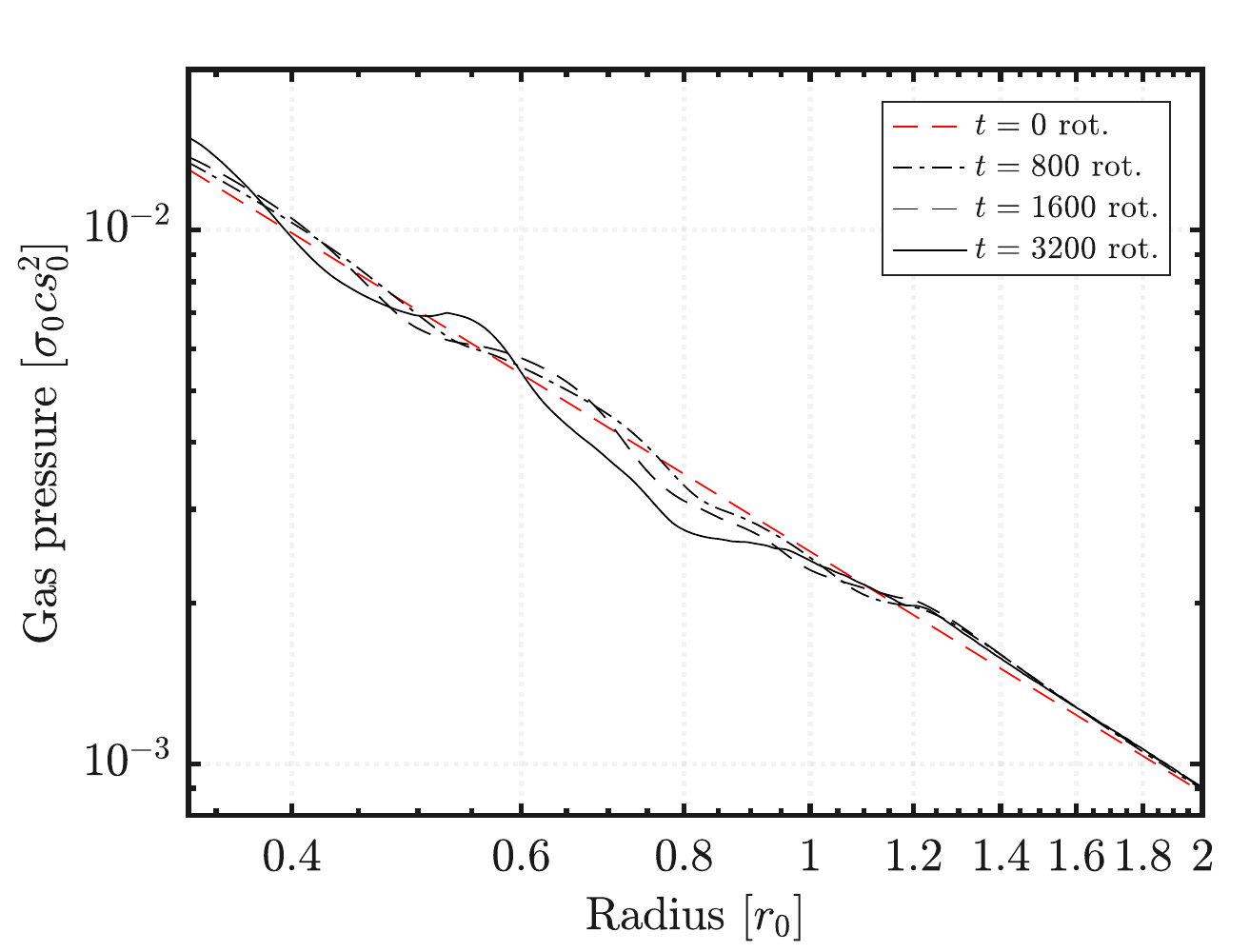}} &
	{\scriptsize{$\:$ \newline \newline (c)}}
	\end{tabular}
	
	\caption{\label{Fig_P_Rho_ave} Radial profiles of gas density and pressure, left and right respectively, at different moments of the disk evolution. Results for the different masses of the embryo $M_p=20, \: 13$ and $10\: M_e$ are shown top (a), middle (b), and bottom (c) rows, respectively. }
  \end{center}
\end{figure*}

	The dust rings that arise in the disk are not all dust traps. Despite the velocity is in average close to being Keplerian, the flux of pebbles varies along with the evolution of the disk. One of the aim of this study is to infer the efficiency of stopping the flux of pebble through the disk by a planet embryo. By stopping the pebble flux, one would avoid loosing the solid material available for planet formation. It also avoid the solids to flow towards the inner parts of the disk. This aspect is one of the constrains that the analysis of the solar system's material shows. Finally, stopping the pebble flux outside the planet orbit can stall its growth by pebble accretion, and could explain the high frequency of Super Earth planets in exoplanetary systems. We will see that this isolation from pebbles is time dependent, and not only dependent on the planet mass.

	We have seen that as the planet orbits in the disk, it perturbes density and pressure at large distances, because of the interaction between the density waves and the surrounding gas. This is responsible for the carving of a gap. The pressure profile affects the flow of solids, as it creates a discrepancy between the velocity field of the gas and of the solids, due to the pressure support in the gas fluid. Moreover, the pressure profile will be affected by the thermodynamics of the gas, as the shocks can heat the disk. As a result, density and pressure can evolve separately, in particular when the gas is far from isothermal equilibrium. In our model, the gas relaxes toward thermal equilibrium on timescale of 100 local orbits. It is more realistic than isothermal equilibrium, as the gas at 5 au in a disk has an optical depth around 100, which justifies a long cooling timescale. The effect of the thermal relaxation will be covered in a future study.

	We show Figure \ref{Fig_P_Rho_ave} the azimuthal average of gas density (left column) and pressure (right column) at different moments, for the three runs. The most obvious observation is that the deviation of the profiles from the initial state (red lines) amplifies as time goes on. This is due first to the accumulation of energy in the disk, sustained by the wave-disk interaction, and second to the migration of the planet, which displaces the gap region. The second most obvious inference is that the pressure profiles are flatter than the density ones. It means that the amplitude of gaps and bumps are more pronounced in density than in pressure. This is a consequence of deviation from thermal equilibrium. As a result, it can be noted that the local density gradient flattens and reverses more rapidly than the pressure gradient. We recall that the radial pressure gradient can be associated with the mergence of a dust trap.

	 Amount the three runs, the case with $M_p=20 \: M_e$ is the most trivial. Because the perturbations are stronger than in the other runs, the pressure gradient cancels quickly around $r=1.2 \: r_0$, creating a dust trap. This is the reason why this mass is often referred to as the pebble isolation mass. In fact, the dust flow from the outer disk is strongly reduced, or even cancelled at this region, avoiding solids to cross the planet orbit. There is a clear gap, even in the pressure, in particular where ${0.8<r<1.2 \: r_0}$. On the opposite side, where ${0.5<r<0.8 \: r_0}$, a bump is created. The amplitude of this bump increases with time, while its location slowly migrates inwards due to the migration of the planet. At $t=1800$ disk rotations, the pressure gradient is already canceled at the top this bump; a dust trap was created. At $t=3200$ disk rotations, this bump is sharper and has a kind of flat top shape.

	From this description we can identify two different regimes in two different regions around the planet: {\it{(i)}} outside the planet orbit, the gap edge where pressure decreases on the inner side of the potential dust trap, {\it{(ii)}} inside the planet orbit, the bump where pressure increases, somehow like a compression. Planet migration affects these two regions by moving the region of excitation to inner part of the disk. These two regions where identified in the Section \ref{Sect_Results} as places where the pebble surface density is the largest, and where turbulent dust rings can form. The existence of these two regions was already mentioned in \cite{Meru2019}, with isothermal 2D models, and migrating planet at a fixed inward rate. Here, we confirm that with a more realistic model, with thermal relaxation and self-consistant migration of the planet.

	If we focus on the first region, we can compare the evolution for different masses of the planet. While the pressure gradient cancels quickly when $M_p=20 \: M_e$, it is not the case for smaller planets. During the first $1600$ disk rotations, the radial profile of pressure flattens compared to the initial profile. This is more evident for $M_p=13 \: M_e$. The width of the region where this effects is at play increases with time, and is related to the displacement of the planet. As explained in the previous section, the 'excitation zone' moves inward with the planet migration. However, the key point here is that pressure does not relax toward the background profile. This is due to the exchange of momentum between solids and the gas. This back-reaction of the drag changes the flow of the gas compared to models without dust (or without back-reaction). Recently, \cite{Meru2019} did a study in similar regimes of planet mass and pebble size, with FARGO, but neglect the back-reaction of the solids onto the gas. As a result, they show that pressure relaxes almost to the initial profile when the planet migrates sufficiently inward. They also do not see the KHI of the high density dust ring that forms at the gap edge.

	The reason for this difference is that, when back-reaction is properly considered, gas velocity field is modified in the dust ring and becomes close to Keplerian motion, as shown previously. This is almost an equilibrium state of the dusty flow, which explains why it survives a long time. Thus no force exists to relax pressure to the initial profile. Moreover, the new dusty equilibrium state has a pressure profile that differs from the initial (or background) pressure, which is an equilibrium only in dust-free flow. As a consequence, when the planet moves inward, the pressure profile inherits the slope that was created, and becomes flatter and flatter. At late stage of the evolution, $t=3200$ disk rotations, we see that the region ${0.8<r<1.2 \: r_0}$ has a flat pressure profile, in particular in the case with $M_p=13 \: M_e$. In the case with $M_p=10 \: M_e$, the pressure gradient cancels at $r=0.8 \: r_0$. The density profile has even a positive slope in part of this region.

	To resume, the region outside the planet orbit is strongly affected by the drag back-reaction. A kind of rarefaction occurs, which reduces the pressure gradient gradually while the planet migrates inwards. The reduction of the averaged radial pressure gradient favors the triggering of ring instabilities, and explains why several of them can form when $M_p<20 \: M_e$.

	The second region, around $r=0.5 \: r_0$, evolves differently. The excitation of a bump, in pressure as well as in density, is visible in the three runs. However the amplitude of the bump is not proportional to the planet mass. The location of the bump shifts to the inner part of the disk, but the plots amplify this impression as the scaling of the radius is logarithmic. The width of the bump reduces while its amplitude increases with time, as a kind of compression of this region by the planet-disk interaction coupled with planet migration. As a result, the radial pressure gradient cancels at the top of the bump. This effect in not influenced by the presence of drag, and exists even in models with only gas. The more massive the planet, the faster the dust trap appears. In the case with $M_p=10 \: M_e$, the inner bump becomes a dust trap after $3000$ disk rotations. It is even earlier than the creation of a dust trap at the outer part of the disk, at $r=0.8 \: r_0$.

\begin{figure}[t]
	\begin{center}
	\begin{tabular}{c}
	\includegraphics[height=6.5cm, trim=4mm 0mm 2mm 3mm, clip=true]{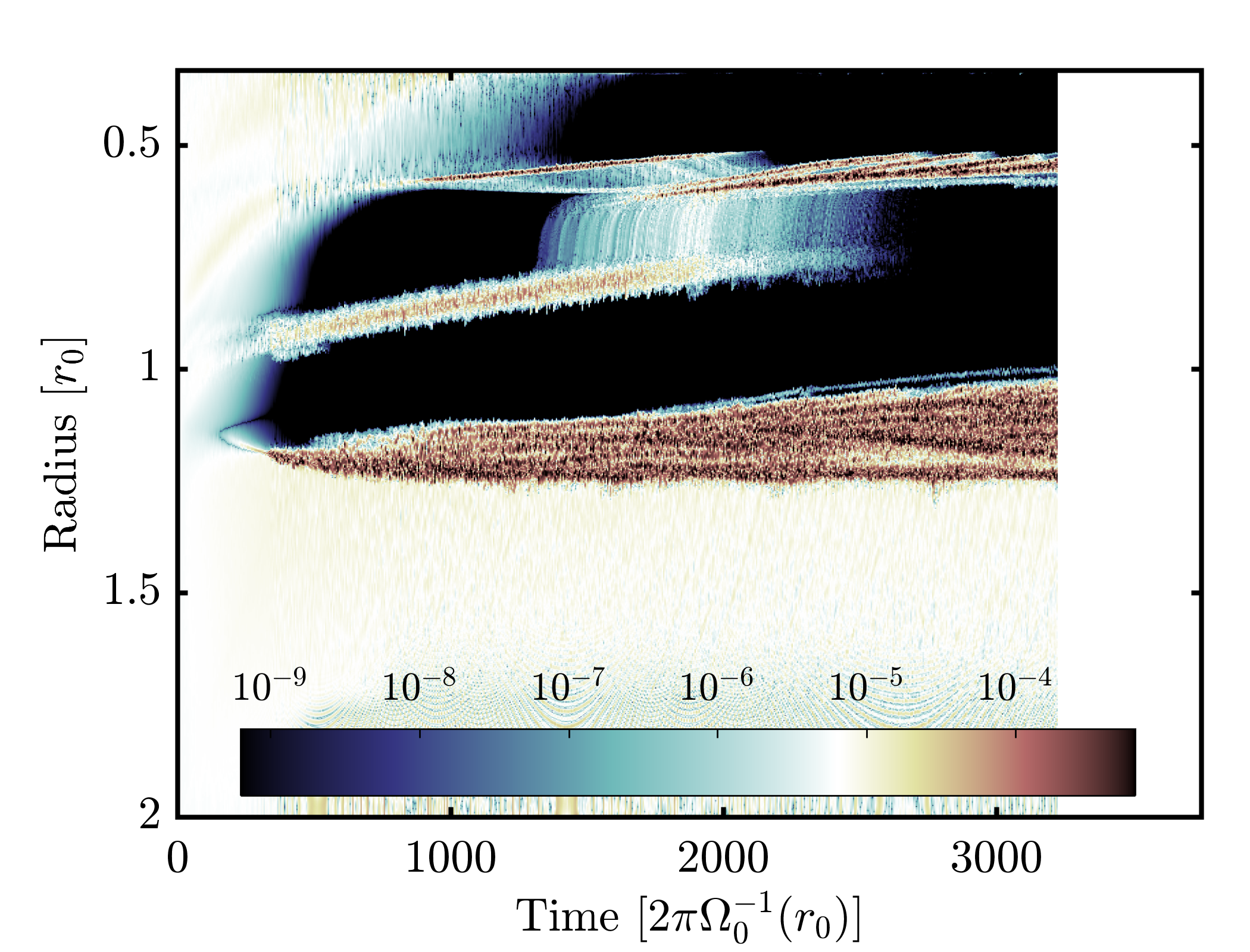} \\
	\includegraphics[height=6.5cm, trim=1mm 0mm 0mm 3mm, clip=true]{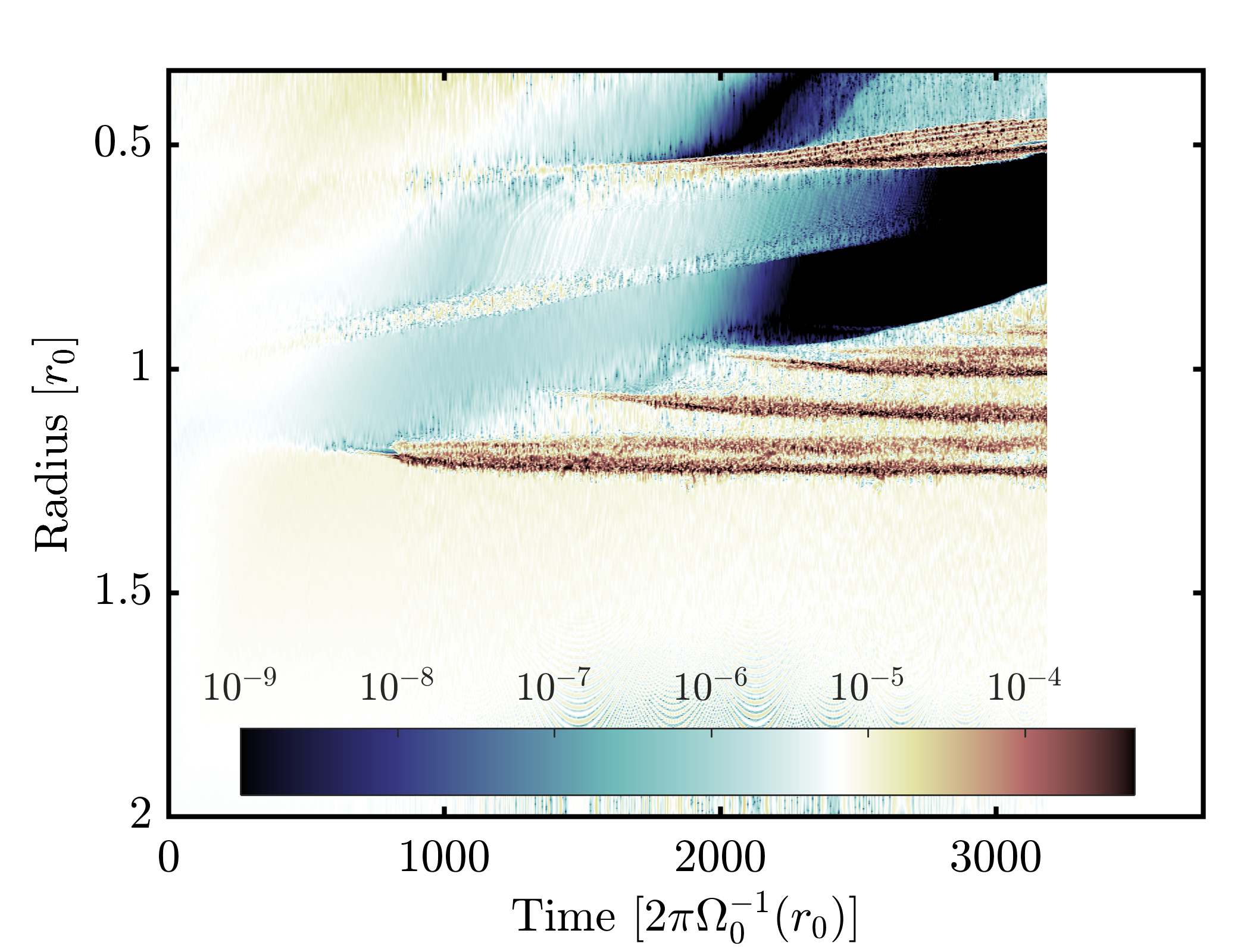} \\
	\includegraphics[height=6.5cm, trim=4mm 0mm 2mm 3mm, clip=true]{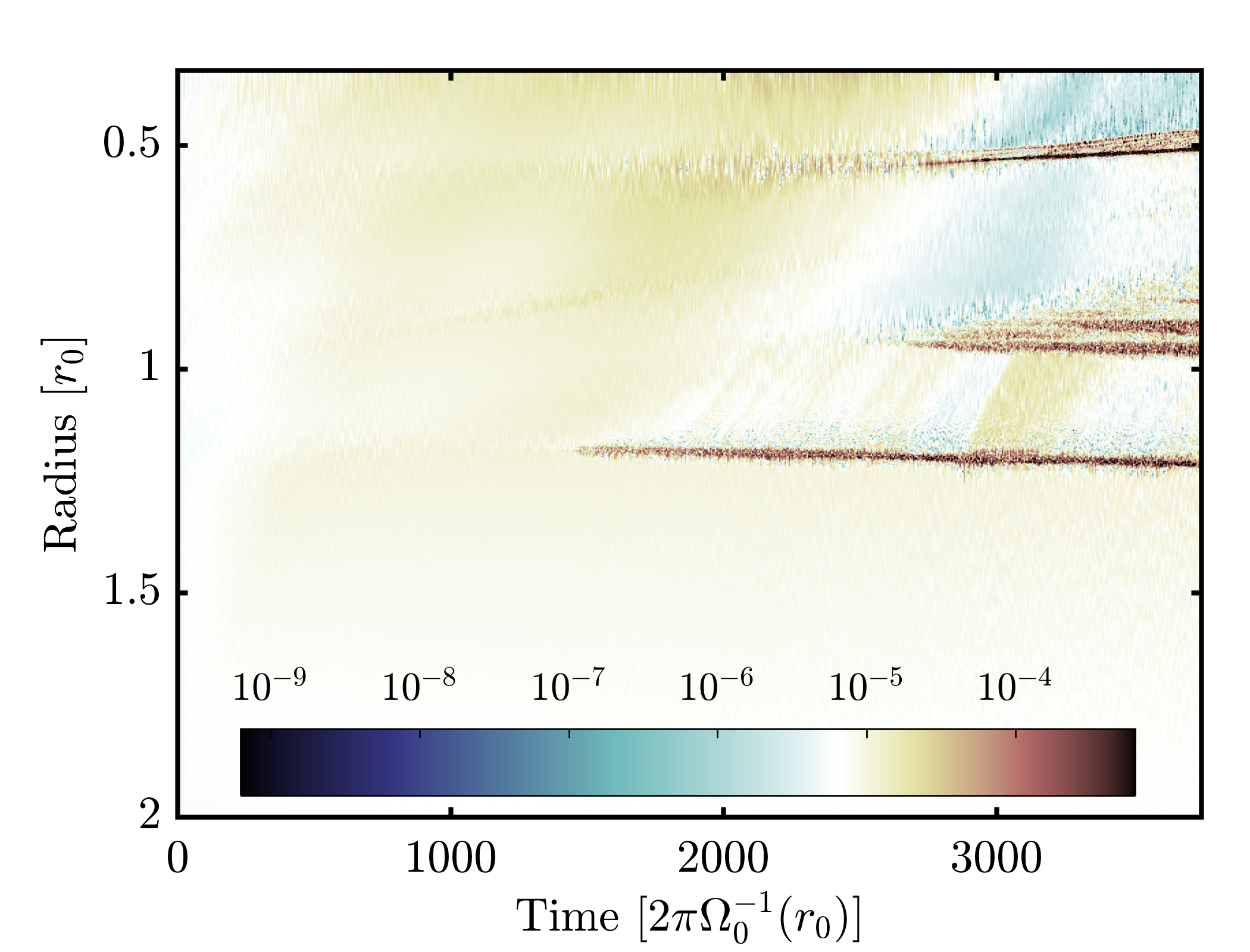}
	\end{tabular}
	\caption{\label{Fig_Time_evo_flux_pebbles} Time evolution of the radial pebble flux, as defined by the azimuthal integral of the radial momentum, $\int \sigma_p V_r d\theta$. Results of the different masses of the embryo $M_p=20, \: 13$ and $10\: M_e$ are shown top, middle, and bottom rows, respectively. }
  \end{center}
\end{figure}

	The chronology of formation of dust traps, and thus of reduction of the flux of pebbles through the disk, has a great impact on the models of planet formation and disk evolution. We show Figure \ref{Fig_Time_evo_flux_pebbles} the time evolution of the integrated radial flux of the pebbles for the three runs. The top panel shows the case with $M_p=20 \: M_e$. We can see that within the first hundreds of disk rotations, the radial flux of pebbles is efficiently reduced at ${r=1.2 \: r_0}$, and drops quickly below $10^{-9}$, i.e. more than four orders of magnitude smaller than the initial flux. During the first thousand disk rotations, the flux of pebbles is almost null inside the gap, but stays close to the normal level where ${r<0.6 \: r_0}$. This is because this region becomes a dust trap only after the bump of pressure excited there is large enough (see previous section). Latter on, at $t=1400$ disk rotations, the flux of pebbles cancels in turn in this region. The only region where the flux of pebbles does not reduce quickly is close to the planet. In fact, locally, the flux stays at a nominal level during almost $2000$ disk rotations, before it drops down under $10^{-9}$. However, this effect results probably from the presence of pebbles everywhere in the disk, and in particular in the vicinity of the planet, at the beginning of the simulation. In a realistic disk, a planet of such a mass would have depleted the solids from this region much earlier. The fact that pebble flux cancels very quickly at $r= 1.2 \: r_0$ reinforce this scenario.

	In the case with $M_p = 13 \: M_e$, middle panel, the evolution is different. During the first thousand disk rotations, the flux reduces in the co-rotation region of the planet, i.e. the small gap it carves. The formation of the turbulent dust ring is not sufficient to stop the flux of pebble. Until $t=2000$ disk rotations, this reduction is steady, with a flux $10$times smaller than the nominal one. This reduction results from the flattening of the pressure gradient in the disk. There is also a king of filtering by the turbulent dust ring, where solids can stay in the eddies, rather than flowing inward. At ${r=0.6 \: r_0}$, the reduction of the flux happens latter, at $t=1500$ disk rotations. Around $t=2000$ disk rotations, the two regions of dust trap formation become efficient, and a drastic reduction of the flux appears. The efficiency of stopping seems larger at the inner bump, than at the outer edge of the gap. It also appears slightly earlier. Then, until the end of the run, the flux of pebbles is negligible inside the gap edge, which moves inward with the planet.

	In the case with the lightest planet, $M_p= 10 \: M_e$, the variation of the flux of pebbles is not noticeable until $2000$ disk rotations. Even the formation of the first dust ring, at ${r= 1.2 \: r_0}$ has a little effect on the flux. Its only after $t= 2500$ disk rotations, that the two regions of dust trap formation produce a reduction by a factor of $10$ of the flux. In this case, the efficiency of the region at $r= 0.6 \: r_0$ is dominant.

	We can summarize the evolution of the flux of pebbles throughout the disk, as function of time, and of the mass of the planet as follows:
\begin{itemize}
\item $M_p <= 10 \: M_e$: the flux reduces slowly by the effect of the planet, and the inner dust trap region is more efficient. The reduction is only by a factor of 10.
\item $10 < M_p <= 20 \: M_e$: a reduction by a factor a $10-100$ is sustained during $2000$ disk rotations until stopping is complete both for the inner disk, and outside the planet gap. 
\end{itemize}

\begin{figure}[t]
	\begin{center}
	\includegraphics[height=6.5cm, trim=0mm 0mm 0mm 0mm, clip=true]{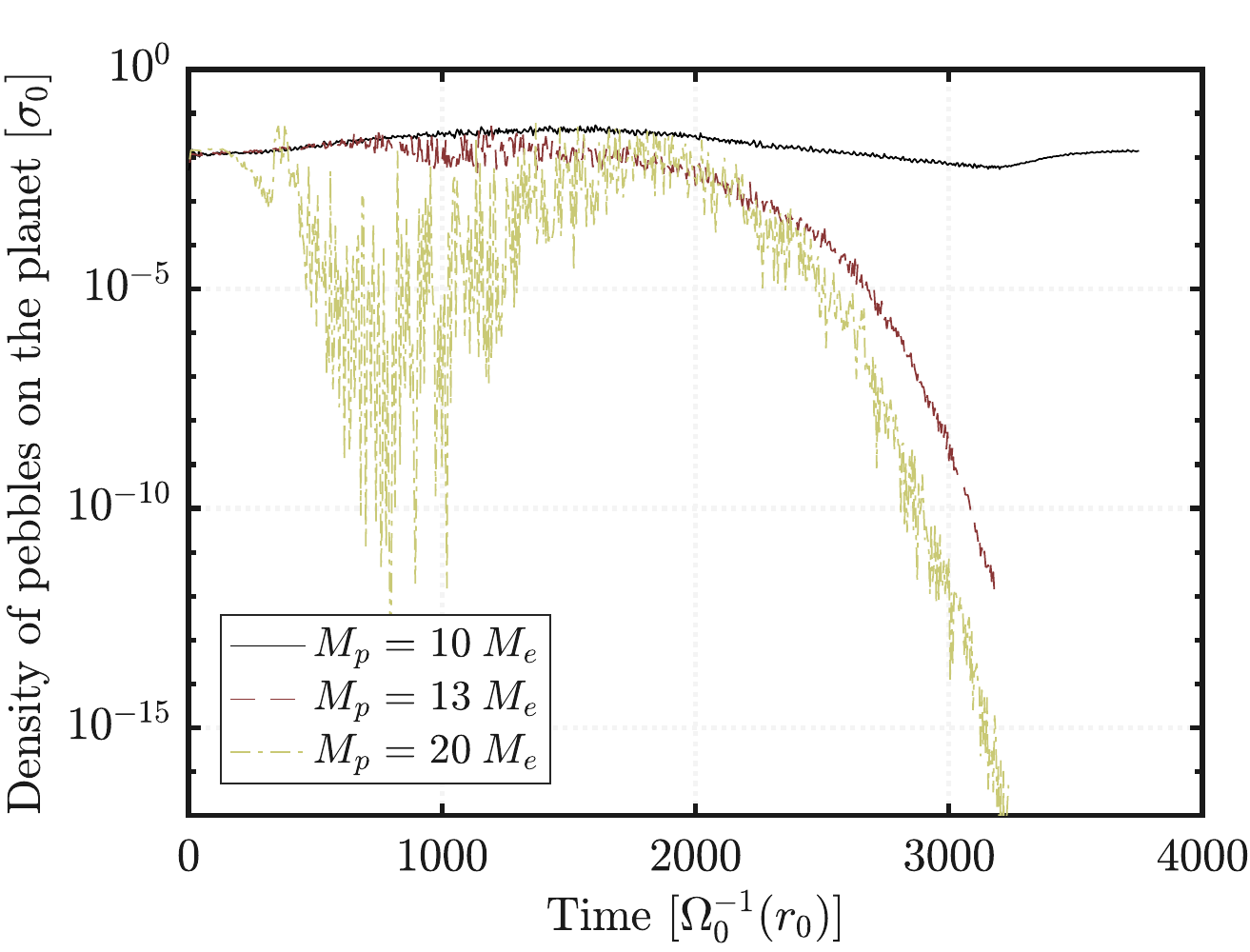}
	\caption{\label{Fig_Rho_p_planet} Evolution of the pebble density at the planet location. Comparison of the runs with $M_p = 10$, $13$, and $20\: M_e$, solid black, dashed red, and dashed-dot yellow lines, respectively. }
  \end{center}
\end{figure}

	To finish, we look at the possible accretion of pebbles on the planet, and if such an accretion can stall. We show Figure \ref{Fig_Rho_p_planet} the surface density of pebbles at the planet location, which gives an estimate of the amount of solids available for accretion onto the core. We see that while for the most massive planet the mass of solids drops quickly down by several orders of magnitude, the least massive planets have almost the same mass density of pebbles around them during $2000$ disk rotations. It means that they could continue to grow during that amount of time, which however is only 20 000 years at $5$ au. The reduction of the surface density happens only if the mass of the planet is at least $13 \: M_e$.

	Most importantly, we find evidence that the variation of the flux of pebbles in the disk is proportional to the planet mass. In the core-accretion scenario, in which a small seed solid core might grow by pebble accretion, the flux would be stopped well before the planet reach $20 \: M_e$. As a result, according to our results, the 'isolation mass' for pebble accretion is smaller than usually found (eg \cite{Bitsch2015}), likely in the range $12-13 \: M_e$, and depends on planet migration and on time. A caveat is that the models explored in this paper are inviscid, except for a small residual numerical viscosity. The effect of viscosity on pebble accretion was studied by \cite{Bitsch2018}, who found that a larger viscosity in the flow promotes pebble accretion. In addition, \cite{Ataiee2018} studied the influence of turbulence on the pebble isolation mass, and found that high levels of turbulents (associated with large viscosity) led to an increase of the pebble isolation mass. However, if we compare the case for low viscosity of \cite{Bitsch2018}, which is the most comparable to ours ( residual numerical $\alpha$-viscosity $<\sim 10^{-6}$ ) we can see that the pebble isolation mass is still higher than what we indicate here ($15-20 \: M_e$). This comparison should be taken with caution, though, because, due to the back reaction, the gas effectively feels a drag force, which acts as a dissipative force as viscosity would, an effect that could correspond to a higher viscosity in the simple parametrization of \cite{Bitsch2018}. In the latter case, our indication for a pebble isolation mass lower than previously found in the literature becomes even stronger.

	The evidence of a dust trap and the stopping of the pebble flux at the inner side of the planet orbit is another new result. It has a strong implication on the distribution of solids in young disk, during the growth of planet embryos. In particular, it can explain the dichotomie in the composition of chondrule, which happens early in the history of the Solar system \citep{Leya2008, Trinquier2007}. This also imply that eventual planetary embryos orbiting in the inner disk regions would be affected by such an effect. Their growth through pebble accretion would stall even if their mass are bellow the isolation mass.

\subsection{ The planetesimal reservoir }
\label{Sect_Planetesimals}

\begin{figure*}
	\begin{center}
	\begin{tabular}{ccp{15mm}}
	\scriptsize{Mass in cell rings} & \scriptsize{Integrated mass} & \\
	\imagetop{\includegraphics[height=6.5cm, trim=0mm 0mm 2mm 3mm, clip=true]{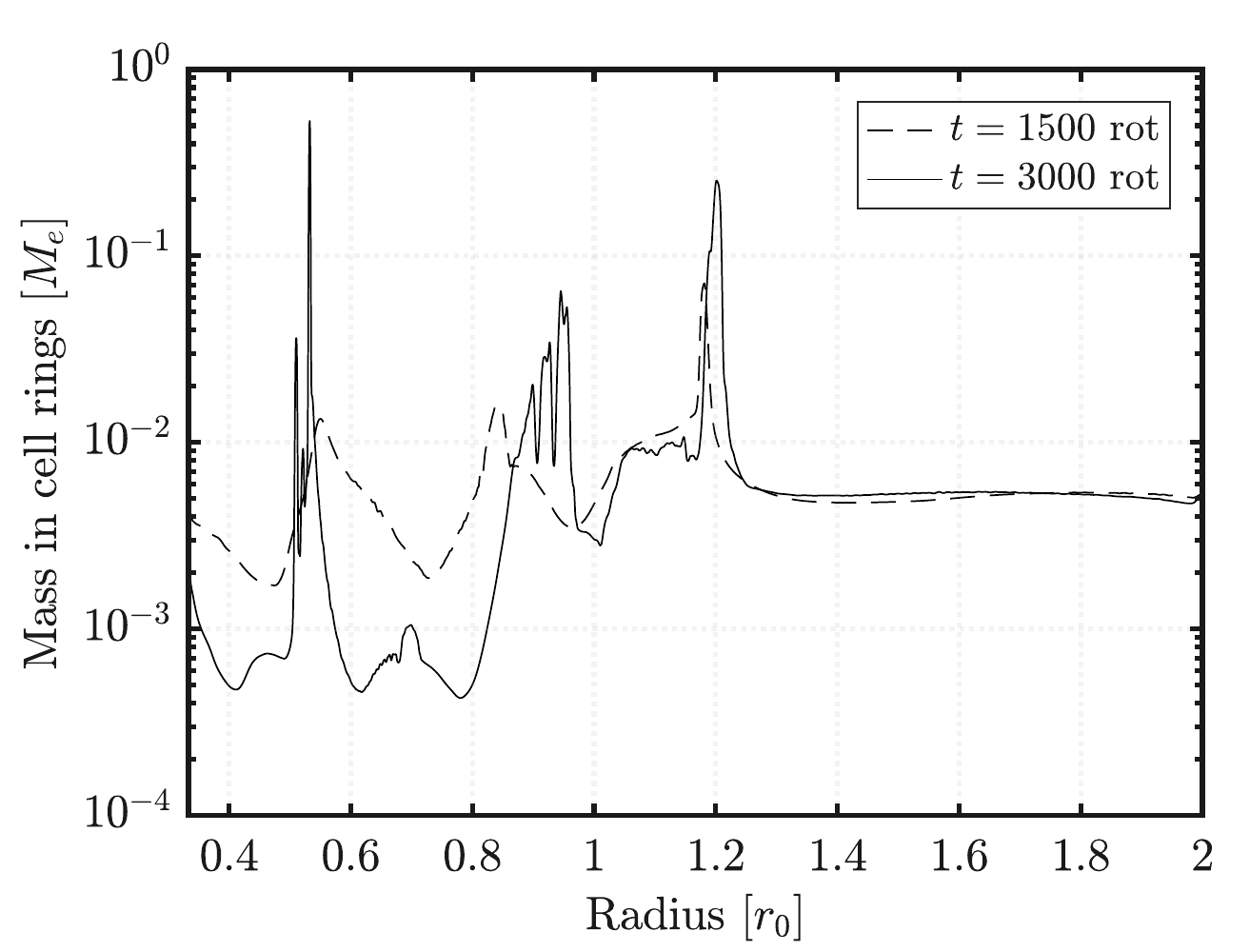}} &
	\imagetop{\includegraphics[height=6.5cm, trim=3mm 0mm 2mm 3mm, clip=true]{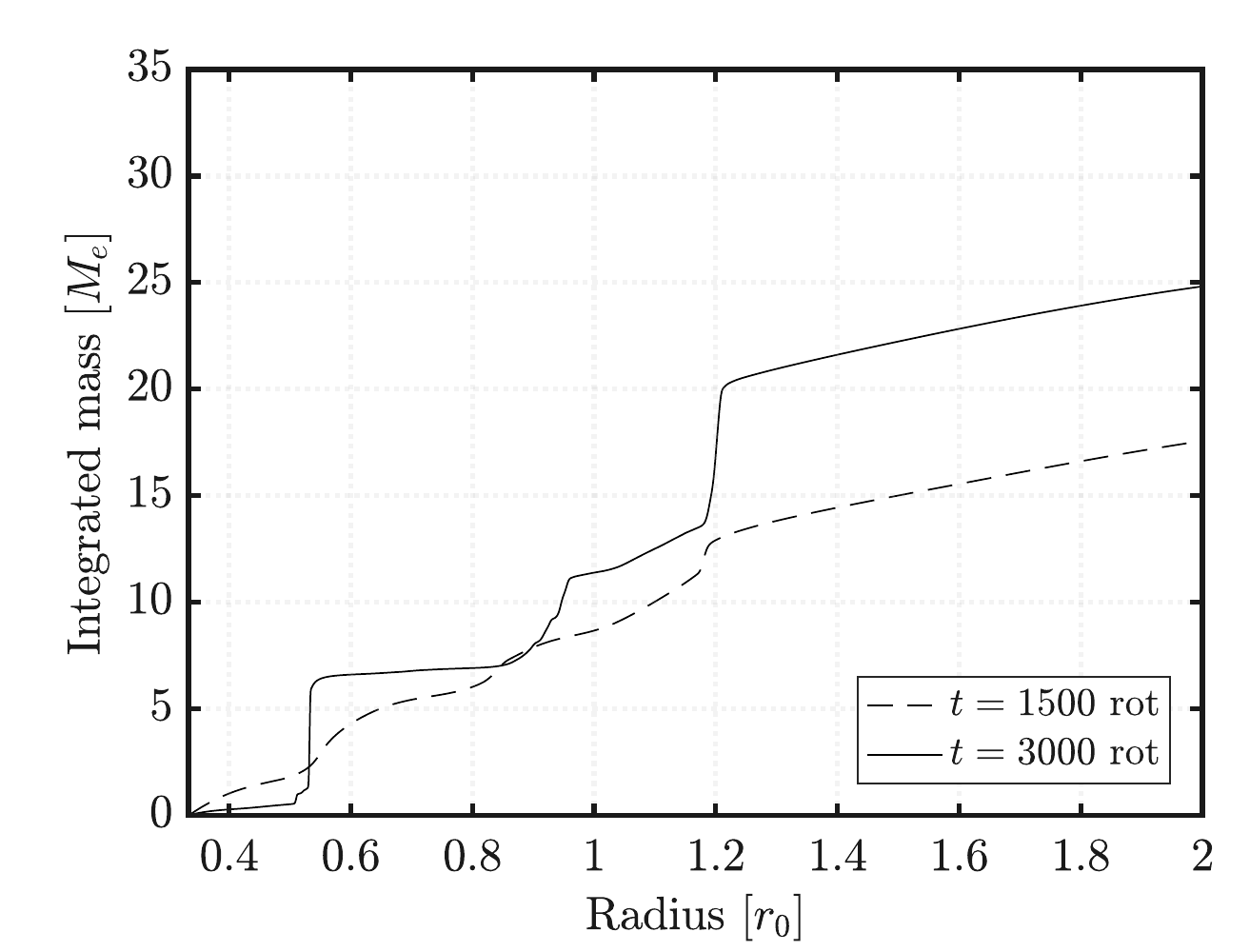}} &
	{\scriptsize{$\:$ \newline \newline (a)}} \\ 
	\imagetop{\includegraphics[height=6.5cm, trim=0mm 0mm 2mm 3mm, clip=true]{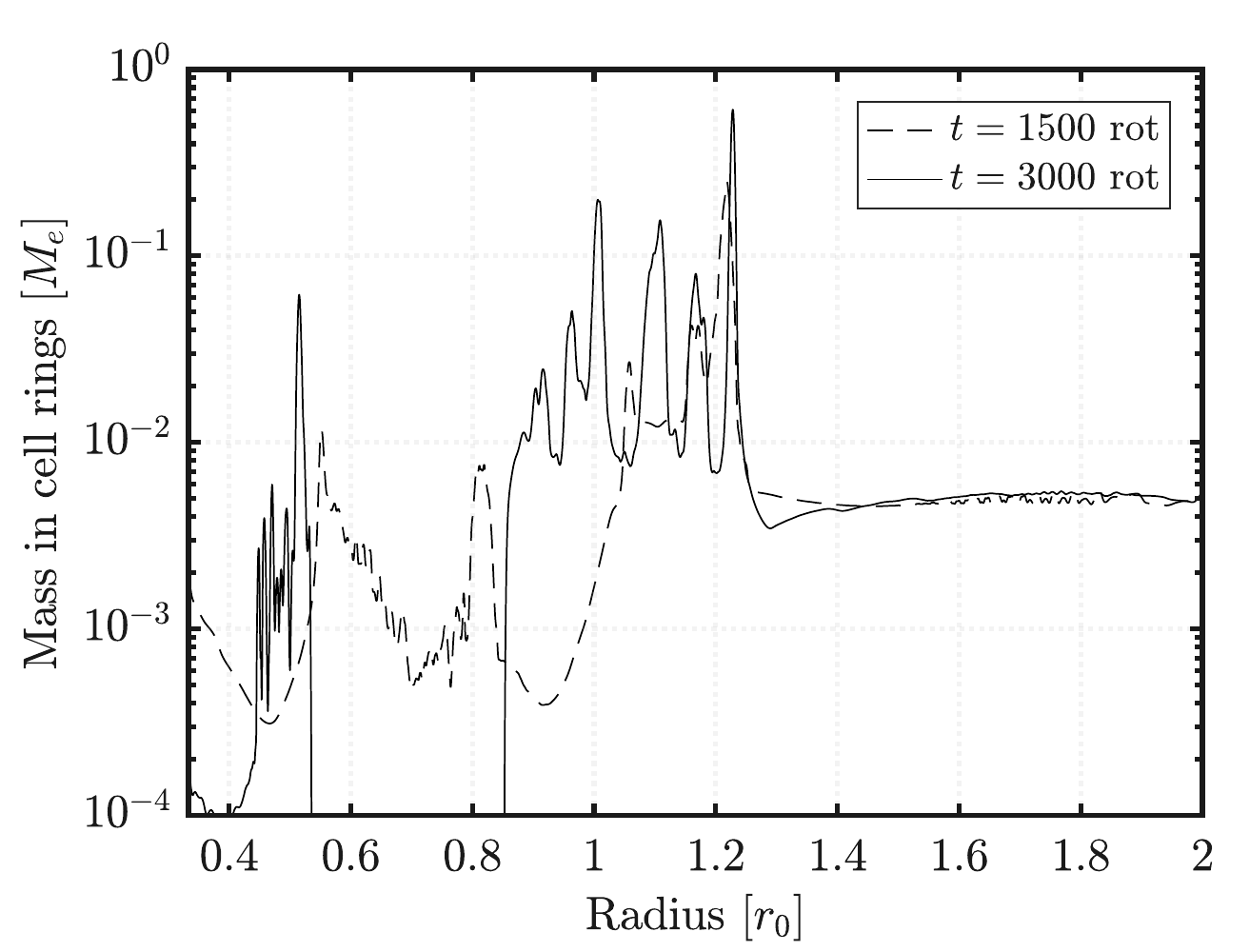}} &
	\imagetop{\includegraphics[height=6.5cm, trim=3mm 0mm 2mm 3mm, clip=true]{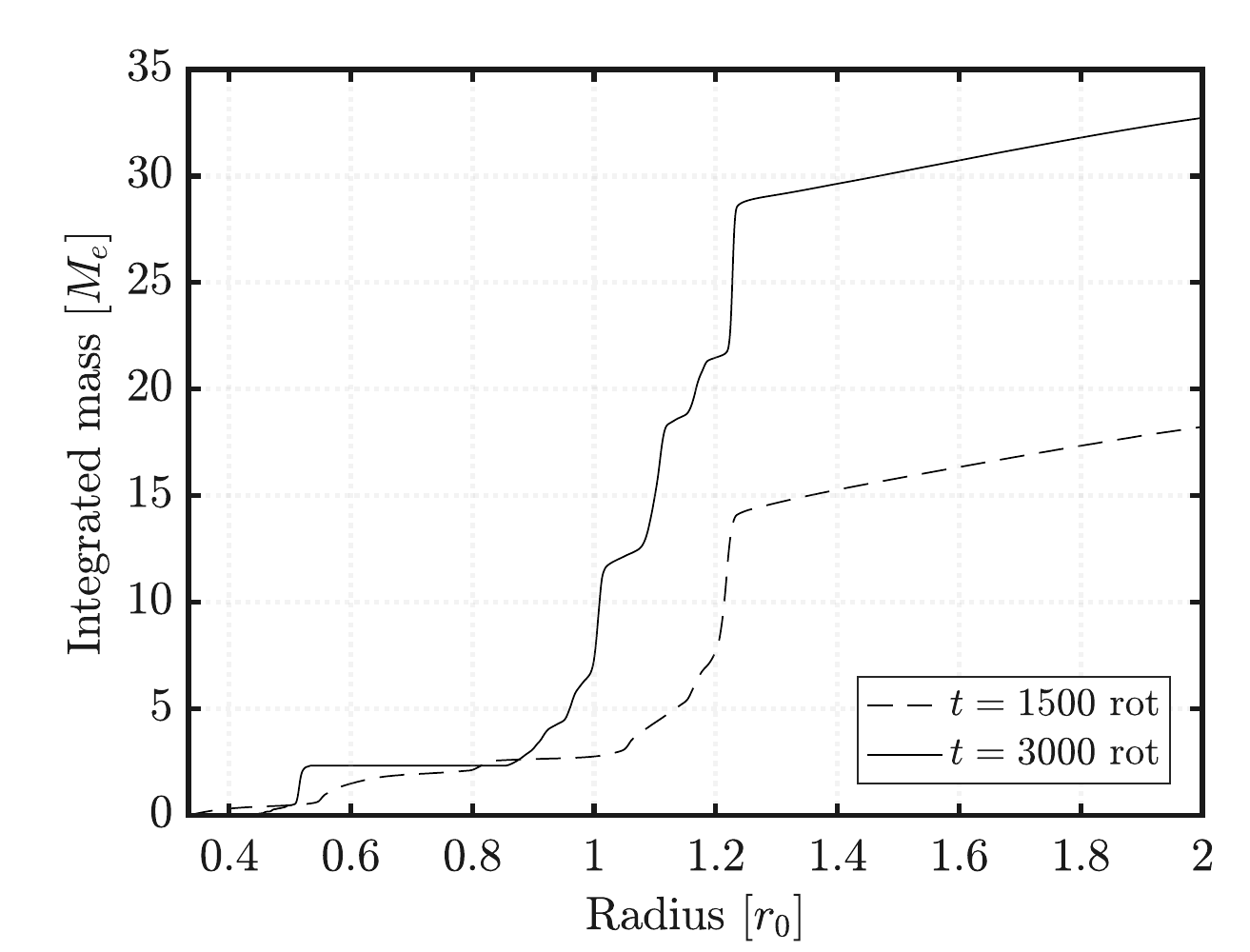}} &
	{\scriptsize{$\:$ \newline \newline (b)}} \\ 
	\imagetop{\includegraphics[height=6.5cm, trim=0mm 0mm 2mm 3mm, clip=true]{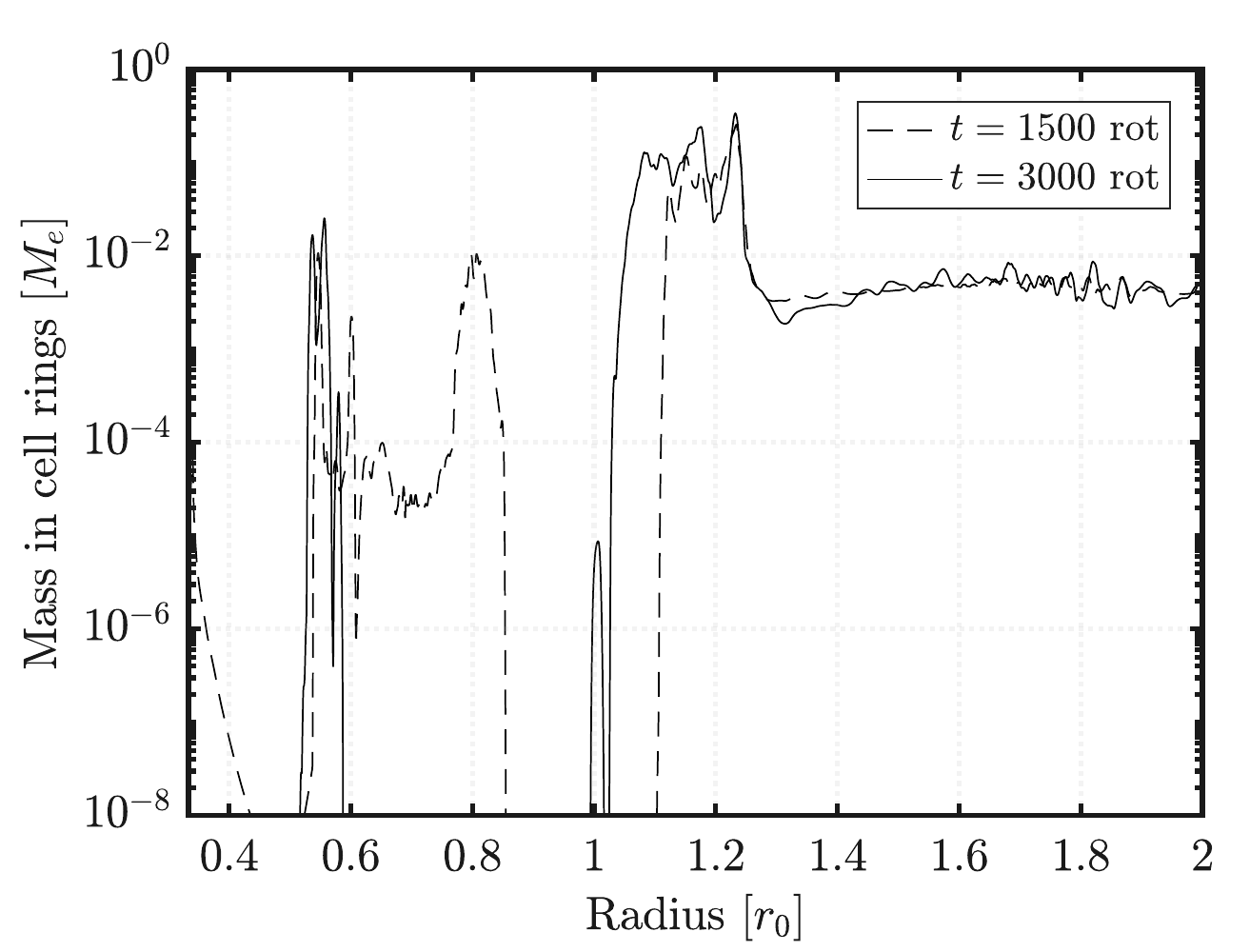}} &
	\imagetop{\includegraphics[height=6.5cm, trim=3mm 0mm 2mm 3mm, clip=true]{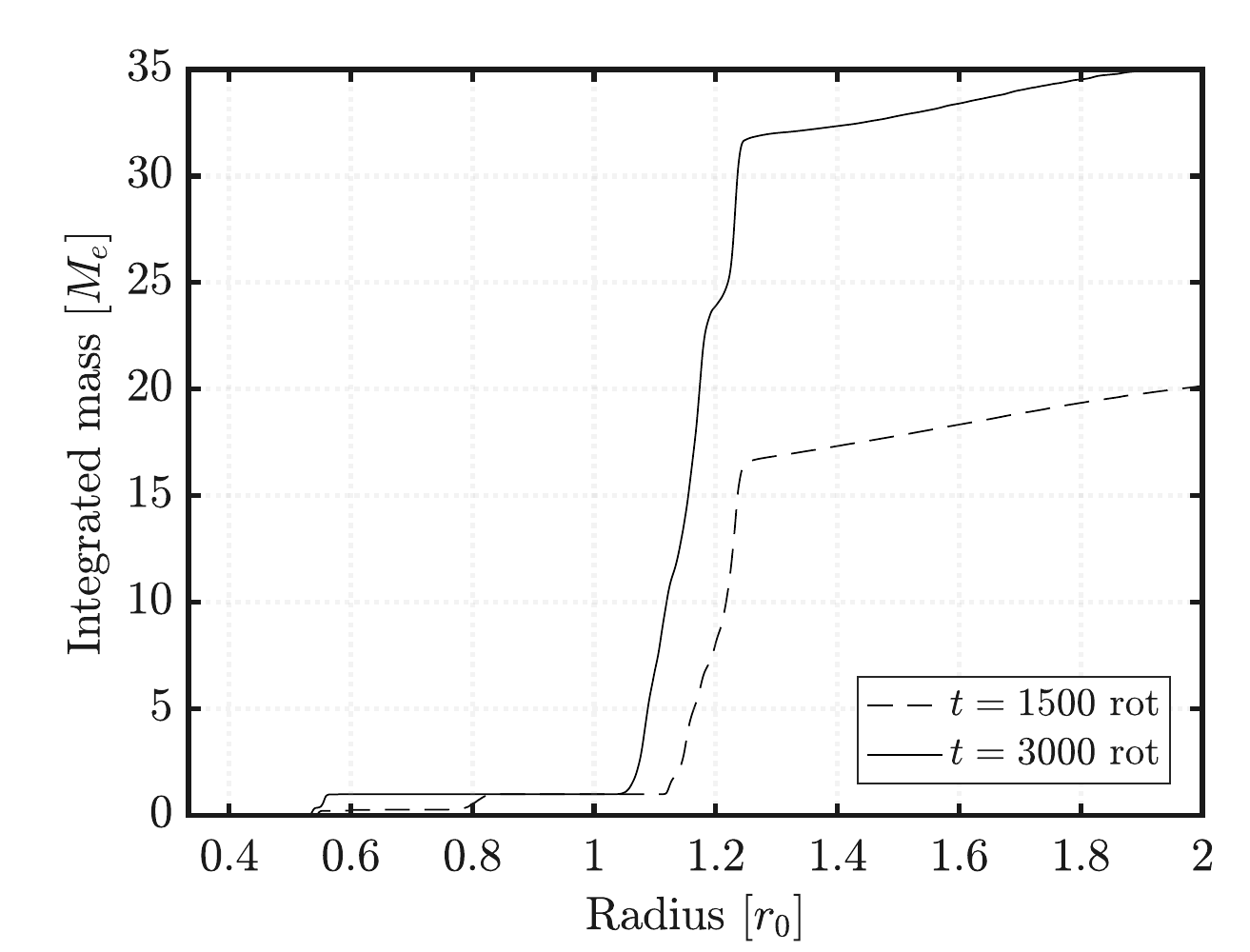}} &
	{\scriptsize{$\:$ \newline \newline (c)}} \\ 
	\end{tabular}
	
	\caption{\label{Fig_Mass_in_disk} Mass of pebbles in the disk at $t=1500$, and $t=3000$ rotations, dashed, and solid lines, respectively. Left: Total mass of pebbles contained at each radius of the grid, $dm(r)=\int \sigma_d r d\theta dr$. Right: Integrated mass of pebbles from the inner disk edge, as function of radius (see text for the formulation). Results for the different masses of the embryo $M_p=10, \: 13$ and $20\: M_e$ are shown top (a), middle (b), and bottom (c) rows, respectively. }
  \end{center}
\end{figure*}

\begin{figure}[t]
	\begin{center}
	\begin{tabular}{cp{15mm}}
	\imagetop{\includegraphics[height=6.5cm, trim=4mm 0mm 2mm 3mm, clip=true]{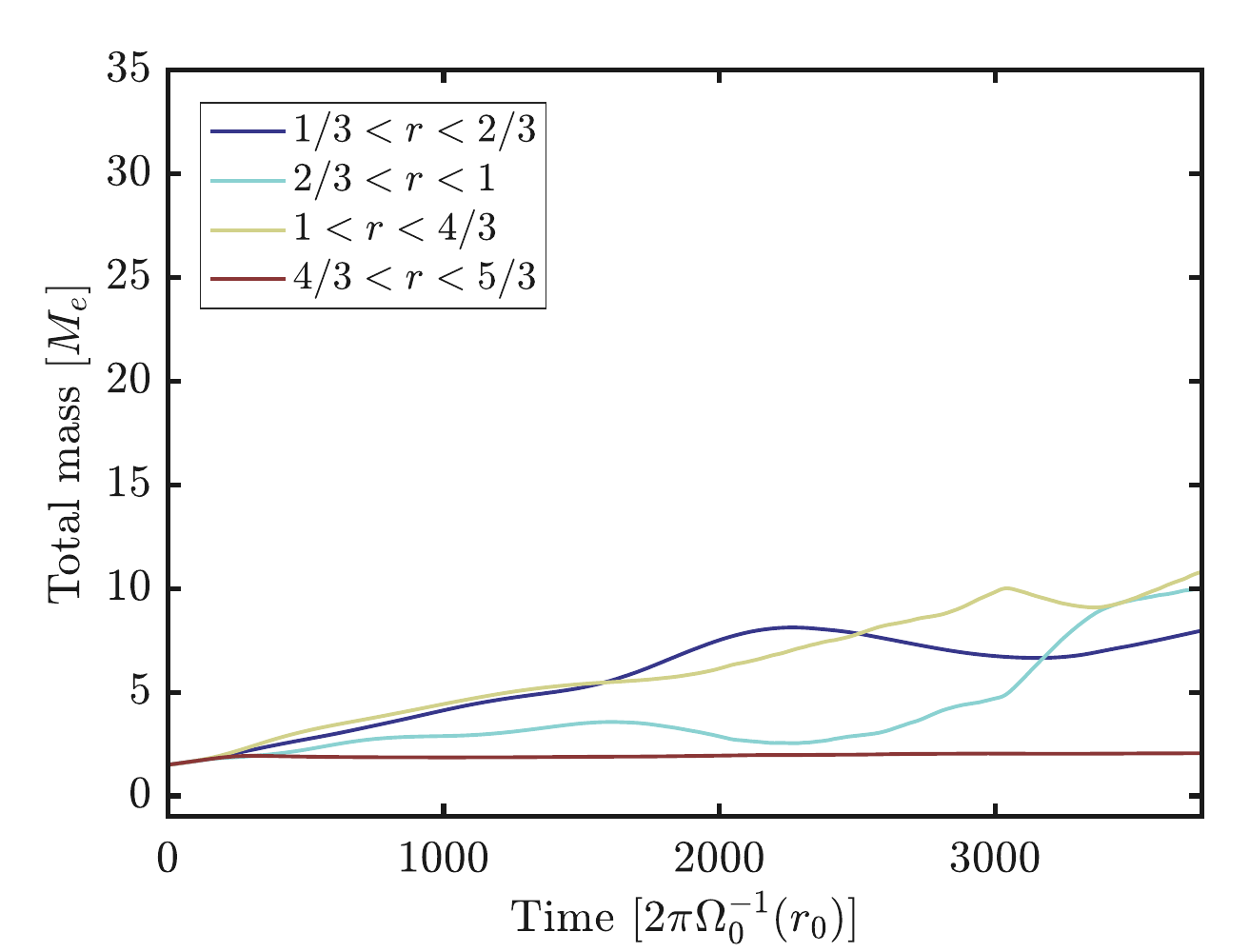}} &
	{\scriptsize{$\:$ \newline \newline (a)}} \\
	\imagetop{\includegraphics[height=6.5cm, trim=4mm 0mm 2mm 3mm, clip=true]{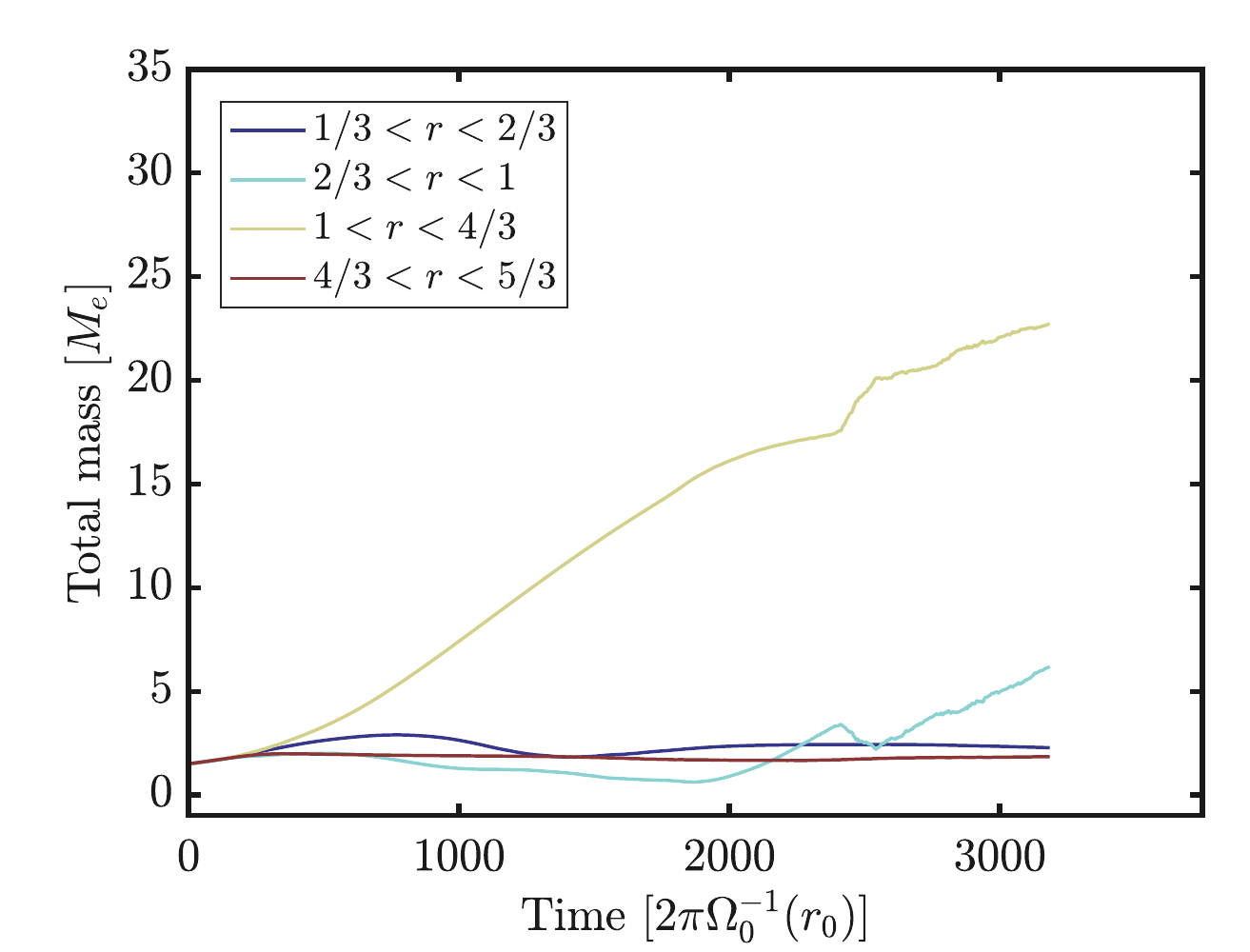}} &
	{\scriptsize{$\:$ \newline \newline (b)}} \\ 
	\imagetop{\includegraphics[height=6.5cm, trim=4mm 0mm 2mm 3mm, clip=true]{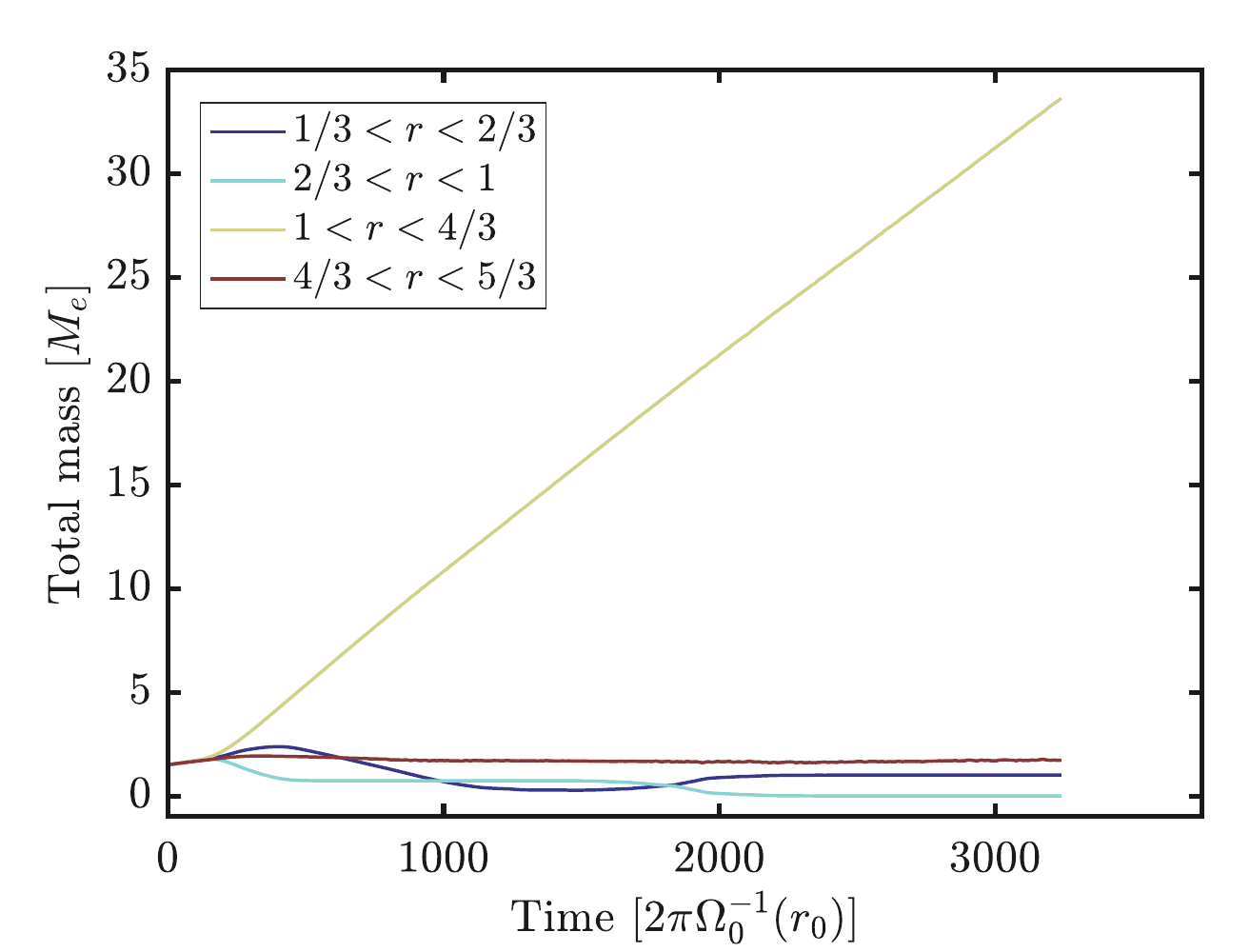}} &
	{\scriptsize{$\:$ \newline \newline (c)}} \\ 
	\end{tabular}
	
	\caption{\label{Fig_Mass_in_disk_rings_dust} Total mass of pebbles in different sections of the disk as function of time, for planet masses of $M_p=10$, $13$, and $20\:M_e$, (a), (b), (c), respectively. The yellow line corresponds to the dust ring formation region.}
  \end{center}
\end{figure}

	The formation of turbulent and long lived dust rings is systematic for the planet mass range we investigate. Their persistence and the high density of solids in them suggest that they could contribute to the formation of planetesimals see also Shibaike and Alibert (submitted) for a non-hydrodynamic approach of the problem of planetesimals formation from the flux of pebble) or even more massive small objects. While the eventual gravitational collapse is not captured by our models, and it will be the topic of another publication, the analysis of the mass distribution in the dust rings can give some insights on this possible path.

	The sensitivity to gravitational collapse depends on the mass and the size of a clump. Rather than looking at individual eddies, we analyse the azimuthally averaged surface density of the pebbles. We can construct different quantities: {\it{(i)}} the mass of pebbles in a ring of the computational domain (of radial width $\Delta r$), $dm(r) = 2 \pi \sigma r \Delta r$, {\it{(ii)}} the integrated mass contained between the inner edge of the disk, and a given location, $M(r) = \int_{r_{in}}^r dm(x) dx$. We show the profiles of these quantities for the different runs, at two given times Figure \ref{Fig_Mass_in_disk}.

	On the left, the mass contained in a grid ring recreates the structures we observe in the disk snapshots, with the presence of gaps in the pebble distribution and highly loaded rings. If we look at the high mass regions, corresponding to the turbulent dust rings, we see that the mass in a cell annulus pics above $0.1 \: M_e$. In fact, while in the outer parts of the disk the mass is $\sim 5 \times 10^{-3} \: M_e$, we observe that the rings are almost 100 times more massive. This would affect the local gravitational instability threshold, i.e. the Toomre parameter, and it is highly possible that the dust rings are unstable. The 2D nature of the rings, with several eddies, may boost this succeptibility and produce compact objects of mass in the range $10^{-2}-10^{-1} \: M_e$. Obviously, this has to be confirmed by calculations including self-gravitation.

	The main difference of distribution between the three planet masses, is at the inner ring. For the $M_p=10 \: M_e$ model, the mass of this ring is larger than in the other rings, and also larger than for the other planet masses. This is due to the efficiency of the dust trapping at $r=0.6 \: r_0$ for this case (see previous section). As a result, the $M_p=10 \: M_e$ case is prone to trigger the formation of planetesimals or even larger bodies in the inner regions of the disk. This could contribute to an outside-in planet formation scenario.

	Concerning the outer region, where multiple dust rings form, the pic mass is similar for the three cases, but the distribution is broader as planet mass increases, meaning that a larger amount of pebbles is contained outside the planet orbit for the most massive planets. This can be seen on the right column of Figure \ref{Fig_Mass_in_disk}. Each variation of the value of the curves represent the amount of solids contained within a section of the disk. As a result, each jump is due to the presence of a dust ring, and the height of the jump corresponds to the mass of the dust ring.

	We observe that for $M_p = 10-13 \: M_e$, the mass of individual dust rings is around $5 M_e$, at $t=3000$ disk rotations. However, in the case of the isolation mass, $M_p = 13 \: M_e$, the amount of pebbles stored outside the gap ($r>0.8$) is close to $30 \: M_e$. Interestingly, it is also the amount of pebbles contained in the dust rings of the model with $M_p = 20 \: M_e$. This matching is due to the fact that the flux of pebbles can stop for these two cases, and the rings retain the pebbles, while for $M_p = 10 \: M_e$, the solids can flow slowly inside, and accumulate at the inner dust trap, at $r= 0.6 \: r_0$.

	This highlights that the time evolution of the mass of pebbles in the disk is important. We show Figure \ref{Fig_Mass_in_disk_rings_dust} the time evolution of the mass of pebbles contained in different annulus of the disk. For $M_p = 10 \: M_e$, top panel, we clearly see that the mass of pebbles contained in ${1/3<r<2/3 \: r_0}$ (the region where the inner dust trap forms) is the largest, and growth with the same linear rate as the mass contained in ${1<r<4/3 \: r_0}$ (the region of the multiple dust ring formation). For the two other case, when the planet is bigger than the isolation mass, the amount of pebbles grows the most in ${1<r<4/3 \: r_0}$. More than $25 \: M_e$ of solids can convert to planetesimals in this region.

	The linear growth of the mass of solids indicates that a huge amount of material can be stored in a limited region of the disk, if the process continues much beyond the duration of our models. This reservoir of solids could contribute to the formation of planetesimals, and eventually of astroid belts. It could also contribute to the formation of massive planets, by the feeding of planetesimal accretion of the embryo that triggered it (this will be discussed in the next section). Finally, such a reservoir could produce a fast accretion event if another planet coming form the outer parts of the disk enters this region because of radial migration. This could support an inside-out planet formation scenario. These implications need to be confirmed or inferred by additional work and future publications.

\subsection{ Implications on the formation of Jupiter }
\label{Sect_Jupiter}

\begin{figure}[t]
	\begin{center}
	\begin{tabular}{cp{15mm}}
	\imagetop{\includegraphics[height=6.5cm, trim=2mm 0mm 2mm 3mm, clip=true]{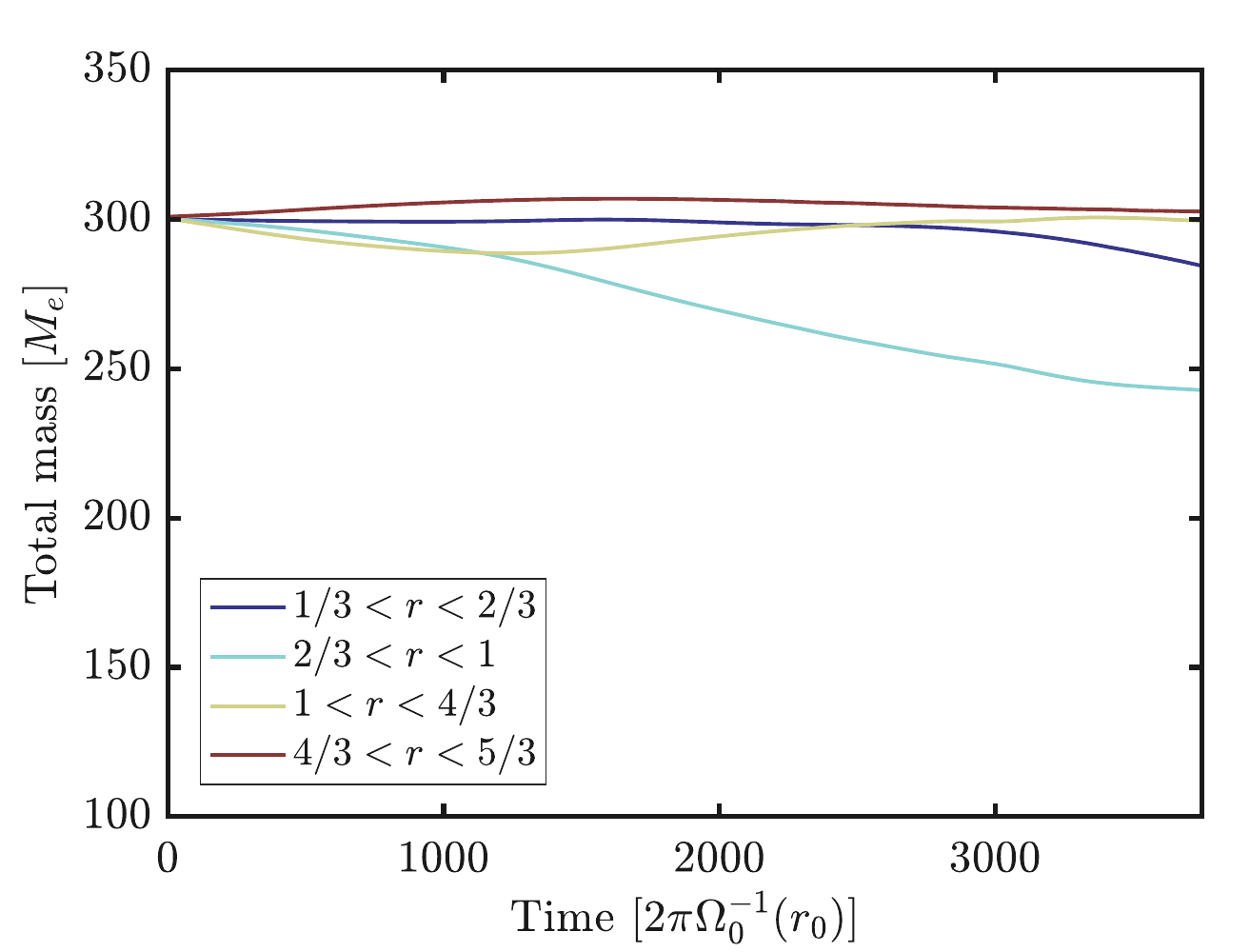}} &
	{\scriptsize{$\:$ \newline \newline (a)}} \\ 
	\imagetop{\includegraphics[height=6.5cm, trim=2mm 0mm 2mm 3mm, clip=true]{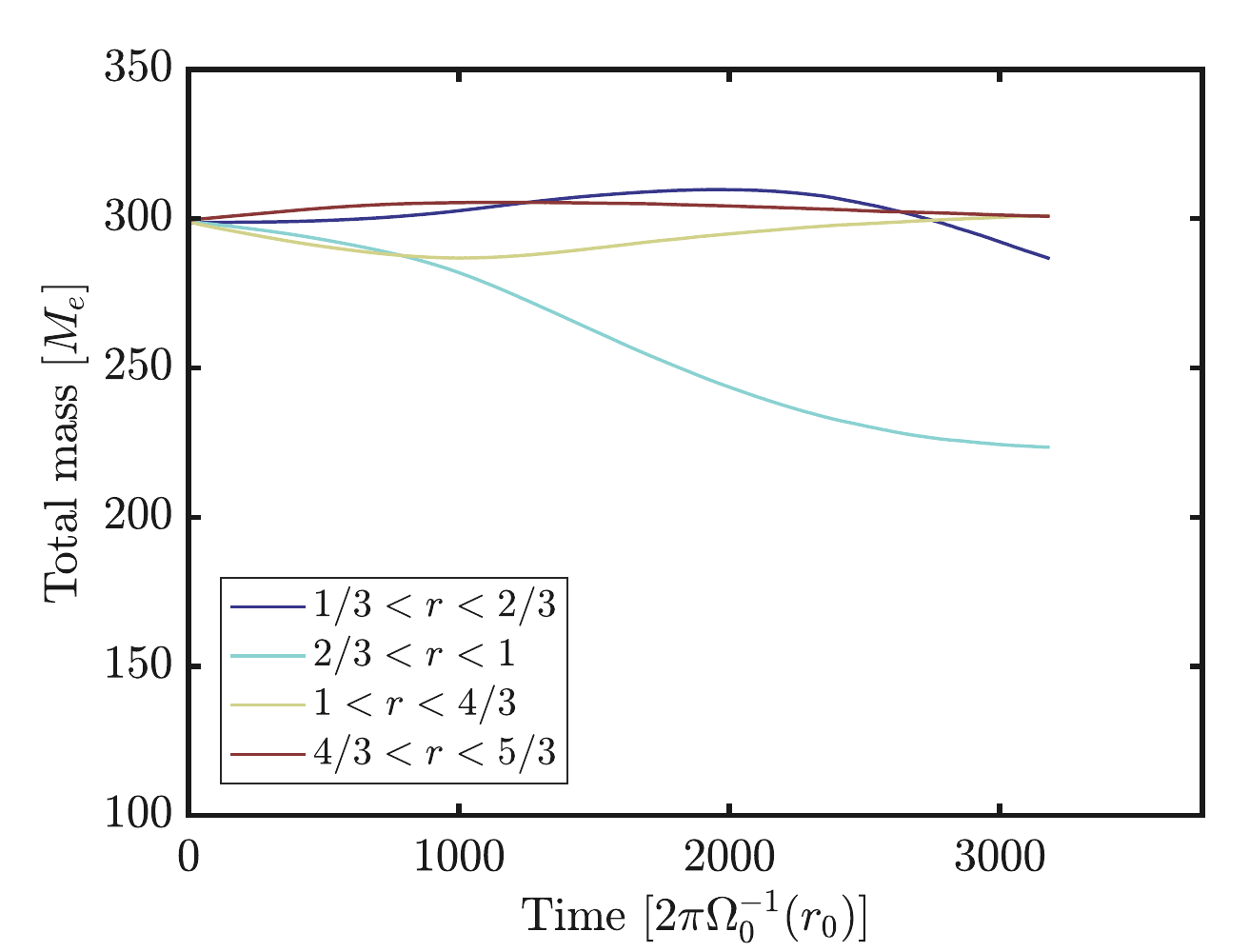}} &
	{\scriptsize{$\:$ \newline \newline (b)}} \\ 
	\imagetop{\includegraphics[height=6.5cm, trim=2mm 0mm 2mm 3mm, clip=true]{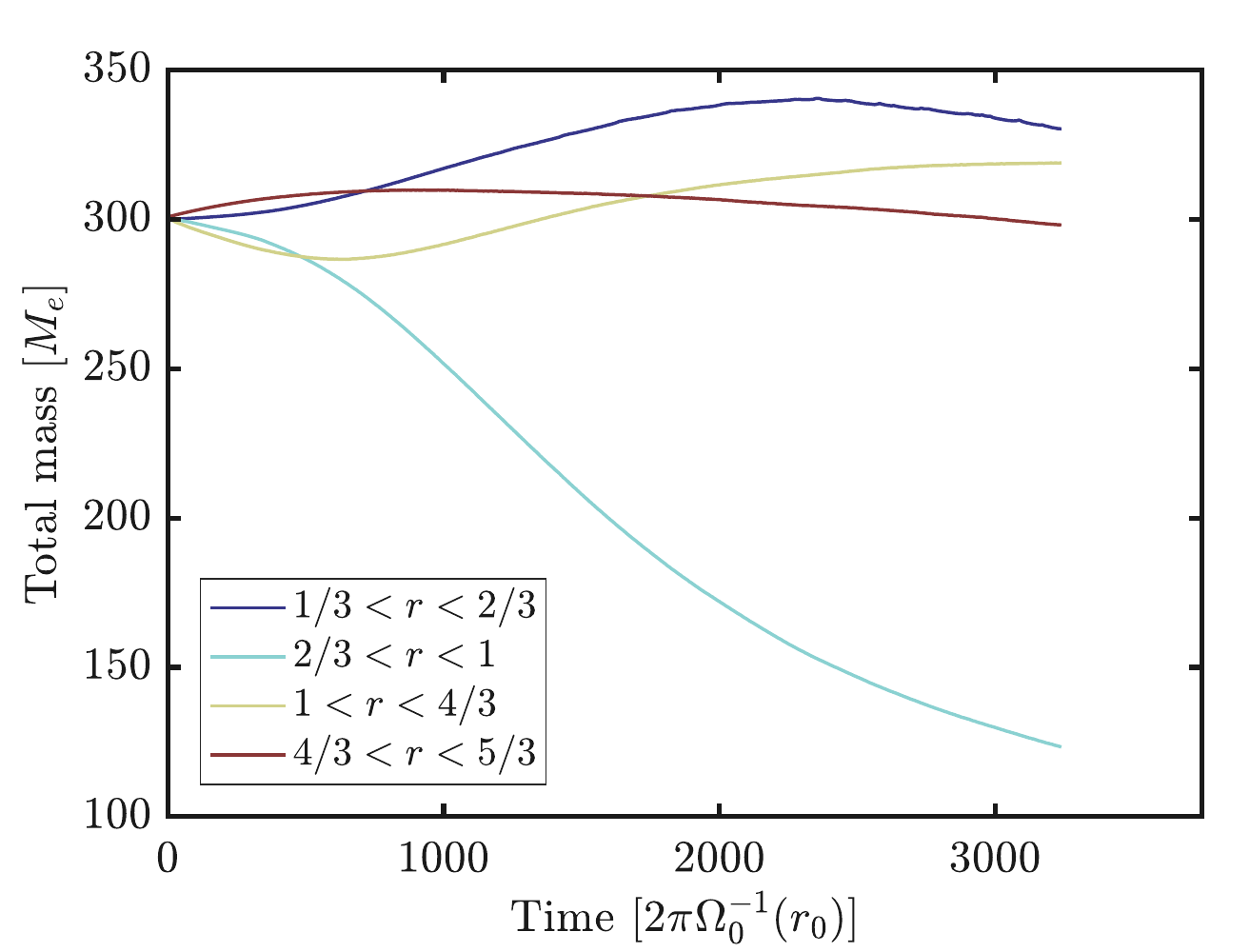}} &
	{\scriptsize{$\:$ \newline \newline (c)}} \\ 
	\end{tabular}
	
	\caption{\label{Fig_Mass_in_disk_rings_gas} Total mass of gas in different sections of the disk as function of time, for planet masses of $M_p=10$, $13$, and $20\:M_e$, (a), (b), (c), respectively. }
  \end{center}
\end{figure}

	The lower pebbles isolation mass, combined with the evolution of the gas reservoir in the vicinity of the planet observed in our simulations, has important implications for the formation of Jupiter, and massive gas giants in general. The dichotomy between classes of chondritic meteorites reservoirs in the Solar System, distinguished based on isotopic ratios, has been recently interpreted as reflecting the timing of gap opening by Jupiter, after it reaches its full mass, which would have prevented material transport and mixing between the two classes \citep{Kruijer2017}. This separation should have lasted between 1 and 3 Myr after the formation of the Solar System. 

	\cite{Alibert2018} have considered a detailed model of Jupiter formation which explores the possibility of a hybrid growth history in which an early phase of pebble accretion is followed by a later phase of planetesimal accretion. The latter introduces a slower second stage for Jupiter's accretion, which terminates after 1 Myr, a long enough timescale to match the expected timescale of separation between the two classes of meteorites (gap opening is assumed to happen as soon as the planet reaches the final mass). In this model the core of Jupiter grows efficiently above $20 \: M_e$ masses by pebble accretion, in a fraction of a Myr, then planetesimals become instrumental to heat up the envelope and slow down accretion.

	Based on our new findings the conditions for slow growth would be naturally attained by growing only slightly above 10 Earth masses, Jupiter would not undergo a prompt fast runaway gas accretion phase, even without the heating effect due to planetesimal accretion. Gas accretion could start when the core is relatively light, remaining in a slow mode for 1-2 Myr before enterring the runaway phase, a situation more reminiscent of the older version of core-accretion, as in \cite{Pollack1996}. Of course planetesimal accretion could still happen, as the hybrid mode of growth is a likely scenario. The assembly history with both pebbles and planetesimal accretion will then have to be recomputed accounting for the lower pebble isolation masses reported here. The presence of turbulent dust rings, in the outer disk region, could produce a reservoir of planetesimals for the second growth stage (about $20-30 \: M_e$ of solids, see Figure \ref{Fig_Mass_in_disk}).

	We note also that cores of $10-13 \: M_e$ appear to have a sufficient amount of gas in their vicinity to grow to Jupiter masses ($200-300 \: M_e$ even at late times, see cyan lines in Figure \ref{Fig_Mass_in_disk_rings_gas}, which refer to the annulus closest to the planet after substantial inward migration has already occurred), while by the time the core has grown to 20 Earth masses, migration torques are already strong enough to repel gas away and reduce the local gas reservoir (in the bottom panel of Figure \ref{Fig_Mass_in_disk_rings_gas} the cyan line reaches barely above 100 Earth masses). This implies it could be a problem to grow a Jupiter-mass planet out of a large core in the first place. On the other end, the latter large core would seem likely to grow into an intermediate mass planet with a larger ratio between the core and the gaseous envelope mass, more akin the ice giants in the Solar System.

	This points to the possible consequence that massive gas giants should be rarer than intermediate mass planet, which is a well known result of exoplanet surveys in slight tension with population synthesis models in standard versions of the core-accretion scenario \citep{Mordasini2018}. Furthermore, if the gas accretion phase is longer, in some cases disk dissipation, which also occurs on a few Myr timescale \citep{Alexander2014, Ercolano2015}, could reduce the gas reservoir further, which would favor even more intermediate mass planets, in the most extreme case even Super-Earths and Mini-Neptunes, as opposed to gas giants, as the natural outcome of the growth of these cores. All this being said, uncertainties exist as disk masses and density profiles in the few au region can vary a lot \citep{Mohanty2013, Andrews2013}. The masses of gas and solids we mention can vary by a factor $2-3$ depending on the type of disk.

\section{ Conclusions }
\label{Sect_Conclusions}

	We carried out a numerical study of the dynamical evolution of a planet in a protoplanetary disk of gas and pebbles, considering masses of the embryos in the range $10-20 \: M_e$, namely in the Super-Earth and Neptune-like regime, which is the most common in exoplanetary systems.
The concurrent action of the different forces at play generates structures in the gas disk as well is in its pebble component yielding new insights on the dynamical evolution of the solids in a disk, on the growth mechanisms for planetary embryos, and on planetesinal formation. We believe that the strength and originality of our study lies in its self-consistent nature. Indeed we treat all the effects resulting from the combined action of planet migration, allowing the planet to actually move through the disk, pebbles subject to drag forces, their very important back-reaction on the gas, and also radiative cooling of the gas, albeit this is included with a rather approximate method. Finally, while most of past similar studies were carried out with the FARGO code, like \cite{Bitsch2018}, our simulations were carried out with RoSSBi, which employes high order finite volume technique equipped with a well-balanced scheme, whose high accuracy is particularly suited to capture even mild perturbations resulting from the planet-disk interaction and the gas-pebbles interaction. Being a different numerical technique, it also provides the opportunity to revisit and validate previous results appeared in the literature.

	We have demonstrated the importance of the interplay between the role of the planet in changing the disk gas profile as it migrates, as well as the fine-grained motion of the gas at large distances, with the response of the pebbles to such changes, through the action of gas drag and its back reaction. This is our most important result. As we have shown, such interplay has several important consequences. These reflect several effects that are linked together by the different forces included in the model:

\begin{itemize}
\item The planetary embryo excites sound waves that drive the gravitational torque responsible for its migration (in Type-I regime most of the time given the moderate masses of the embryos in this paper). This is, of course, a known result.
\item Near the planet, a depression in the gas density is carved, even for the least massive objects, which triggers a zonal flow at $r=1.2 \: r_0$;
\item At large distances from thee planet, the wake it generates interacts in a nonlinear manner with the background disk, and triggers a pressure bump, in particular at $r=0.5 \: r_0$;
\item At these locations, dust rings form as a result of a KHI triggered by the back reaction of the drag on the gas.
\item We found a timescale of this formation as $\propto M_p^{-2}$
\end{itemize}

	The most important new physical effect that we discovered is the concurrent influence of planet migration and drag back-reaction on the disk structure. First, as the planet migrates, the favourable region of dust ring formation, i.e. the outer edge of the horseshoe region, moves inward. As a result, several dust rings can form when $M_p < 20 \: M_e$. The seperation of the multiple dust rings scales as ${\Delta r \propto 1/M_p}$, which could, in principle. be used to infer the mass of a planet from observations. Secondly, while the planet migrates inward, and after the triggering of dust rings, the drag back reaction induces a flattening of the outer part of the horseshoe region, producing a reduction of the flux of pebbles. Finally, at the inner wake region, $r = 0.5 \: r_0$, the accumulation of momentum combined with the planet migration generate a pressure bump, like a compression, which also acts as a dust trap.

	The consequences of these combined findings on planet formation models are the following:
\begin{itemize}
\item Planets of mass $<10\: M_e$ can reduce substantially, or even stop, the pebble flux through the disk at the inner wake, i.e. for regions $r< 0.5$ the orbit of the planet. This effect could help explaining the dichotomy of the chondritic meteorites between the inner and outer Solar system.
\item Pebble accretion onto the planet, namely the accretion of solids coming from the outer disk, is sustained only if $M_p< 10 \: M_e$.
\item The pebble isolation mass is expected in the range $10-13 \: M_e$, because the pebble flux stops due to the flattening of the outer gap edge.
\item The wide dust ring region that forms is a favorable place for planetesimal formation, with a large reservoir of mass ($\sim 20 \: M_e$). Here planetesimals could form via the a two-stage process involving first the streaming instability, or perhaps directly via gravitational instability, an important aspect that will have to be investigated by including self-gravity in even higher resolution studies.
\end{itemize}

	This study yields new insight on the core-accretion scenario, in particular at the crucial regime of Super-Earth-sized embryos. By simultaneously determining the pebble isolation mass and triggering planetesimal formation, an intermediate mass planetary embryo could grow in a bimdodal fashion as envisioned in the scenario of Jupiter formation proposed in \cite{Alibert2018}, in which growth occurs via both pebbles and planetesimal accretion. If planetesimal accretion becomes necessary to reach masses high enough to trigger runaway gas accretion, the growth timescale of the embryo can be substantially delayed relative to a pure pebble accretion channel, as shown by \cite{Alibert2018}. If the gas disk dissipates in the meantime, the end result will not be a gas giant, but a lower mass planet, in the Super-Earth to Neptune mass range, perhaps with a moderate gas ennvelope. This pathway should be somehow included in population synthesis models of planet formation to investigate whether it can increase substantially the fraction of intermediate mass planets and bring thus theory in closer agreement with exoplanetary observations.

	However, we should also stresse certain limitations of our work that should be overcome in the future.
First, the limit of very low turbulent viscosity favors the formation of pressure bumps, and inevitable yields a lower value of the isolation mass than with a finite $\alpha$-viscosity. However, as discussed Section \ref{Sect_Pebble_flux}, the process of gap edge flattening due to the combined action of migration and drag back-reaction, should be valid even in context of turbulent viscosity. As a result, the isolation mass range we propose should be adequate for different regimes of turbulence. Furthermore, as there is mounting observational evidence that disks, at least in the Class II stage, are very weakly turbulent, the regime that we assumed is actually quite realistic at least for the late phase of disk evolution.

	Second, as we miss to include the effect of self-gravity we can only speculate that planetesimal formation will occur in dust rings, but not prove it. Third, the radiative cooling model we adopted is quite simplified. Temperature variations due to variations in optical depth as the gas and dust density vary could play a role in the dissipation of the waves triggered by the planet, Future simulations adopting a simple radiative transfer scheme well suited for finite volume methods, such as flux limited diffusion or the M1 approximation, or the '$\kappa$-cooling' law we used in \cite{Bodenan2019}, will be needed to investigate this aspect.

	Finally, we have argued that the growth of the planetary embryo, by both pebbles and planetesimals, will be severely impacted by our findings but cannot address this important implication quantitatively. More specifically, we considered different models with a given mass of the planet, and found a possible transition of growth regime between $10$ and $13$ Earth masses. The nature of the processes we demonstrate suggests that what the growth of the planet could change drastically at this transition. Considering the difficulties to convert a realistic model of this growth into the RoSSBi code, one could implement a phenomenological growth prescription based on the Hill radius criterion as done in \cite{Kley1999}, or more recently in \cite{Robert2018}. The accretion of pebbles could also be computed with this recipe in a more self-consistant way than with a strictly semi-analytical technique. These effects will be investigated in a future publication.

\begin{acknowledgements}

	This work has been carried out within the frame of the National Center for Competence in Research {\it{PlanetS}} supported by the Swiss National Science Foundation (SNSF). The authors acknowledge the financial support of the SNSF. Numerical simulations were performed on the {\it{Piz Daint}} Cray XC30 system of the Swiss National Supercomputing Center (CSCS).

\end{acknowledgements}

\bibliographystyle{aa}
\bibliography{Biblio}

\begin{thebibliography}{53}
\expandafter\ifx\csname natexlab\endcsname\relax\def\natexlab#1{#1}\fi

\bibitem[{Alexander {et~al.}(2014)Alexander, Pascucci, Andrews, Armitage, \&
  Cieza}]{Alexander2014}
Alexander, R., Pascucci, I., Andrews, S., Armitage, P., \& Cieza, L. 2014, in
  Protostars and Planets VI, ed. H.~Beuther, R.~S. Klessen, C.~P. Dullemond, \&
  T.~Henning, 475

\bibitem[{Alibert {et~al.}(2018)Alibert, Venturini, Helled, Ataiee, Burn,
  Senecal, Benz, Mayer, Mordasini, Quanz, \&
  Sch{\"{o}}nb{\"{a}}chler}]{Alibert2018}
Alibert, Y., Venturini, J., Helled, R., {et~al.} 2018, NatAstr, 2, 873

\bibitem[{{ALMA Partnership} {et~al.}(2015){ALMA Partnership}, Brogan,
  P{\'{e}}rez, Hunter, Dent, Hales, Hills, Corder, Fomalont, Vlahakis, Asaki,
  Barkats, Hirota, Hodge, Impellizzeri, Kneissl, Liuzzo, Lucas, Marcelino,
  Matsushita, Nakanishi, Phillips, Richards, Toledo, Aladro, Broguiere, Cortes,
  Cortes, Espada, Galarza, Garcia-Appadoo, Guzman-Ramirez, Humphreys, Jung,
  Kameno, Laing, Leon, Marconi, Mignano, Nikolic, Nyman, Radiszcz, Remijan,
  Rod{\'{o}}n, Sawada, Takahashi, Tilanus, {Vila Vilaro}, Watson, Wiklind,
  Akiyama, Chapillon, de~Gregorio-Monsalvo, {Di Francesco}, Gueth, Kawamura,
  Lee, {Nguyen Luong}, Mangum, Pietu, Sanhueza, Saigo, Takakuwa, Ubach, van
  Kempen, Wootten, Castro-Carrizo, Francke, Gallardo, Garcia, Gonzalez, Hill,
  Kaminski, Kurono, Liu, Lopez, Morales, Plarre, Schieven, Testi, Videla,
  Villard, Andreani, Hibbard, \& Tatematsu}]{ALMAPartnership2015}
{ALMA Partnership}, Brogan, C., P{\'{e}}rez, L., {et~al.} 2015, ApJL, 808, L3

\bibitem[{Andrews {et~al.}(2018)Andrews, Huang, P{\'{e}}rez, Isella, Dullemond,
  Kurtovic, Guzm{\'{a}}n, Carpenter, Wilner, Zhang, Zhu, Birnstiel, Bai,
  Benisty, Hughes, {\"{O}}berg, \& Ricci}]{Andrews2018}
Andrews, S.~M., Huang, J., P{\'{e}}rez, L.~M., {et~al.} 2018, ApJL, 869, L41

\bibitem[{Andrews {et~al.}(2013)Andrews, Rosenfeld, Kraus, \&
  Wilner}]{Andrews2013}
Andrews, S.~M., Rosenfeld, K.~A., Kraus, A.~L., \& Wilner, D.~J. 2013, ApJ,
  771, 129

\bibitem[{Ataiee {et~al.}(2018)Ataiee, Baruteau, Alibert, \& Benz}]{Ataiee2018}
Ataiee, S., Baruteau, C., Alibert, Y., \& Benz, W. 2018, A{\&}A, 615, A110

\bibitem[{Ayliffe \& Bate(2011)}]{Ayliffe2011}
Ayliffe, B.~A. \& Bate, M.~R. 2011, MNRAS, 415, 576

\bibitem[{Bae {et~al.}(2017)Bae, Zhu, \& Hartmann}]{Bae2017}
Bae, J., Zhu, Z., \& Hartmann, L. 2017, ApJ, 850, 201

\bibitem[{Bitsch {et~al.}(2015)Bitsch, Lambrechts, \& Johansen}]{Bitsch2015}
Bitsch, B., Lambrechts, M., \& Johansen, A. 2015, A{\&}A, 582, A112

\bibitem[{Bitsch {et~al.}(2018)Bitsch, Morbidelli, Johansen, Lega, Lambrechts,
  \& Crida}]{Bitsch2018}
Bitsch, B., Morbidelli, A., Johansen, A., {et~al.} 2018, A{\&}A, 612, A30

\bibitem[{Bod{\'{e}}nan {et~al.}(2019)Bod{\'{e}}nan, Surville, Szul{\'{a}}gyi,
  Mayer, \& Sch{\"{o}}nb{\"{a}}chler}]{Bodenan2019}
Bod{\'{e}}nan, J.-D., Surville, C., Szul{\'{a}}gyi, J., Mayer, L., \&
  Sch{\"{o}}nb{\"{a}}chler, M. 2019, arXiv e-prints, arXiv:1912.09732

\bibitem[{Carrera {et~al.}(2020)Carrera, Simon, Li, Kretke, \&
  Klahr}]{Carrera2020}
Carrera, D., Simon, J.~B., Li, R., Kretke, K.~A., \& Klahr, H. 2020, arXiv
  e-prints, arXiv:2008.01727

\bibitem[{Crida {et~al.}(2006)Crida, Morbidelli, \& Masset}]{Crida2006}
Crida, A., Morbidelli, A., \& Masset, F. 2006, Icarus, 181, 587

\bibitem[{Dong {et~al.}(2017)Dong, Li, Chiang, \& Li}]{Dong2017}
Dong, R., Li, S., Chiang, E., \& Li, H. 2017, ApJ, 843, 127

\bibitem[{Dong {et~al.}(2018)Dong, Li, Chiang, \& Li}]{Dong2018}
Dong, R., Li, S., Chiang, E., \& Li, H. 2018, ApJ, 866, 110

\bibitem[{Dong {et~al.}(2011)Dong, Rafikov, Stone, \& Petrovich}]{Dong2011a}
Dong, R., Rafikov, R.~R., Stone, J.~M., \& Petrovich, C. 2011, ApJ, 741, 56

\bibitem[{Engler {et~al.}(2020)Engler, Lazzoni, Gratton, Milli, Schmid,
  Chauvin, Kral, Pawellek, Th{\'{e}}bault, Boccaletti, Bonnefoy, Brown, Buey,
  Cantalloube, Carle, Cheetham, Desidera, Feldt, Ginski, Gisler, Henning,
  Hunziker, Lagrange, Langlois, Mesa, Meyer, Moeller-Nilsson, Olofsson, Petit,
  Petrus, Quanz, Rickman, Stadler, Stolker, Vigan, Wildi, \&
  Zurlo}]{Engler2020}
Engler, N., Lazzoni, C., Gratton, R., {et~al.} 2020, A{\&}A, 635, A19

\bibitem[{Ercolano \& Rosotti(2015)}]{Ercolano2015}
Ercolano, B. \& Rosotti, G. 2015, MNRAS, 450, 3008

\bibitem[{Garufi {et~al.}(2014)Garufi, Quanz, Schmid, Avenhaus, Buenzli, \&
  Wolf}]{Garufi2014}
Garufi, A., Quanz, S., Schmid, H., {et~al.} 2014, A{\&}A, 568, A40

\bibitem[{Gomes {et~al.}(2015)Gomes, Klahr, Uribe, Pinilla, \&
  Surville}]{Gomes2015}
Gomes, A.~L., Klahr, H., Uribe, A.~L., Pinilla, P., \& Surville, C. 2015, The
  Astrophysical Journal, 810, 94

\bibitem[{Huang {et~al.}(2020)Huang, Li, Isella, Miranda, Li, \&
  Ji}]{Huang2020}
Huang, P., Li, H., Isella, A., {et~al.} 2020, ApJ, 893, 89

\bibitem[{Kanagawa {et~al.}(2015)Kanagawa, Tanaka, Muto, Tanigawa, \&
  Takeuchi}]{Kanagawa2015}
Kanagawa, K., Tanaka, H., Muto, T., Tanigawa, T., \& Takeuchi, T. 2015, MNRAS,
  448, 994

\bibitem[{Klahr \& Hubbard(2014)}]{Klahr2014}
Klahr, H. \& Hubbard, A. 2014, ApJ, 788, 21

\bibitem[{Kley(1999)}]{Kley1999}
Kley, W. 1999, MNRAS, 303, 696

\bibitem[{Kley \& Nelson(2012)}]{Kley2012}
Kley, W. \& Nelson, R. 2012, ARA{\&}A, 50, 211

\bibitem[{Kruijer {et~al.}(2017)Kruijer, Burkhardt, Budde, \&
  Kleine}]{Kruijer2017}
Kruijer, T.~S., Burkhardt, C., Budde, G., \& Kleine, T. 2017, PNAS, 114, 6712

\bibitem[{Lambrechts \& Johansen(2012)}]{Lambrechts2012}
Lambrechts, M. \& Johansen, A. 2012, A{\&}A, 544, A32

\bibitem[{Lambrechts \& Johansen(2014)}]{Lambrechts2014a}
Lambrechts, M. \& Johansen, A. 2014, A{\&}A, 572, A107

\bibitem[{Lambrechts {et~al.}(2014)Lambrechts, Johansen, \&
  Morbidelli}]{Lambrechts2014}
Lambrechts, M., Johansen, A., \& Morbidelli, A. 2014, A{\&}A, 572, A35

\bibitem[{Leya {et~al.}(2008)Leya, Sch{\"{o}}nb{\"{a}}chler, Wiechert,
  Kr{\"{a}}henb{\"{u}}hl, \& Halliday}]{Leya2008}
Leya, I., Sch{\"{o}}nb{\"{a}}chler, M., Wiechert, U., Kr{\"{a}}henb{\"{u}}hl,
  U., \& Halliday, A.~N. 2008, E{\&}PSL, 266, 233

\bibitem[{Li {et~al.}(2009)Li, Lubow, Li, \& Lin}]{Li2009}
Li, H., Lubow, S., Li, S., \& Lin, D. 2009, ApJL, 690, L52

\bibitem[{Lin \& Papaloizou(1986)}]{Lin1986}
Lin, D. \& Papaloizou, J. 1986, ApJ, 309, 846

\bibitem[{Lyra \& Klahr(2011)}]{Lyra2011a}
Lyra, W. \& Klahr, H. 2011, A{\&}A, 527, A138

\bibitem[{Lyra {et~al.}(2016)Lyra, Richert, Boley, Turner, {Mac Low}, Okuzumi,
  \& Flock}]{Lyra2016}
Lyra, W., Richert, A.~J., Boley, A., {et~al.} 2016, ApJ, 817, 102

\bibitem[{Malik {et~al.}(2015)Malik, Meru, Mayer, \& Meyer}]{Malik2015}
Malik, M., Meru, F., Mayer, L., \& Meyer, M. 2015, ApJ, 802, 56

\bibitem[{Meru {et~al.}(2019)Meru, Rosotti, Booth, Nazari, \&
  Clarke}]{Meru2019}
Meru, F., Rosotti, G.~P., Booth, R.~A., Nazari, P., \& Clarke, C.~J. 2019,
  MNRAS, 482, 3678

\bibitem[{Mohanty {et~al.}(2013)Mohanty, Greaves, Mortlock, Pascucci, Scholz,
  Thompson, Apai, Lodato, \& Looper}]{Mohanty2013}
Mohanty, S., Greaves, J., Mortlock, D., {et~al.} 2013, ApJ, 773, 168

\bibitem[{Mordasini(2018)}]{Mordasini2018}
Mordasini, C. 2018, in Handbook of Exoplanets, ed. H.~J. Deeg \& J.~A. Belmonte
  (Cham: Springer International Publishing), 2425--2474

\bibitem[{Muro-Arena {et~al.}(2020)Muro-Arena, Benisty, Ginski, Dominik,
  Facchini, Villenave, van Boekel, Chauvin, Garufi, Henning, Janson, Keppler,
  Matter, M{\'{e}}nard, Stolker, Zurlo, Blanchard, Maurel, Moeller-Nilsson,
  Petit, Roux, Sevin, \& Wildi}]{Muro-Arena2020}
Muro-Arena, G., Benisty, M., Ginski, C., {et~al.} 2020, A{\&}A, 635, A121

\bibitem[{Ogilvie \& Lubow(2002)}]{Ogilvie2002}
Ogilvie, G. \& Lubow, S. 2002, MNRAS, 330, 950

\bibitem[{Paardekooper(2014)}]{Paardekooper2014}
Paardekooper, S.~J. 2014, MNRAS, 444, 2031

\bibitem[{Paardekooper {et~al.}(2010)Paardekooper, Baruteau, Crida, \&
  Kley}]{Paardekooper2010a}
Paardekooper, S.~J., Baruteau, C., Crida, A., \& Kley, W. 2010, MNRAS, 401,
  1950

\bibitem[{Pollack {et~al.}(1996)Pollack, Hubickyj, Bodenheimer, Lissauer,
  Podolak, \& Greenzweig}]{Pollack1996}
Pollack, J.~B., Hubickyj, O., Bodenheimer, P., {et~al.} 1996, Icarus, 124, 62

\bibitem[{Rafikov(2002)}]{Rafikov2002}
Rafikov, R. 2002, ApJ, 569, 997

\bibitem[{Richert {et~al.}(2015)Richert, Lyra, Boley, Low, \&
  Turner}]{Richert2015}
Richert, A. J.~W., Lyra, W., Boley, A., Low, M.-M.~M., \& Turner, N. 2015, ApJ,
  804, 95

\bibitem[{Robert {et~al.}(2018)Robert, Crida, Lega, M{\'{e}}heut, \&
  Morbidelli}]{Robert2018}
Robert, C., Crida, A., Lega, E., M{\'{e}}heut, H., \& Morbidelli, A. 2018,
  A{\&}A, 617, A98

\bibitem[{Surville \& Mayer(2019)}]{Surville2019}
Surville, C. \& Mayer, L. 2019, ApJ, 883, 176

\bibitem[{Surville {et~al.}(2016)Surville, Mayer, \& Lin}]{Surville2016}
Surville, C., Mayer, L., \& Lin, D. N.~C. 2016, ApJ, 831, 82

\bibitem[{Takeuchi {et~al.}(1996)Takeuchi, Miyama, \& Lin}]{Takeuchi1996}
Takeuchi, T., Miyama, S.~M., \& Lin, D. 1996, ApJ, 460, 832

\bibitem[{Tanaka {et~al.}(2002)Tanaka, Takeuchi, \& Ward}]{Tanaka2002}
Tanaka, H., Takeuchi, T., \& Ward, W.~R. 2002, ApJ, 565, 1257

\bibitem[{Trinquier {et~al.}(2007)Trinquier, Birck, \&
  All{\`{e}}gre}]{Trinquier2007}
Trinquier, A., Birck, J.-L., \& All{\`{e}}gre, C.~J. 2007, ApJ, 655, 1179

\bibitem[{Yang \& Zhu(2020)}]{Yang2020}
Yang, C.-C. \& Zhu, Z. 2020, MNRAS, 491, 4702

\bibitem[{Zhu {et~al.}(2018)Zhu, Petrovich, Wu, Dong, \& Xie}]{Zhu2018}
Zhu, W., Petrovich, C., Wu, Y., Dong, S., \& Xie, J. 2018, ApJ, 860, 101

\end{thebibliography}

\end{document}